\documentclass[aps,pra,superscriptaddress,nofootinbib]{revtex4}
\pdfoutput=1
\usepackage[intlimits]{amsmath}
\usepackage{amsfonts}

\usepackage{psfrag}
\usepackage{array}

\usepackage{float}
\usepackage{tikz}

\advance\voffset by -1cm
\advance\hoffset by -0.5cm
\newcommand\beq            {\begin{equation}}
\newcommand\be            {\begin{equation}}
\newcommand\bea           {\begin{equation}\begin{array}l\displaystyle}
\newcommand\ee            {\end{equation}}
\newcommand\eeq            {\end{equation}}
\newcommand\bes           {\begin{subequations}}
\newcommand\esu           {\end{subequations}}

\renewcommand{\[}{\left[}

\newcommand{\bigx}[1]{\bBigg@{#1}}

\def\3pt#1#2#3{{\langle{#1}\vert{#2}\vert{#3}\rangle}}

\newcommand\doi[2]        {\href{http://dx.doi.org/#1}{#2}}

\newcommand{\EQ}{\begin{equation}}
\newcommand{\EN}{\end{equation}}
\usepackage{epsfig}
\usepackage{psfrag}
\usepackage{amsmath}
\usepackage{graphicx}
\usepackage{amsfonts}
\usepackage{amssymb}

\begin{document}
\bibliographystyle{plainnat}

\title{{\Large {\bf Yang-Lee Zeros of the Yang-Lee Model}}}
%{\large {\em Integrable Quantum Field Theories}}}
\author{G. Mussardo}
\affiliation{SISSA and INFN, Sezione di Trieste, via Bonomea 265, I-34136, 
Trieste, Italy}
\affiliation{International Centre for Theoretical Physics (ICTP), I-34151, Trieste, Italy}
\author{R. Bonsignori}
\affiliation{SISSA and INFN, Sezione di Trieste, via Bonomea 265, I-34136, 
Trieste, Italy}
%\affiliation{International Institute of Physics, Natal, Brasil} 
\author{A. Trombettoni}
\affiliation{CNR-IOM DEMOCRITOS Simulation Center, Via Bonomea 265, I-34136
Trieste, Italy} 
\affiliation{SISSA and INFN, Sezione di Trieste, via Bonomea 265, I-34136, 
Trieste, Italy}

\begin{abstract}
\noindent
To understand the distribution of the Yang-Lee zeros in quantum integrable field theories we analyse the simplest of these systems 
given by the two-dimensional Yang-Lee model. The grand-canonical partition function of this quantum field theory, as a function of the fugacity $z$ and 
the inverse temperature $\beta$, can be computed in terms of the Thermodynamics Bethe Ansatz based on its exact $S$-matrix. We extract the Yang-Lee zeros in the complex plane by using a sequence of polynomials of increasing order $N$  in $z$ which converges to the grand-canonical partition function. We show that these zeros are distributed along curves which are approximate circles as it is also the case of the zeros for purely free theories. There is though an important difference between the interactive theory and the free theories, for the radius of the zeros in the interactive theory goes continuously to zero 
in the high-temperature limit $\beta \rightarrow 0$ while in the free theories it remains close to 1 even for small values of $\beta$, jumping to 0 only at $\beta = 0$.

\vspace{3mm}
\noindent
Pacs numbers: 11.10.St, 11.15.Kc, 11.30.Pb

\end{abstract}
\maketitle

\section{Introduction}
\noindent
Many physical quantities reveal their deeper structure by going to the complex plane. This is the case, for instance, of the analytic properties of the scattering amplitudes where the angular momentum is not longer restricted to be an integer but allowed to take any complex value giving rise in this way to the famous Regge poles \citep{Regge}. Another famous example is the Yang-Lee theory of equilibrium phase transitions \citep{YL1,YL2} based on the zeros of the grand-canonical partition function in the complex plane of the fugacity: in a nutshell, the main observation of Yang and Lee was that the zeros of the grand-canonical partition functions in the thermodynamic limit usually accumulate in certain regions or curves of the complex plane, with their positive local density $\eta(z)$ which 
changes by changing the temperature; if at a critical value $T_c$, the zeros accumulate and pinch a positive value of the real axis, this is what may mark the onset of a phase transition. 

\begin{figure}[t]
\begin{center}
%\vspace{8mm}
\psfig{figure=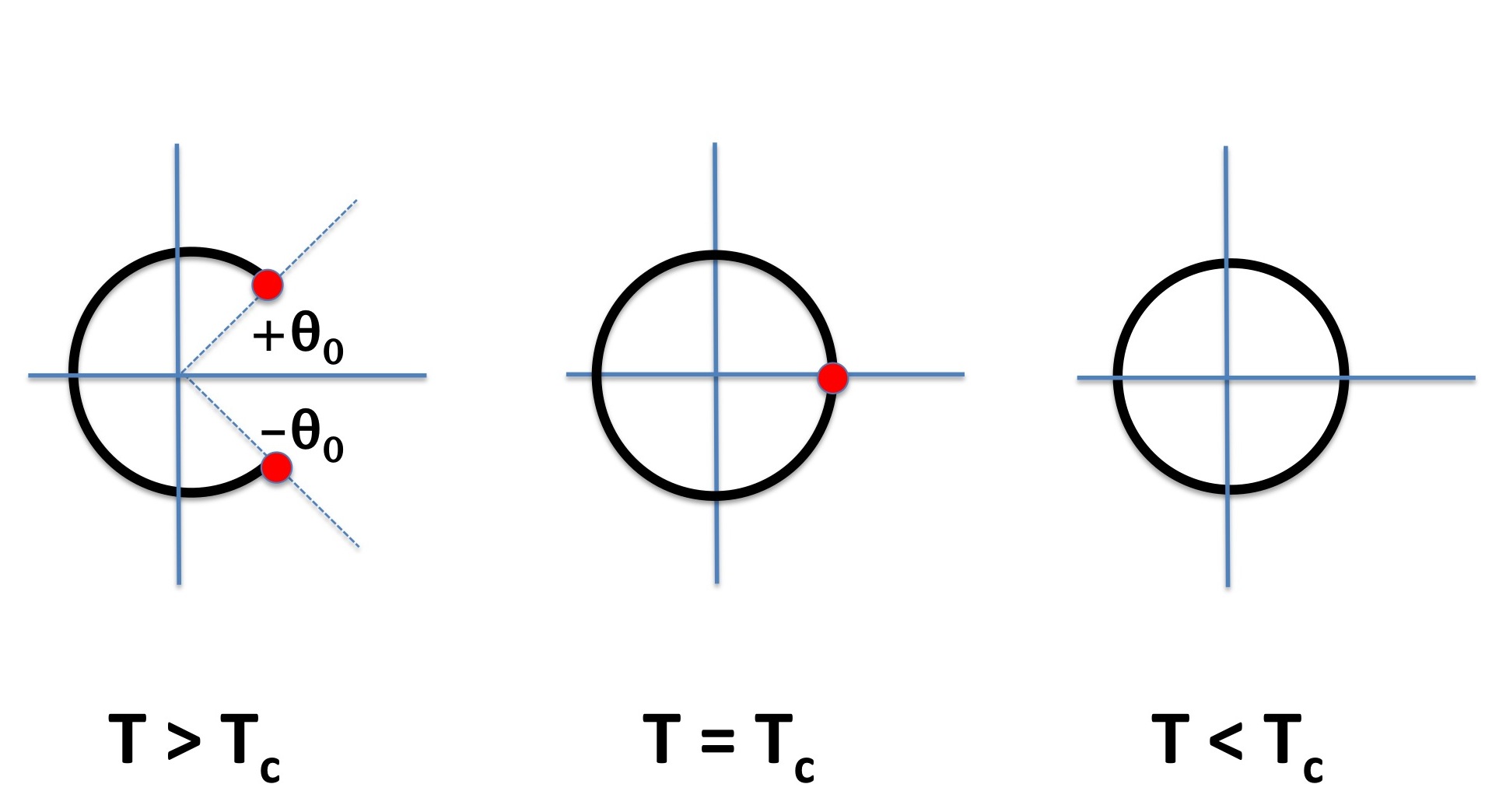,height=6cm,width=11cm}
\caption{Distribution of the Yang-Lee zeros for the Ising model in the complex plane of the fugacity $z$.}
\label{circleYL}
\end{center}
\end{figure}

As we are going to discuss extensively through the rest of the paper, the pattern of zeros of grand-canonical partition functions can be generally quite interesting and this study alone is a source of many stimulating physical and mathematical questions. If the study of the patterns of Yang-Lee zeros is then the first topic of this paper, the Yang-Lee model (and its zeros!) is our second main topic. In order to introduce such a model and present the work of this paper in its proper perspective, we need to talk about the pattern of zeros of just one particular statistical system: the Ising model. In ref.\,\citep{YL2} Yang and Lee showed that for ferromagnetic Ising-like models, independently on the dimensionality and regularity of the lattice and also largely independently on the nature of the couplings, the zeros of the Ising model lie on the unit circle\footnote{This circle-theorem was later extended by many authors to ferromagnetic Ising model of arbitrarily high spin and with many-body spin interactions \citep{Asano, Suzuki,Griffiths,SuzukiFisher}.}  in the complex plane of the variable $z = e^{-2 \beta h}$ (where $\beta = 1/(k T)$ and $h$ is the external magnetic field): posing $z = e^{i \theta}$, they have the following structure (see Figure \ref{circleYL})
\begin{itemize}
\item for $T > T_c$ the zeros are placed along a "C", namely a symmetric arc around $\theta = \pi$ whose edges are at $\pm \theta_0(T)$; 
\item at $T=T_c$ these edges move to the real axis and pinch it;
\item for $T< T_c$ the zeros densely cover the entire circle. 
\end{itemize} 

\noindent
Kortman and Griffiths \citep{KortmanGriffiths} were the first to notice that the density of the Yang-Lee zeros of the Ising model nearby the edges $\pm \theta_0(T)$ 
gives rise to a problem which has its own interest since such a density presents an anomalous behaviour with a scaling law ruled by a critical exponent $\sigma$  
\be
\eta(\theta,T) \sim |\theta - \theta_0(T)|^\sigma 
\,\,\,\,\,\,
,
\,\,\,\,\,\,
T > T_c  \,\,\,.
\label{YLedgeexponent} 
\ee 
Such a behavior is closely analogous to the usual critical phenomena (although in this case triggered by a purely imaginary magnetic field $i h$) and therefore  Fisher \citep{FisherEdge} posed the question about its effective quantum field theory and argued that, in sufficiently high dimension $d$, this 
consists of a $\phi^3$ Landau-Ginzburg theory for the scalar field $\phi(x)$ with euclidean action given by 
\be 
{\cal A} \,=\, \int d^dx \left[\frac{1}{2} (\partial \phi)^2 +  i (h-h_0) \phi + i g\,\phi^3 \right] \,\,\,.
\label{YLaction}
\ee
This action is what defines the Yang-Lee model, i.e. the quantum field theory studied in this paper, and the imaginary couplings present in such an action is what makes the Yang-Lee model a non-hermitian theory; the theory is however invariant under a CP transformations \citep{CM,Bender} and therefore its spectrum is real. At criticality, i.e. when $h = h_0$, the corresponding fixed point of the Renormalization Group presents only one relevant operator, namely the field $\phi(x)$ itself: in two dimensions Cardy showed \citep{CardyYL} that all these properties are encoded into the simplest non-unitarity minimal model of Conformal Field Theories which has central charge $c = - 22/5$ and only one relevant operator of conformal dimension $\Delta = - 1/5$. Still in two dimensions, Cardy and Mussardo \citep{CM} then argued that the action (\ref{YLaction}), regarded as deformation of this minimal conformal model by means of its relevant operator $\phi(x)$ corresponds to an integrable quantum field theory. This fact has far-reaching consequences: indeed, the infinite number of conservation laws -- the fingerprint of any integrable field theory \citep{ZamZam,Zam1,GM} -- implies that scattering processes of the Yang-Lee model are completely elastic and factorizable in terms of the two-body scattering matrix which was exactly computed in \citep{CM}. The spectrum can be easily extracted by the poles of the $S$-matrix and turns out to consist of only one massive particle which may be regarded as a bound state of itself. Exact Form Factors and two-point correlation functions of this massive model were later computed  and discussed in \citep{ZamFFYL}.  

Based on the $S$-matrix scattering theory and the Thermodynamic Bethe Ansatz introduced by Al.B. Zamolodchikov \citep{ZamTBA}, one can then in principle determine exactly (although numerically) the free-energy of the integrable models as function of the inverse temperature $\beta$ and the fugacity $z$ \citep{Fendley,KM}. This is indeed what we have done in this paper for the Yang-Lee model, i.e. one of the simplest representative of integrable quantum field theories, for the purpose of studying what kind of distribution emerges for the zeros of the grand-canonical partition function of these systems. We hope these brief introductory considerations were useful to clarify the general aims of this paper and the "recursive" use of Yang-Lee names, both to denote the zeros and the 
eponymous model, and the zeros of the model itself! 

A comment is in order on the logic of this work. In the Yang-Lee theory of phase transitions one relates thermodynamical properties, such the free energy 
and the magnetization, to the density of Yang-Lee zeros $\eta(z)$. Usually the focus is on the (often approximate) determination of $\eta(z)$ in order then  
to extract the equilibrium quantities of interest. However, the point of view we adopt to deal with integrable quantum field theories as, for instance the 
Yang-Lee model, is rather the opposite, since the free energy of these models are already known in terms of the Thermdoynamics Bethe Ansatz equations, 
and this gives the possibility to determine the properties of the zeros. We have applied such a procedure to the Yang-Lee model but it is clear that it 
can be applied as well to other integrable quantum field theories.
  
The rest of the paper is organised as follows. Section \ref{seconda} is devoted to a brief recap of the main points of the Yang-Lee theory of the phase transition and the importance to control the distribution of the zeros of the grand-canonical partition function in order to study the thermodynamics. Section \ref{SectionZeros} contains a detailed discussion on the nature of the roots of polynomials which may be regarded as grand-canonical partition functions of some appropriate physical system. In Section \ref{quarta} we present the closed formulas of the partition functions of free bosonic and fermionic theories, both in the non-relativistic and relativistic case, and also in the presence of an harmonic trap. Sections \ref{quinta} and \ref{freerel} contain a detailed discussion of the zeros of the grand-canonical partition functions of the non-relativistic and relativistic free theories respectively: apart from some peculiar features emerging from this study, the basic 
purpose of these Sections is to set the stage for the analysis of the interacting integrable models discussed in the later Sections. In particular, in Section \ref{sesta} we start recalling the basic properties of integrable quantum field theories, namely the $S$-matrix formulation and the Thermodynamics Bethe Ansatz which allows us to recover the partition function of the integrable models. In Section \ref{settima} we study in detail the Yang-Lee zeros of the simplest quantum integrable field theories, namely the Yang-Lee model.  Our conclusions are finally collected in Section \ref{finale}.   

\section{Yang-Lee theory of phase transitions}\label{seconda}
\noindent
In order to overcome some inadequacies of the Mayer method \citep{Mayer} for dealing with the condensation of a gas and also to understand better the underlying mathematical reasons behind the occurrence of phase transitions, in 1952 Yang and Lee proposed to analytically continue the grand-canonical partition function to the complex plane of its fugacity and determine the pattern of zeros of this function. Although only real values of the fugacity determine the physical value of the pressure, the magnetization or other relevant thermodynamical quantities, the overall analytic behavior of these observables can only be understood by looking at how the zeros move as a function of an external parameter such as the temperature. In this Section we are going to simply write down, mainly for future reference, 
the basic formulas of the Yang-Lee formalism with few extra comments. 

Concerning general references on this topic, in addition to the original papers \citep{YL1,YL2} and others previously mentioned \citep{Asano,Suzuki,Griffiths,SuzukiFisher,KortmanGriffiths,FisherEdge,CM,CardyYL}, the reader may also 
benefit of some standard books \citep{Huang,Mattis} or reviews such as \citep{Bena,Blythe} and references therein. As a matter of fact, the literature on the subject is immense, even ranging across several fields of physics and mathematics. For this reason we cannot definitely do justice to all authors who contributed to the development of the subject but we would like nevertheless to explicitly mention few more references which we have found particularly useful, such as the series of papers by Ikeda \citep{Ikeda1,IkedaF,IkedaB} or by Abe \citep{Abe1,Abe2}, the papers by Katsura on some analytic expressions of the density of zeros \citep{Suzuki}, the paper by Fonseca and Zamolodchikov \citep{FonsecaZam} on the analytic properties of the free energy of the Ising field theory and some related references on this subject such as \citep{ZamTBA,YurovZam,Mossa,Wydro,Jacobsen,Ana}, and finally some references on the experimental observations of the Yang-Lee zeros \citep{Binek}, in particular those based on the coherence of a quantum spin coupled to an Ising-type thermal bath \citep{Wei,Peng}.
 
\vspace{3mm}
\noindent
{\bf Yang-Lee formulation.} Consider for simplicity the grand-canonical partition function $\Omega_N(z)$ of a gas made of $N$ particles with hard cores $b$ in a volume $V$, of activity $z$ and at temperature $T$ is given by 
\be 
\Omega_N(z) \,=\ \sum_{k=0}^N \frac{1}{k!} Z_k(V,T) z^k \,=\,\prod_{l=1}^N \left(1 - \frac{z}{z_l}\right) 
\,\,\,,
\label{basicdefinitionZ}
\ee 
where $N = V/b$ is the largest number of particles that can be contained in the volume, the coefficients $Z_k(V,T)$ are the canonical partition functions of a system of $k$ particles and the $z$ is the fugacity. As a polynomial of order $N$, $\Omega_N(z)$ has  $z_1, z_2, \ldots z_N$ zeros in the complex plane. 
The thermodynamics of the system is recovered by defining, in the limit $V \rightarrow \infty$, the pressure $p(z)$ and the density $\rho(z)$ of the system 
as 
%\begin{eqnarray}
%&& \frac{p(z)}{k T} \,\equiv \, \hat f(z) \,=\, \lim_{V \rightarrow \infty} \frac{1}{V} \,\log  \Omega_N(z) \,\,\,, 
%\label{pressurezeros}\\
%&& 
%\rho(z) \,\equiv\, z \hat f'(z) \,=\, \lim_{V \rightarrow \infty} \frac{1}{V} z \frac{d}{d z}\,\log  \Omega_N(z) \,\,\,.
%\label{densityzeros}
%\end{eqnarray}
\begin{eqnarray}
&& \frac{p(z)}{k T} \,\equiv \, \hat f(z) \,=\,\lim_{V \rightarrow \infty} \frac{1}{V} \,\log  \Omega_N(z) \,=\, \lim_{V,N \rightarrow \infty} \frac{1}{V} \sum_{l=1}^N\,\log\left(1 - \frac{z}{z_l}\right)  \,\,\,, 
\label{pressurezeros}\\
&& 
\rho(z) \,\equiv\, z \hat f'(z) \,=\, \lim_{V \rightarrow \infty} \frac{1}{V} z \frac{d}{d z}\,\log  \Omega_N(z) \,=\, 
\lim_{V,N \rightarrow \infty} \frac{1}{V} \sum_{l=1}^N\,\frac{z}{z -z_l} \,\,\,.
\label{densityzeros}
\end{eqnarray}
For extended systems the limit $V \rightarrow \infty$ also enforces $N \rightarrow \infty$ and therefore there will be an infinite number 
of zeros: these may become densely distributed in the complex plane according to their positive density function $\eta(z)$, which can be 
different from zero either in a region ${\cal A}$ of the complex plane (such a situation we will call later {\em area law} for the zeros) or 
along a curve, ${\cal C}$ (to which we refer to as a {\em perimeter law}). Apart from these two generic cases, it can also be that the zeros may 
remain isolated points in the complex plane or, rather pathologically, accumulated instead around single points.

As shown in the original papers by Yang and Lee \citep{YL1,YL2}, the entire thermodynamics can be recovered in terms of the density $\eta(z)$ of the zeros of the grand-canonical partition function. Notice that, for the reality of the original $\Omega_N(z)$, this function must satisfy the property 
\be 
\eta(z) \,=\, \eta(z^*) \,\,\,, 
\label{realityeta}
\ee
i.e. must be symmetric with respect to the real axis. The region ${\cal A}$ or the curve ${\cal C}$ depend on the temperature $T$ 
and they change their shape by changing $T$. Let's initially assume that the zeros are placed on an extended area ${\cal A}$: 
in this case, for all points outside this region one can analytically extend the definition of the pressure $p(z)$ as 
\be
\frac{p(z)}{k T} \, =\, \int_{\,\,\,\cal A} d\xi \, \eta(\xi)\,\log\left(1-\frac{z}{\xi}\right) \,\,\,. 
\ee 
We can split this function into its real and imaginary part 
\be
\frac{p(z)}{k T} \equiv P(z) \,=\, \varphi(z) + i \psi(z) \,\,\,, 
\ee
where its real part 
\be
\varphi(z) \,=\, \int d\xi \, \eta(\xi)\,\log\left| 1-\frac{z}{\xi}\right| \ \,\,\,
\ee
involves $\log |z|$, i.e. the Green function of the two-dimensional Laplacian operator $\Delta$. Therefore $\varphi(z)$ satisfies the Poisson equation 
\be
\Delta \varphi(z) \,=\, 2 \pi \, \eta(z) \,\,\,. 
\label{Poisson}
\ee
Drawing an electromagnetic analogy, this equation implies that $\varphi(z)$ can be thought as the electrostatic potential generated by the 
the (positive) distribution of charges with density $\eta(z)$. Posing $z = x + i y$, the corresponding components of the electric field are given by 
\be 
E_1 = - \frac{\partial\varphi}{\partial x} 
\,\,\,\,\,
,
\,\,\,\,\,
E_2 = - \frac{\partial\varphi}{\partial y}
\,\,\,.
\ee
Since in any region not occupied by the charges both $\varphi(z)$ and its companion $\psi(z)$ are analytic functions related by the Cauchy-Riemann equations, 
we have 
\be 
\frac{d P}{dz} \,=\, \frac{d\varphi}{d x} + i \frac{d \psi}{d x} =  \frac{d\varphi}{d x} - i \frac{d \psi}{d y}
\,=\, - \overline{E} \,\,\,,
\ee
where $E =E_1 + i E_2$ is the complex electric field ($ \overline{E} = E_1 - i E_2$). 

\begin{figure}[t]
\centering
\,\,\,\includegraphics[width=0.55\textwidth]{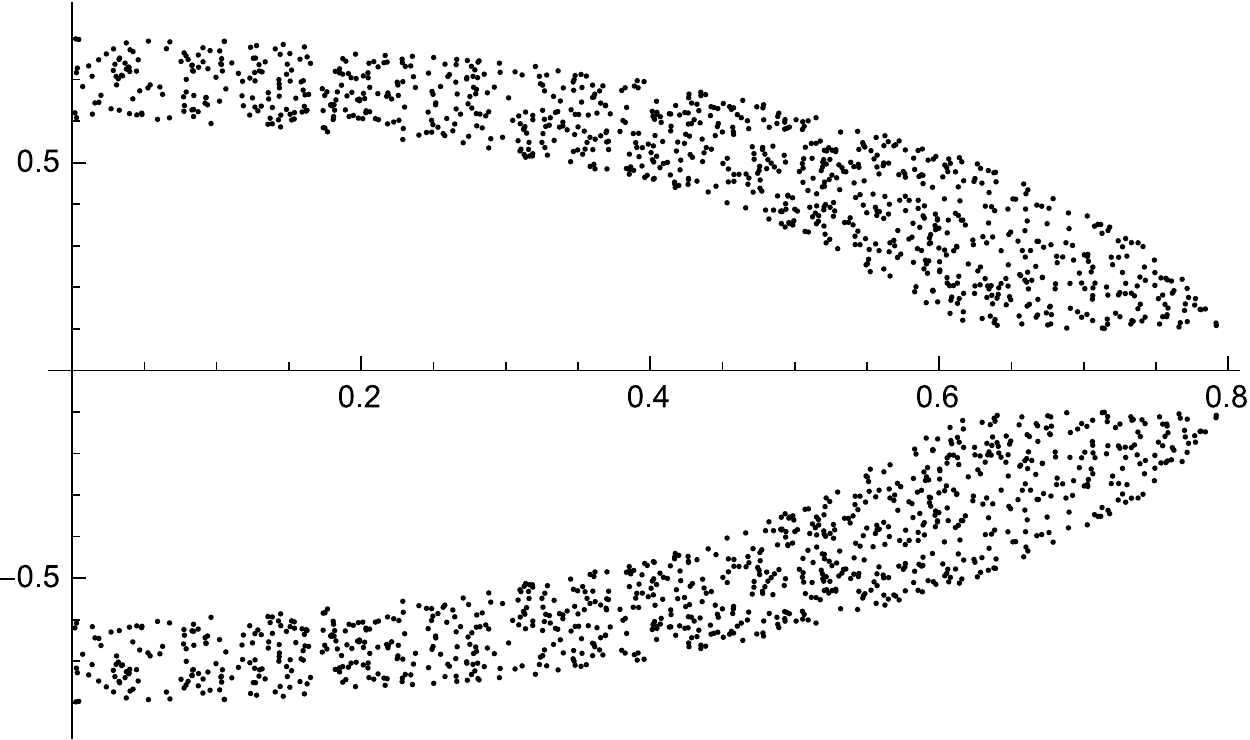}
\,\,\,\,\,\,\,\,\,\,\,\,\,\,\,\,\,\,\,\,\,\,\,

\vspace{3mm}

\includegraphics[width=0.45\textwidth]{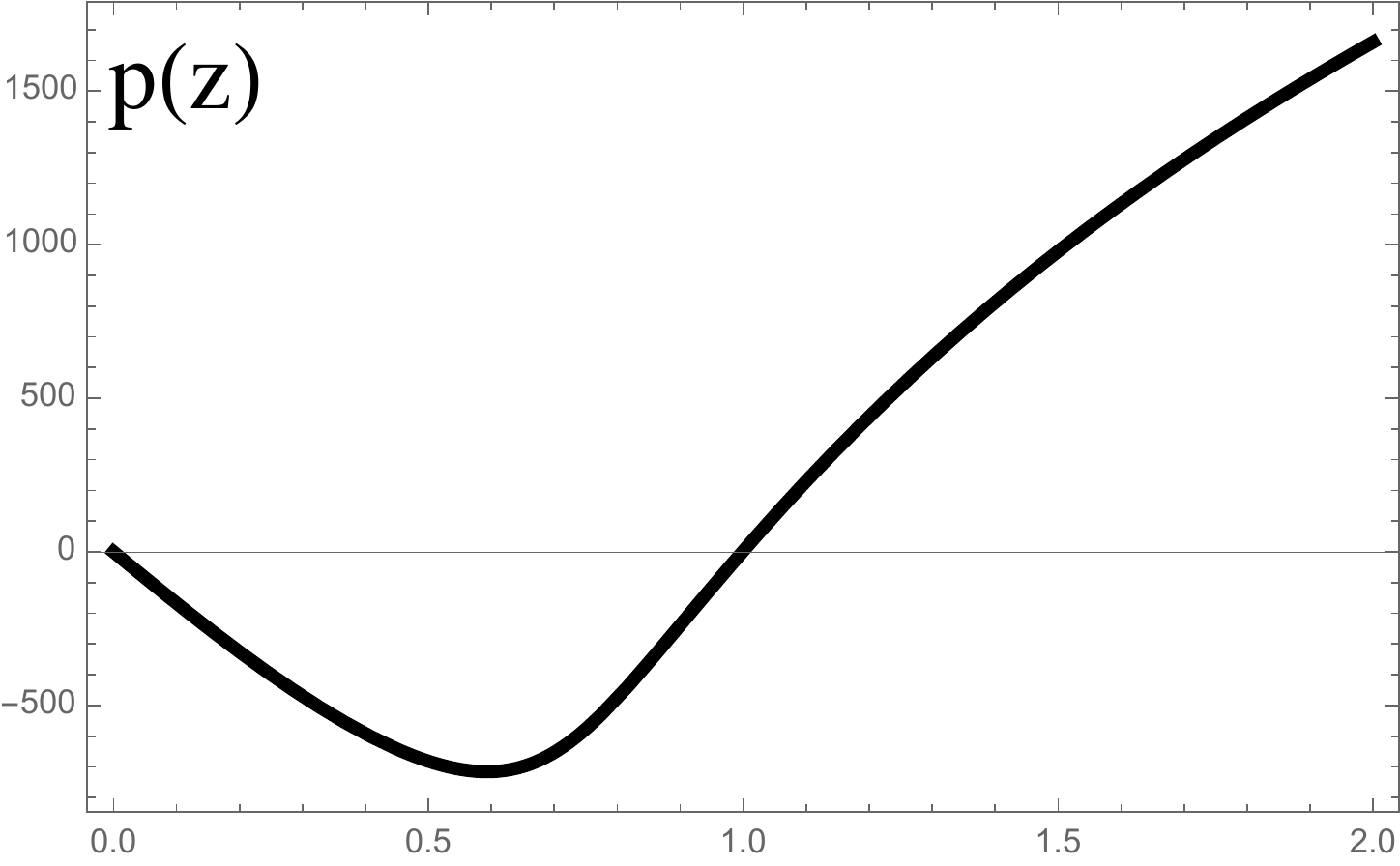}
\includegraphics[width=0.45\textwidth]{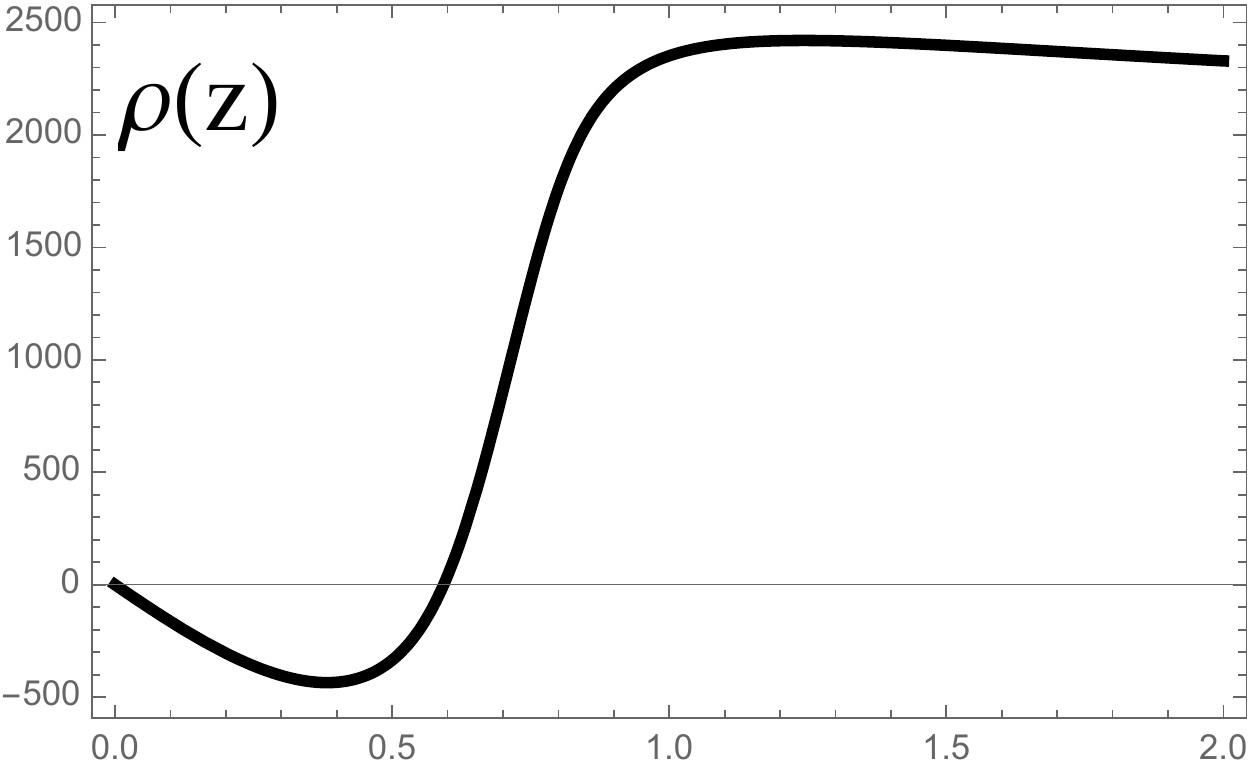}
\caption{A random distribution of zeros (on top) which gives rise to a non positive and non monotonic behavior both of the pressure $p(z)$ 
and the density $\rho(z)$ along the positive real values of the fugacity $z$.}
\label{randomsetzeroo}
\end{figure}

This electrostatic analogy is pretty appealing but it is important to realise that {\em not} all charge densities are appropriate for the statistical mechanics problem. For this purpose, there are indeed certain requirements to fulfil, such as: its pressure $p(z)$, computed according to eq.\, (\ref{pressurezeros}), must be 
necessarily a continuous and positive function of $z$, monotonically increasing, along the real axis of the variable $z$; at the same time, its density $\rho(z)$, computed according to eq.\,(\ref{densityzeros}), must also be a positive but not necessarily a continuous function, increasing too along the real axis 
of $z$. It may be stressed that these conditions alone may be not sufficient to define a physical systems but if they are violated the system at hand is 
surely unphysical. For instance, for the random distributions of zeros shown in the top of Figure \ref{randomsetzeroo}, the corresponding pressure and density, shown on the bottom of the same figure, do not fulfil the physical conditions of positivity and monotonicity: therefore, this set of zeros shown does not correspond to any physical statistical system. We will come back again to this issue later at the end of this Section.

\begin{figure}[t]
\begin{center}
%\vspace{8mm}
\psfig{figure=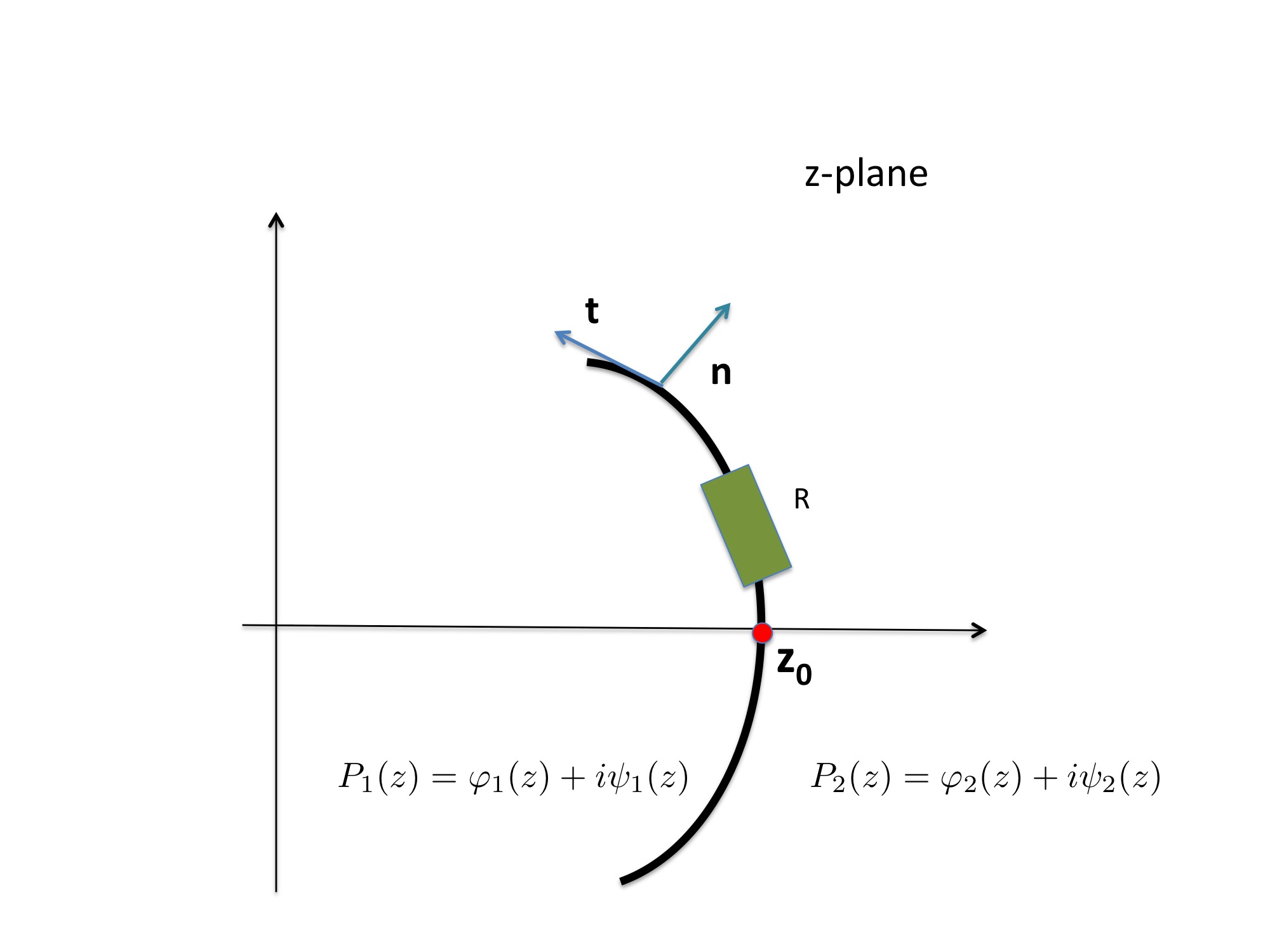,height=8cm,width=11cm}
\caption{Linear distribution of the Yang-Lee zeros nearby the real zero $z_0$ 
and the Gauss law for the rectangle $R$ across the line of zeros. 
{\bf n} and {\bf t} are respectively the normal and the tangential unit vectors of the the curve ${\mathcal C}$. }
\label{Gausslaww}
\end{center}
\end{figure}

\vspace{3mm}
\noindent
{\bf Ising models.} The formulas (\ref{pressurezeros}) and (\ref{densityzeros}) for the pressure and the density of a gas are also useful to 
express the parametric equations of state for the Ising model whose spins can assume values $ \pm 1$. This can be easily done in terms of the mapping 
which exists between the quantities of the Ising model with $N$ spin and the lattice gas in a volume $V$ \citep{YL2}. In the following $N_{\pm}$ denotes the number of positive/negative spins, $M$ the magnetization of the system $M = (N_+ - N_-)/N$, $F$ the free energy of the Ising model and $H$ an external magnetic field. So, making the substitutions in eqs. (\ref{pressurezeros}) and (\ref{densityzeros}) according to the following dictionary
\be 
\begin{array}{lll}
{\rm {\bf Ising \,\,model}} & & {\rm {\bf Lattice \,\, gas}} \\
&&\\
N & = & \,\,\,\,V \\
&& \\
1 - M & = & \,\,\,\, 2 \rho \\
&& \\
F + H & = & \,\,\,\, -$p$ 
\end{array}
\ee  
one can easily write down the free energy and the magnetization of the Ising spin in terms of the zeros in the variable $z = e^{-2 \beta H}$. 

\vspace{3mm}
\noindent
{\bf Singular behavior.} It may happen that in the thermodynamic limit the zeros of the grand-canonical partition function do not spread on an area but lie instead 
along a particular curve ${\mathcal C}$ thus satisfying a {\em perimeter law}. Such a curve has not to be necessarily a circle but in any case the previous formulas for the pressure and the density become in this case 
\begin{eqnarray}
&& \frac{p(z)}{k T} \,\equiv \, \hat f(z) \,=\, \int_{\mathcal{C}} \eta(s) \, \log\left(1-\frac{z}{z(s)}\right)\, ds \,\,\,,
\label{curvezerosss1}\\
&& \rho(z) \,=\, z \hat f'(z) \,=\, z \,\int_{\mathcal{C}} \frac{\eta(s)}{z - z(s)} \, ds \,\,\,.
\label{curvezerosss2}
\end{eqnarray}
These equations are particularly useful to characterise the nature of the phase transition which occurs when the zeros pinch the real axis at same point $z_0$. Let's indeed suppose, as it is often the case, that nearby this point the zeros lie on a smooth curve ${\mathcal C}$: using the electrostatic analogy, in this case we are in presence of a line distribution of the charges and therefore there will be a discontinuity in the electric field, 
given by the gradient of $\varphi(z)$, across the line charge distribution. Denoting by $P_1(z)$ and $P_2(z)$ the pressure (up to $k T$ factor) across the line of the zeros and using the parameter $s$ to move along this line, applying the Gauss law to the rectangle $R$ shown in Figure \ref{Gausslaww}, one sees that \citep{Bena,Blythe}
\be
(\nabla \varphi_2 - \nabla \varphi_1)\cdot n \left|_{\mathcal C} \right. \,=\, 2 \pi \eta(s)  
\ee
Using the Cauchy-Riemann equation, this relation can be also expressed as 
\be
\frac{d}{ds}(\psi_2 - \psi_1) \left|_{\mathcal C} \right. \,=\, 2\,\pi \eta(s) \,\,\,.
\ee
The nature of the phase transition at $z = z_0$ is then determined by the behavior of the density of the zeros nearby this point. Particularly important are 
two cases: 
\begin{enumerate}
\item 
When $ \eta(0) \neq 0$ we have a jump in the derivative of the pressure at $z = z_0$, therefore we are in presence of a first order phase transition. 
\item 
When $\eta \sim | s |$,  the density of zeros vanishes at $z = z_0$ and therefore the first derivative of the pressure is continuous at $z = z_0$ while there is 
a discontinuity in its second derivative. In this case we are in presence of a second order phase transition.  
\end{enumerate}

\vspace{3mm}
\noindent
{\bf Zeros on a circle.} Notice that the equations (\ref{curvezerosss1}) and (\ref{curvezerosss2}) further simplify when the curve ${\mathcal C}$ is a circle, say of radius equal to 1: in this case the density $\eta(z)$ becomes an even function of the angle $\theta$, normalised as
\be
\int_{-\pi}^\pi \eta(\theta) \,=\, 1 \,\,\,, 
\label{normalization}
\ee 
and the pressure and density are expressed as
\begin{eqnarray}
&& \frac{p(z)}{k T} \,=\, \int_0^\pi \eta(\theta) \, \log(z^2 - 2 z \cos\theta +1) \, d\theta \,\,\,, 
\label{pressurezeroscircle}\\
&& 
\rho(z) \,=\,2 z \, \int_0^\pi \eta(\theta) \, \frac{z -\cos\theta}
{z^2 - 2 z \cos\theta +1} \, d\theta \,\,\,. 
\label{densityzeroscircle}
\end{eqnarray}
For a circle distribution of the zeros, it is easy to find a condition on the density $\eta(\theta)$ which ensures that both the pressure and the density are positive monotonic functions of $z$: as shown in \citep{Ikeda1}, it is in fact sufficient that the density $\eta(\theta)$ is bounded and continuous, while its derivative $\eta'(\theta)$ is a bounded, continuous and positive function. Indeed, taking the derivative of $\rho(z)$ with respect to $z$ we have 
\be
\frac{d\rho}{dz} \,=\, 2  \int_0^\pi \eta(\theta) \, \frac{2 z - (1+z^2) \cos\theta}
{(z^2 - 2 z \cos\theta +1)^2} \, d\theta\,\,\,, 
\ee
which, with the change of variable $\xi= \tan(\theta/2)$ and an integration by part, can be expressed as 
\be
\frac{d\rho}{dz} \,=\, - 4 \left[\frac{\xi  \,\eta(2 \arctan \xi)}{(z+1)^2 \xi^2 + (z-1)^2}\right]^{\infty}_0 + 
8  \int_0^\pi \, \frac{\xi \,\eta'(2 \arctan\xi)} 
{(1+\xi)^2 ((z+1)^2 \xi^2 + (z-1)^2}  \, d\xi\,\,\,.
\ee 
With the hypothesis that $\eta(\theta)$ is a bounded function, the first term of this expression vanishes while the second term, as far as 
$\eta'(\theta) > 0$, is positive. Once established that the function $\rho(z)$ is then an increasing monotonic function, to show that it is always positive is sufficient to 
calculate its value at the origin and, if not negative, the function $\rho(z)$ will be indeed always positive. 
Since at $z = 0$ we have $\rho(0) =0$, this is sufficient to show the 
positivity of $\rho(z)$. 

Under the same hypothesis for $\eta(\theta)$, we can also conclude that the pressure $p(z)$ is a positive and increasing function of $z$: since 
\be 
\rho(z) \,= \, z \frac{d p}{d z} \,=\, \frac{d p}{d \log z} \,\,\,,
\ee
the positivity of $\rho(z)$ implies that $p(z)$ is an increasing function of $\log z$, i.e. of $z$ itself since the logarithmic function is a monotonic function. 
So, also for $p(z)$ to prove that this is always a positive function it is sufficient to compute its value at the origin and check that it is not negative. Since $p(0) =0$, this concludes the argument. 

The importance of these considerations becomes more clear once one realises that even assuming the most favourable distribution of the zeros, i.e. along a circle, the analytic expression of the densities $\eta(\theta)$ is largely unknown. The very few cases where we have such an information %simply reduce to 
include the one-dimensional Ising model with nearest-neighbor interaction \citep{YL2} and various versions of the mean field solutions of the same model \citep{Katsura}. As a further intriguing remark, it seems that if one is able to point out what are the conditions in order to have a proper physical density $\eta(\theta)$, there is a plenty of room to define new statistical model by inverting somehow the theory of Yang and Lee. We will further comment on this point in the conclusions of the paper.

\vspace{3mm}
\noindent
{\bf Yang-Lee edge singularities.} We can take advantage of the known expression of the density of zeros of the one-dimensional ferromagnetic Ising model for presenting the simplest example of Yang-Lee edge singularities. The absence of a phase transition in this one-dimensional model of course implies that $\eta(\theta)$ must vanish in an interval around the origin. This is indeed the case and the exact expression of the distribution is given by \citep{YL2} 
%\be 
%\eta(\theta) \,=\, 
%\left\{
%\begin{array}{cll}
%0 & {\rm if} & |\theta| < \theta_0  \\
%\frac{1}{2 \pi} \frac{\sin\frac{\theta}{2}}{\sqrt{\sin^2\frac{\theta}{2} - \sin^2\frac{\theta_0}{2}}} 
%& {\rm if} & |\theta| > \theta_0 
%\end{array}
%\right.
%\label{1deta}
%\ee
\be 
\eta(\theta) \,=\, 
\frac{1}{2 \pi} \frac{\sin\frac{\theta}{2}}{\sqrt{\sin^2\frac{\theta}{2} - \sin^2\frac{\theta_0}{2}}} 
\,\,\,\,\,\,\,
,
\,\,\,\,\,\,\,\,
{\rm if} \,\, |\theta| > \theta_0 
\label{1deta}
\ee
otherwise $0$, where $\theta_0 =\arccos(1 - 2 e^{-2 \beta J})$ and $J$ is the coupling constant between two next-neighbor spins. The zeros take then the 
$C$-shape of the plot on the left in Figure \ref{circleYL}. In this example $\theta_0$ plays the role of an edge singularity and this value pinches the origin 
only when $\beta J \rightarrow \infty$, namely at $T = 0$. Nearby $\theta_0$ the density of zeros behaves anomalously as 
\be 
g(\theta) \sim \frac{C}{\sqrt{\theta - \theta_0}} \,\,\,,
\ee
($C = 1/(2 \pi) \sqrt{\tan\frac{\theta_0}{2}}$), therefore for the one-dimensional Ising model the Yang-Lee edge singularity exponent 
defined in eq.\,(\ref{YLedgeexponent}) is equal to $\sigma = - 1/2$. Consider now the expression of the magnetization in terms of the 
density of the zeros using the dictionary previously established
\be 
M(z) \,=\,1 - 4 z \,\int_0^\pi \eta(\theta) \, \frac{z -\cos\theta}
{z^2 - 2 z \cos\theta +1} \, d\theta \,\,\,.
\ee
Performing the integral we have 
\be 
M(z) \,=\, \frac{z-1}{\sqrt{\left(z - e^{i \theta_0}\right) \left(z - e^{-i \theta_0}\right)}} \,\,\,. 
\ee 
Posing $z = e^{-2  \beta h}$ and $\theta_0 = -2 \beta h_c$, and expanding this formula around $h = i h_c$, we have 
\be
M(h) \sim \frac{M_0}{(h - i h_c)^{1/2}}\,\,\,,
\ee
i.e. the value $i h_c$ can be considered as a singular point. 

\vspace{3mm}
\noindent
{\bf Polynomials vs Series}. The discussion made so far concerned with the singular behavior which emerges by increasing the order $N$ of a sequence of genuine  polynomials, in particular when $N \rightarrow \infty$. But what about is the partition function $\Omega(z)$ is {\em ab-initio} given in terms not of a polynomial but an infinite series? Let's say $\Omega(z)$ in a neighborough of $z=0$ is given by the infinite series 
\be
{\cal S}(z) \,=\, \sum_{k=0}^\infty \alpha_n z^n \,\,\,. 
\label{infiniteseries}
\ee
In this case one must be aware that analysing the Yang-Lee zeros of an expression such as in eq.\,(\ref{infiniteseries}) there may be a condensation of zeros along some positive value $z_0$ of the fugacity $z$ which however does not necessarily signal a phase transition of the physical system but rather the finite radius of convergence of the series itself! Imagine in fact that the series (\ref{infiniteseries}) could be analytically continued and that the corresponding function $\Omega(z)$ has the closest singularity nearby the origin at a negative value $z = - R$. This automatically fixes the radius of convergence of the series (\ref{infiniteseries}) to be $R$ and therefore if we would use such a series to define the partition function, the corresponding Yang-Lee zeros may also condensate at the {\em positive} value $z = R$, even though this point is not associated to a phase transition of the actual function $\Omega(z)$. A simple example of this phenomena is worked out in detail in Appendix A. We will see that similar cases also emerge in discussing the Yang-Lee zero distributions of fermionic theories whose corresponding series $S(z)$ has alternating sign and therefore a singularity at a negative value of $z$: their zeros however also condensate at a positive real value of $z$.

\section{Playing with polynomials}\label{SectionZeros}
\noindent
In this Section we are going to deal extensively with the properties of the main mathematical object of this paper, namely the class of {\em real} polynomials $\Omega_N(z)$ of order $N$ in the variable $z$ 
\be 
\Omega_N(z) \,=\, \gamma_0 + \gamma_1 z + \gamma_2 \,z^2 + \ldots \gamma_N \, z^N \,\,\,.  
\label{part1}
\ee 
The coefficients $\gamma_n$ of these polynomials are real and we choose hereafter $\gamma_0 =1$. Since we are going to interpret $\Omega_N(z)$ as a generalised grand-canonical partition function of a statistical model, either classical or quantum, we pose 
\be 
\Omega_N(z) \equiv e^{\mathcal{F}_N(z)}\,\,\,, 
\label{freefreefree}
\ee
where we have define the so-called free-energy $\mathcal{F}_N(z)$ of the system, directly related to the pressure of the system, see eq.\,(\ref{pressurezeros}). 
In light of this statistical interpretation of $\Omega_N(z)$ in the following we will also express it with a different normalization of the coefficients   
\be
\Omega_N(z) \,=\, \sum_{k=0}^N \frac{a_k}{k!}z^k \,\,\,,
\label{part2}
\ee
with {\bf $\gamma_k=a_k/k!$} and $a_0=1$ while the higher coefficients $a_k$'s assume the familiar meaning of canonical partition functions of $k$ particles. Formulas which we derive below are sometimes more elegantly expressed in terms of the $a_k$'s although we will switch often between the two equivalent expressions ({\ref{part1}) and (\ref{part2}), hoping that this will not confuse the reader. 

%\vspace{3mm}\noindent{\bf Sign of the coefficients}. 
\subsection{Sign of the coefficients}
\noindent
For models coming from classical statistical physics, it is easy to argue that the coefficients $a_k$ of the relative polynomial $\Omega_N(z)$ are 
generically all positive, $a_k > 0$. In this respect, consider for instance two significant examples: 
\begin{itemize}
\item {\bf Classical Gas.} A classical model of a gas in $d$-dimension, made of $N$ particles of mass $m$ and Hamiltonian of the form 
\be
H \,=\,\sum_{i=1}^N \frac{p^2_i}{2 m} + \sum_{i,j} u(r_{ij}) \,\,\,, 
\ee
where $r_{ij}$ is the distance between the $i$-th and $j$-th particle. In this case the grand-canonical partition function of the system assumes 
the form (\ref{part2}),  where the variable $z$ expressed by  
\be 
z = e^{\beta \mu} \,\left(2 \pi m \beta/h^2\right)^{\frac{d}{2}}\,\,\,,
\ee 
where $\mu$ is the fugacity, $\beta = 1/k T$ where $T$ is the temperature and $h$ is the Planck constant here introduced to normalise the phase-space integral.  
Therefore in this example the coefficients $a_k$ are given the positive integrals 
\be
a_k \,=\,\int \cdots\int dr_1\cdots dr_k \,\exp\left[-\beta \sum_{i,j} u(r_{ij})\right] > 0 \,\,\,. 
\ee
\item {\bf Ising in a magnetic field.} 
As a second example of classical statistical mechanics, consider the ferromagnetic Ising model in a magnetic field $H$ studied originally by 
Lee and Yang in \citep{YL2}: in this case, with $\sigma_i = \{\pm 1\}$ and a general two-body ferromagnetic Hamiltonian of $N$ spins 
in a regular lattice in arbitrary $d$-dimensional lattice of the form 
\be
H \,=\, - \sum_{i,j} J_{ij} \sigma_i \sigma_j - H \sum_{i} \sigma_i 
\,\,\,\,\,
,
\,\,\,\,\,
J_{ij} > 0 
\ee 
posing 
\be 
z \,=\, e^{- 2 \beta H} \,\,\,,
\ee
the partition function of the model is expressed by a palindrome polynomial in this variable, i.e. a polynomial of the form (\ref{part1}) 
where the coefficients satisfy the additional condition
\be 
\gamma_k \,=\, \gamma_{N-k} > 0 \,\,\,. 
\ee
This because $\gamma_k$ is the contribution to the partition function of the Ising model in zero magnetic field coming from configurations in which 
the number $N_-$ of spins $\sigma_i = -1$  is equal to $k$; this contribution is evidently the same also for $N_ -=  (N-k)$.    
\end{itemize}
It is however important to stress that for situations which come from quantum statistics and which involve, in particular, fermionic degrees of freedom, the sign of the coefficients $\gamma_k$'s of the polynomial may be not necessarily all positive but rather alternating. However, despite the presence of negative coefficients, 
in the {\em physical domain} of the variable $z$ (alias the positive real axis, $z >0$), the partition function $\Omega_N(z)$ of these systems assumes only positive values and is a monotonic function of $z$. The request of positivity of $\Omega_N(z)$ in its physical domain will play an important role in our future analysis when we are going to truncate our expressions of the partition function. 

%\vspace{3mm}\noindent{\bf Patterns of the zeros}. 
\subsection{Patterns of the zeros}
\noindent
Let's now make a preliminar discussion about the nature of the zeros of a polynomial such as $\Omega_N(z)$ before focusing our attention on a series of examples. Being $\Omega_N(z)$ a real polynomial, its N zeros are either real or grouped in pairs of complex conjugated values in the complex $z$ plane. Hence, any distribution of the zeros has to be symmetric with respect the real axis. Concerning the real zeros, when all coefficients $\gamma_k$ of $\Omega_N(z)$ are positive, they cannot be obviously positive. Hence, for any {\em finite} order $N$ of the polynomial, it should exist a region $R$ which contains the whole positive real axis which is free of zeros: a positive real zero $z_c$ can emerge only in the limit $N \rightarrow \infty$, i.e. as an accumulation point of complex zeros and, in this case, it corresponds to a point of singularity of the grand-canonical partition function \citep{YL1,YL2}.
\begin{figure}[t]
\centering
\includegraphics[width=0.48\textwidth]{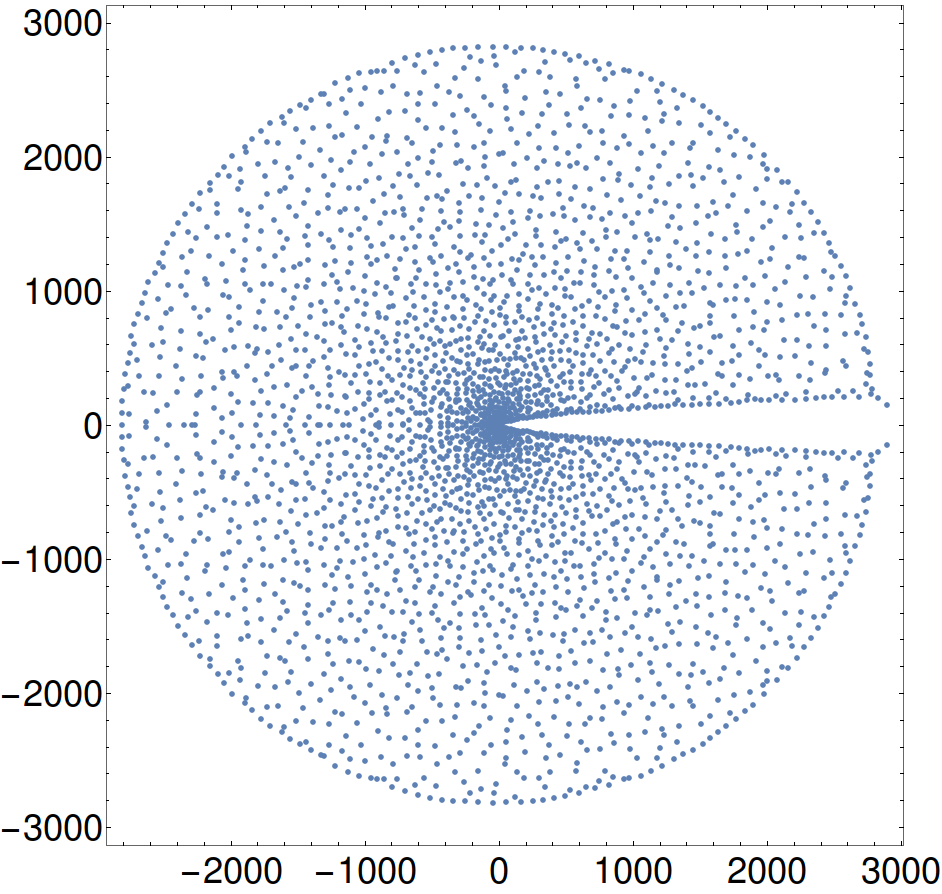}
\includegraphics[width=0.46\textwidth]{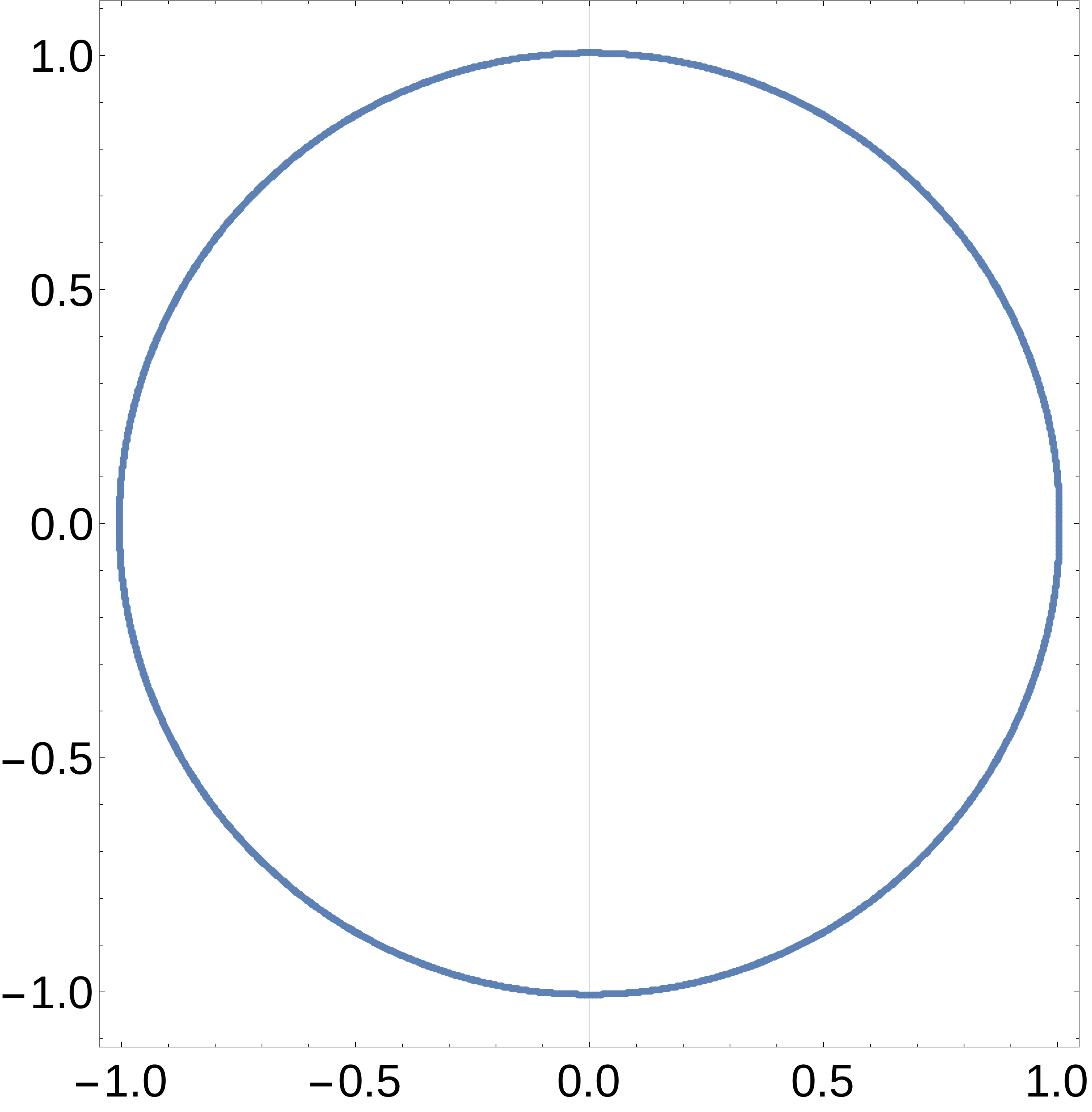}
\caption{a) Distribution of zeros of the partition function where the coefficients $a_n$ are given by prime numbers. b) Distribution of zeros of the partition function where the coefficients are given by $\gamma_n=(n-1)!/n$. Notice the very different scales of the two figures. }
\label{Distributions}
\end{figure}

Clearly, the overall distribution of zeros and its details depend upon the choice of the coefficients $\gamma_k$ and therefore, in general, there may be a wide range of situations. We find however important to try to identify at least two important patterns of zeros, hereafter called {\em area} and {\em perimeter} distributions respectively, for which in Section \ref{areaperimeter} we will find some criteria for their realisation. At the level of terminology, we say 
\begin{itemize}
\item we are in presence of an {\em area distribution} when the the N zeros are distributed in an extended area of the complex plane. This is the case, for instance, of the distribution shown in the first plot in Fig. \eqref{Distributions}, obtained by taking as coefficients $p_k$ of $\Omega_N$ just the $k$-th prime number. We will comment more on this example later. 
\item we are in presence of a {\em perimeter distribution} when the N zeros are distributed along particular lines. A particular interesting case of perimeter distributions is when all the zeros are along a circle. This is the case, for instance, of the distribution of the zeros shown in the second plot in Fig. \eqref{Distributions}, obtained by considering a grand-canonical partition function $\Omega_N(z)$ with coefficients $a_k$ given the sequence $\gamma_k = 1/k^2$.  This and other similar cases will be discussed in more details below. 
\end{itemize}

%\vspace{3mm}\noindent{\bf Bounds on the zeros.} 
\subsection{Bounds on the absolute module of the zeros}
\noindent
Given a polynomial of the form (\ref{part1}), there are several bounds on the magnitudes $|z_i|$ of its roots. Here we simply state some of these bounds without any proof for them (it may be useful to consult the book by Prasolov \citep{prasolov} as a general reference on polynomials and bounds on their roots). Some of these bounds are more stringent than others (and we list them in the order of increasing their level of refinement) but altogether they may help in getting an idea about the distribution of the roots. 
\begin{itemize}
\item {\bf Cauchy bound.} The Cauchy bound states that the roots $z_i$ have an absolute value less than $R_C$, $|z_i | < R_C$, where 
\be
R_C = 1 + \frac{1}{|\gamma_N|} \, {\rm max}\, \{|\gamma_0|, |\gamma_1|, \ldots |\gamma_{N-1}|\}\,\,\,.
\ee 
This is usually the less stringent bound on the module on the roots. 
\item {\bf Sun-Hsieh bound}. The Cauchy bound can be refined in terms of the Sun-Hsieh bound, $|z_i | < R_{SH}$, where 
\be 
R_{SH} \,=\, 1 + \frac{1}{2} \left( (|\gamma_{N-1}/\gamma_N| -1) + \sqrt{(|\gamma_{N-1}/\gamma_N| -1)^2 + 4 a}\right) 
\,\,\,\,\,
,
\,\,\,\,\,
a = {\rm max} \{|\gamma_k/\gamma_N|\}
\,\,\,.
\ee
\item {\bf Fujiwara bound}. A further refinement comes from the Fujiwara bound where all roots are within the disc of radius $R_F$, $|z_i| < R_F$, where 
\be 
R_K = 2\, {\rm max}\, \left\{ \left| \frac{\gamma_{N-1}}{\gamma_N}\right|,
\left|\frac{\gamma_{N-2}}{\gamma_N}\right|^{\frac{1}{2}},
\cdots , 
\left|\frac{\gamma_{1}}{\gamma_N}\right|^{\frac{1}{N-1}},
\frac{1}{2} \left|\frac{\gamma_{0}}{\gamma_N}\right|^{\frac{1}{N}}
\right\}\,\,\,.
\label{Fujiwarabound}
\ee
\item {\bf Enestr\"{o}m bounds}. When all $\gamma_k > 0$, we can have a lower and an upper bound of the magnitude of the roots given by 
\be
R_d \leq |z_i| \leq R_u \,\,\,,
\label{Kakeyabounds1}
\ee
where 
\be 
R_d = {\rm \min}\left\{\frac{\gamma_k}{\gamma_{k+1}}\right\} 
\,\,\,\,\,\,
,
\,\,\,\,\,\,
R_u = {\rm \max}\left\{\frac{\gamma_k}{\gamma_{k+1}}\right\} 
\,\,\,\,\,\,\,
,
\,\,\,\,\,\,\,
k=0,1,\ldots N-1\,\,\,.
\label{Kakeyabounds2}
\ee
The Enestr\"{o}m bounds are usually the best estimate of the annulus in the complex plane where the roots are located. 
\item {\bf Enestr\"{o}m-Kakeya bound}. Let us also mention the Enestr\"{o}m-Kakeya bound which refers to polynomials where 
the coefficients $\gamma_k$ satisfy the condition 
\be 
\gamma_0 \leq \gamma_1 \leq \gamma_2 \ldots \leq \gamma_N \,\,\,. 
\ee
In this case we have that  $|z_i| \leq (|\gamma_N| - \gamma_0 + |\gamma_0|)/|\gamma_N|$. In particular, if all coefficients $\gamma_k$ are positive, the roots are within the unit disc, $|z_i | \leq 1$.  Of course, when the coefficients satisfy instead the condition 
\be 
\gamma_0 \geq \gamma_1 \geq \gamma_2 \ldots \geq \gamma_N \,\,\,. 
\ee
when $| z_i | \geq 1$. 
\end{itemize}
Let's make a final comment on the information provided by the bounds on the modules of the zeros: these bounds refer to {\em all} roots, so if knew for instance that $|z _i | < 100$, it may happen that just one or few zeros has module $|z| \sim 100$ while all the rest may have a very small module. In other words, bounds usually refer to the {\em largest} of the zeros rather than their overall distribution.

%\vspace{3mm}\noindent{\bf Motion of the zeros: perturbative analysis}. 
\subsection{Motion of the zeros: perturbative analysis}
\noindent
We can study how the distribution of the zeros of a polynomial is modified by the addition of a new root as the order of the polynomial is increased by $1$. To this aim, consider a polynomial of the form 
\be
\Omega_N(x) = \Omega_{N-1}(x) + \gamma_N x^N\,.
\ee
The addition of the new term $\gamma_N x^N$ to $\Omega_{N-1}(x)$ has two effects:
\begin{enumerate}
\item it creates a new zero;
\item it moves the previous ones.
\end{enumerate}
Let us address the first issue by considering the equation for the zeros of the new polynomial 
\be
\Omega_{N-1}(x) + \gamma_N x^N = 0 \,\,\,. 
\ee
Writing it as 
\be
x \,=\, -\frac{\Omega_{N-1}(x)}{\gamma_N x^{N-1}}\, =\, -\frac{\gamma_{N-1}}{\gamma_N} - \frac{\gamma_{N-2}}{\gamma_{N-1}x} + \ldots\,, 
\label{inversepowers}
\ee
we see that, perturbatively in $a_N$, the new root $x^*$ is roughly placed at
\be
x^* \simeq -\frac{\gamma_{N-1}}{\gamma_N}\,\,\,.
\label{firstorder}
\ee
If $\gamma_N$ is infinitesimally small with respect to the other previous coefficients, the value $x^*$ determined from this equation is large and therefore, 
self-consistently, it is justified to neglect its corrections coming from the inverse powers $(1/x^*)^l$ present in the left hand side of (\ref{inversepowers}). 

Let us now address the second issue, i.e. the motion of the other zeros, once again perturbatively in the quantity $a_N$. Let $x_i$ ($i = 1,\ldots, N-1$) be one of the zeros of the polynomial $P_{N-1}(x)$. Let's now write 
\be
x_i + \delta x_i
\ee
as the new position of the root, once the new term $a_Nx^N$ has been added
\begin{eqnarray}
\Omega_N(x_i + \delta x_i) &=& \Omega_{N-1}(x_i + \delta x_i) + \gamma_N (x_i + \delta x_i)^N \,=\, 0\nonumber \\[3pt]
&=& \Omega_{N-1}(x_i) + \delta x_i \frac{d \Omega_{N-1}}{dx_i} + \gamma_N x_i^N + N \gamma_N x_i^{N-1} \delta x_i \,=\, \nonumber\\
&=& \delta x_i \left[ \frac{d\Omega_{N-1}}{dx_i} + N \gamma_N x_i^{N-1} \right] + \gamma_N x_i^N  = 0\,,
\end{eqnarray}
hence  the displacement of the $i-th$ root due to the addition on the term $a_Nx^N$ is given by
\be
\delta x_i = -\frac{\gamma_N x_i^N}{\left(\frac{d\mathcal{Z}_{N-1}}{dx_i} +  \gamma_N N x_i^{N-1}\right)}\,\,\,.
\ee

%\vspace{3mm}\noindent{\bf Statistical approach}. 
\subsection{Statistical approach}
\noindent
In order to understand better the pattern of the zeros and, in particular, to see whether we are able to qualitatively predict if an area or a perimeter law occurs, it is worth setting up a statistical analysis of the zeros. The obvious quantities to look at are the statistical moments $s_k$ of the zeros defined as 
\be
s_k \equiv \sum_{l=1}^N z_l^k \,\,\,,
\ee
where $k$ can be either a positive or a negative integer. Notice that, being the polynomial real, all $s_k$ are real quantities. 

\vspace{3mm}
\noindent
{\bf Negative moments}. Let's see how to relate the negative moments of the zeros 
\be 
s_{-m} \equiv \hat s_m \,=\, \sum_{l=1}^N \left(\frac{1}{z_l}\right)^m \,\,\,,
\ee
to the coefficients of the polynomial $\Omega_N(z)$. In order to do so, let's factorise the partition functions in terms of its zeros as 
\be
\Omega_N \,=\, \prod_{l=1}^N \left(1- \frac{z}{z_l} \right)\,.
\ee
Taking the logarithm of both sides we arrive to the familiar 
cluster expansion of the free-energy $\mathcal F_N(z)$ 
\begin{eqnarray*}
\mathcal{F}_N(z) \,=\, \log \Omega_N(z) &=&  \sum_{l=1}^N \log\left(1- \frac{z}{z_l} \right) 
\,=\, -\sum_{l=1}^N \sum_{m=1}^{\infty} \frac{1}{m} \left( \frac{z}{z_l} \right)^m \,=\,\\
&\equiv&\sum_{m=1}^{\infty} b_m z^m \,\,\,. 
\end{eqnarray*}
The negative moments $\hat s_m$ are then related to the cluster coefficients $b_m$  as \citep{YL1}
\be
b_m = -\frac{1}{m}\sum_{l=1}^N \left( \frac{1}{z_l}\right)^m\,=\, -\frac{1}{m} \, \hat s_{m}\,\,.
\label{bmsm}
\ee
It is also custom to expand the free-energy $\mathcal{F}_N(z)$ in terms of the cumulants $c_k$ defined by 
%\citep{McCullay}:
\be\
\mathcal{F}_N(z)\,=\,\ln \Omega_N \equiv \sum_{k=1}^{\infty} \frac{c_k}{k!}z^k\,\,\,. 
\ee
If we use for $\Omega(z) $ the expression (\ref{part2}) and the coefficients $a_k$, the cumulants are given by   
\be
\begin{split}
c_1 &= a_1\\
c_2 &= a_2 - a_1^2\\
c_3 &= a_3 - 3a_2 a_1 +2 a_1^3\\
\vdots
\end{split}
\ee
As a matter of fact there exists a closed formula which relates the two sets of coefficients $c_k$ and $a_k$. To this aim let's introduce the determinant of a ($k \times k$) matrix $M(k)$ whose entries involve the first $k$ coefficients $a_l$ ($l=1,2,\ldots k$)  
\be 
M(k) \,=\, 
\begin{pmatrix}
a_1 & 1 & 0 & \cdots & 0 \\
a_2 & a_1 & 1 &  & \vdots \\
a_3 & a_2 & \binom{2}{1}a_1 & & 0 \\
\vdots & \vdots & \vdots & &  \vdots \\
a_k & a_{k-1} & \binom{k}{1}a_{k-2} & \cdots & \binom{k-1}{k-2}a_1
\end{pmatrix}\,\,\,.
\ee
The final formula is then 
\be
c_k = (-1)^{k-1}\, \text{det}\,  M(k)\,\,\,.
\label{cumulant}
\ee
The relevant thing to note is that the $k$-{th} cluster coefficient $c_k$, which is of course related to the $k$-{th} moment of the inverse roots as 
\be 
c_k \,=\, - (k-1)! \, \hat s_k \,\,\,, 
\ee
is entirely determined only by the first $k$ coefficients of the original polynomial $\Omega_N$. This implies that, if we increase the order $N\rightarrow N+\tilde{N}$ of the polynomial by adding to $\Omega_N$ the new coefficients $a_{N+1},\ldots, a_{\tilde{N}}$ but keeping fixed all the previous ones, there will be an increasing number of the zeros but their overall positions are constrained by the condition that the first $N$ negative moments 
\be
s_{-k} \equiv \hat{s}_k = \sum_{l=1}^N \left(\frac{1}{z_l}\right)^k \,=\,\sum_{l=1}^{\tilde N} \left(\frac{1}{z_l}\right)^k 
\quad , \quad k=1, \ldots, N 
\ee
before and after adding the new coefficients, remain constant.

\vspace{3mm}
\noindent
{\bf Positive moments}. Let's now discuss the positive moments of the zeros
\be
s_k =  \sum_{l=1}^N (z_l)^k\quad , \quad k>0 \,.
\ee
We can relate them to the coefficients of the polynomial $\Omega_N(z)$ as follows. Let's first introduce the elementary symmetric polynomials 
$\sigma_k(x_1,\ldots, x_N)$ defined as 
\be
\sigma_k(x_1,\ldots,x_N)=\sum_{1 \leq j_1 \leq j_2 \leq \ldots \leq j_k \leq N} x_{j_1}\ldots x_{j_k}\,
\,\,\,\,\,\,
,
\,\,\,\,\,\,
k=0,\ldots,N \,\,\,.
\ee
Notice that $\sigma_0 =1$ and $\sigma_N = x_1x_2\ldots x_N$. 
%\be
%\label{symm}
%\sigma_N(x_1,\ldots,x_N)= x_1x_2\ldots x_N\,.
%\ee
Since 
\be
%\begin{split}
G(z) \equiv \prod_{l=1}^N(z-z_l) \,=\,\sum_{k=0}^N (-1)^k \sigma_k(z_1,\ldots, z_N) z^{N-k}\,\,\,,
%=z^N-\sigma_1(z_1,\ldots,z_N)z^{N-1}+\ldots+(-1)^N\sigma_N(z_1,\ldots,z_N) \,=\, \\
%&= \sum_{k=0}^N (-1)^k \sigma_k(z_1,\ldots, z_N) z^{N-k}\,.
%\end{split}
\ee
we can express the partition function $\Omega_N(z)$ as 
\be
\begin{split}
\Omega_N(z) &= \sum_{m=0}^N \gamma_m\, z^m \equiv \gamma_N\, G(z)=\\
& = \gamma_N\, \sum_{k=0}^N (-1)^k \sigma_k(z_1,\ldots, z_N) z^{N-k}\,.
\end{split}
\ee
Therefore
\be
\gamma_m \,=\, (-1)^{N-m} \gamma_N\, \sigma_{N-m}\,\,\,,
\ee
and, sending $m \rightarrow N - m$, we have the final relation between the symmetric polynomials and the coefficients $\gamma_k$ (or $a_k$) of the partition function $\Omega_N(z)$ 
\be
\label{sim2}
\sigma_m \,=\, (-1)^m \frac{\gamma_{N-m}}{\gamma_N} \,=\, (-1)^{m} \frac{N!}{(N-m)!} \frac{a_{N-m}}{a_N} \,\,\,.
\ee
Moreover, we can relate the moments $s_k$ to the elementary symmetric polynomials thanks to the\textit{ Newton-Girard formula} 
\be
(-1)^m\, m\, \sigma_m(z_1,\ldots, z_N) + \sum_{k=1}^m (-1)^{k+m}\, s_k (z_1,\ldots, z_N) \sigma_{m-k} = 0\,\,\,, 
\ee
so that  
\begin{eqnarray}
%\begin{split}
&& s_1 - \sigma_1 = 0 \,\,\,, \nonumber \\
&& s_2 - s_1\sigma_1 + 2\sigma_2 = 0 \,\,\,,\nonumber \\
&& s_3 - s_2\sigma_1 + s_1\sigma_2 - 3\sigma_3 = 0 \,\,\,, \label{newtongirard}\\
&& \,\,\,\qquad\vdots \nonumber \\ 
&& s_N-s_{N-1}\sigma_1+ s_{N-2}\sigma_2-\ldots+(-1)^N \sigma_N=0\,\,\,.\nonumber
%\end{split}
\end{eqnarray}
A closed solution of these relations can be given in terms of a formula which employs the following determinant   
\be
s_p = 
\begin{vmatrix}
\sigma_1  & 1 & 0 & \cdots & 0 \\
2\sigma_2 & \sigma_1 & 1 & \cdots&  0\\
3\sigma_3 & \sigma_2 & \sigma_1 & \\
\vdots & \vdots & \vdots & \\
p\sigma_p & \sigma_{p-1} & \sigma_{p-2} & \cdots & \sigma_1
\end{vmatrix} \,\,\,.
\ee

\vspace{1mm}
\noindent
The relevant thing to notice in this case is that the positive moment $s_p$ is determined by the first $p$ elementary symmetric polynomials which, on the other hand, are fully determined by the last $(N-p)$ coefficients of the polynomial $\Omega_N$ (see eq.~\eqref{sim2}). 

\vspace{3mm}
\noindent
{\bf Moments: summary}. So far we have seen that the negative moments $\hat{s}_k$ of the zeros are determined by the first $k$ coefficients of the polynomial $\Omega_N(z)$ while the positive moments $s_l$ are determined instead by the last $l$ coefficients of $\Omega_N(z)$  
%\be
%\Omega_N = \underbrace{1+a_1 z+\frac{a_2}{2!}z^2 + \ldots + \frac{a_k}{k!}z^k}_{\hat{s}_k = s_{-k}} + \ldots + \underbrace{\frac{a_{N-l}}{(N-l)!}z^{N-l} + \frac{a_{N-l+1}}{(N-l+1)!}z^{N-l+1} + \ldots+ 
%\frac{a_N}{N!}z^N}_{s_l}\,.
%\ee
\be
\Omega_N = \underbrace{1+\gamma_1 z +\gamma_2 \, z^2 + \ldots + \gamma_k \, z^k}_{\hat{s}_k = s_{-k}} + \ldots + 
\underbrace{
\gamma_{N-l} \, z^{N-l} + \gamma_{N-l+1} \, z^{N-l+1} + \ldots+ 
\gamma_N \, z^N}_{s_l}\,.
\ee
Obviously all the negative and positive moments higher or equal than $N$ are linearly dependent from the previous ones. To show this, let's interpret the polynomial 
$\Omega_N(z)$ as the characteristic polynomial of a $(N \times N)$ matrix ${\bf M}$ which satisfies the same equation satisfied by the zeros themselves of 
$\Omega_N(z)$  
\be
{\bf 1} + \gamma_1\, {\bf M} + \gamma_2\, {\bf M}^2 + \cdots \gamma_N\, {\bf M}^N \,=\, 0\,\,\,, 
\label{characteristic}
\ee
where ${\bf 1}$ is the $(N \times N)$ identity matrix. Since $s_k = {\rm Tr} \,{\bf M}^k$, taking now the trace of the equation above, it is easy to see that
the $N$-th positive moment is linearly dependent from the previous $N-1$ moments 
\be
s_N \,=\,-\frac{1}{\gamma_N} \left(N + \gamma_1 \, s_1 + \gamma_2 \, s_2 + \cdots \gamma_{N-1} \, s_{N-1} \right)\,\,\,. 
\ee
Moreover, to get the linear equation which links the higher positive moment $s_{N+m}$ to the $N$ previous ones, it is sufficient to multiply by ${\bf M}^m$ eq.\,(\ref{characteristic}) and then to take the trace of the resulting expression. 

Concerning instead the linear combinations which involve the negative moments higher or equal to $N$, it is sufficient to multiply eq.\,(\ref{characteristic}) by ${\bf M}^{-N}$ and then repeating the steps described above. For instance, the negative moment $\hat s_N$ depends linearly from the previous ones as 
\be
\hat s_N \,=\,-\gamma_1 \hat s_{N-1} - \gamma_2\, \hat s_{N-2} - \cdots - \gamma_N\, N \,\,\,.
\ee 

\subsection{Area and perimeter laws}\label{areaperimeter}
\noindent
Hereafter we are going to set a certain number of "rules of thumb" that allow us to have a reasonable guess whether the zeros satisfy the area or the perimeter laws. The criteria make use of the bounds on the modules of the zeros, the geometrical and the arithmetic means of the zeros and also of their variance. 
All these quantities are easy to compute in terms of the coefficients of the partition function and therefore they provide a very economical way for trying to anticipate  their distribution. In other words, we must subscribe to a reasonable compromise between the reliability of the prediction and the effort to compute the indicators on the zeros: of course, would one increase the number of computed moments of the zeros, then he/she would narrow better and better the prediction but at the cost of course to engage into the full analysis of the problem! This is precisely the origin of the compromise. One must be aware, however, of the heuristic nature of the arguments we are going to discuss below, which have not at all the status of a theorem, and therefore they must be taken with a grain of salt. Moreover, one must also be aware that not all distributions are either area or perimeter laws, since there exist cases where the roots have simultaneously both distributions or they are made by isolate points.    
  
\vspace{3mm}
\noindent
{\bf Geometrical Mean of the zeros}. From eq.\,\eqref{sim2}, taking $m = N$ we have 
\be
\label{28}
\sigma_N \equiv \prod_{k=1}^N z_k = (-1)^N
\frac{N!}{a_N}\,=\,(-1)^N \frac{1}{\gamma_N} \,\,\,. 
\ee
Let's put 
$
q_N \equiv \frac{N!}{p_N} = \frac{1}{\gamma_N}
$
and take the N-th root of both terms in eq.~\eqref{28}: for the left hand side, we have the geometrical mean of the roots
\be
\langle z \rangle_{\rm geom} \,=\, \left(\prod_{k=1}^N z_k \right)^{1/N}\,\,\,,
\ee
while for the right-hand side we have:
\be
 (-1) (q_N)^{1/N}\,.
\ee
For the nature of the zeros of a real polynomial -- which either pair in complex conjugate values $z_k = \rho_k e^{i\theta_k}$ and $z^*_k = \rho_k e^{-i \theta_k}$ 
or are real $z_a = \rho_a$ (where $\rho_a$ can be also negative) -- the product of all zeros is a real quantity which depends upon only 
the product of the {\em modules} $\rho_k$ of all roots. When the module of the geometrical mean is finite, say $|\langle z \rangle_{\rm geom}| \sim \xi$, 
the zeros basically must be symmetric under the mapping $z \rightarrow \xi^2/z$. Notice that the easiest way to implement such a symmetry is that the 
zeros are placed along a circle of radius $\xi$. 

\vspace{3mm}
\noindent
{\bf Arithmetic Mean of the zeros}. Given the $N$ zeros of a polynomial we can define their arithmetic mean: this is simply 
the first positive moment of the zeros divided by $N$
\be
\langle z \rangle_{\rm arith}  \,=\,\frac{1}{N} \sum_{k=1}^N z_k \,=\,\frac{1}{N} \, s_1\,=\,\frac{1}{N}\, \sigma_1\,\,,
\ee
and therefore, using eqs.\,(\ref{sim2}) and (\ref{newtongirard}), it is easily related to the ratio of the last two coefficients 
\be
\langle z \rangle_{\rm arith} \,=\, -  \frac{a_{N-1}}{a_N}\,=\,- \frac{1}{N} \frac{\gamma_{N-1}}{\gamma_N} \,\,\,.
\label{arithmeticmean}
\ee
For the reality of the polynomials considered in this paper it is obvious that the arithmetic mean of the zeros just depends on their real part and 
therefore it is a real number.

\vspace{3mm}
\noindent
{\bf Variance of the zeros}. The variance\footnote{Notice that what we call here the {\em variance} $\mu^2$ is {\em not} the familiar variance of real random variables, since our definition employs the square of the differences of  complex numbers and therefore $\mu^2$ is not necessarily positive.} 
of the zeros is defined as 
\be
\mu^2 \,=\, 
\frac{1}{N}\,\sum_{k=1}^N \left(z_k - \bar{z}_{\rm arith}\right)^2\,\,\,. 
\ee
Putting $\langle z \rangle_{\rm arith} \equiv a \in \mathbb{R}$ and expressing the zeros in terms of their real and imaginary parts, $z_k = x_k + i y_k$, 
we have
\be 
\mu^2 \,=\,\frac{1}{N} \sum_{k=1}^N \left[(x_k - a)^2 - y_k^2\right] \,\,\,,
\ee
while the imaginary part of $\mu^2$ vanishes for the symmetry $z_k \leftrightarrow z_k^*$ of the zeros. Hence, the variance $\mu^2$ essentially measures the 
unbalance between the imaginary and the real components of the zeros (the latter component measured with respect to the arithmetic mean of the zeros). 
The distributions of the zeros which have small values of $\mu^2$ are those which are almost symmetric for the interchange $(x_k -a) \leftrightarrow y_k$, 
as for instance is the case for zeros placed uniformly on a circle of center at $z =a$. However, notice that another and quite different way to have small 
values of $\mu^2$ is that the zeros are real and grouped very close to their arithmetic mean. 

Using eqs.\,(\ref{sim2}) and (\ref{newtongirard}), one can see that the variance of the zeros can be expressed in terms of the coefficients of the polynomial 
$\Omega_N(z)$ as 
\be
\mu^2 \,=\, \frac{1}{N}\,\left[\left(1-\frac{1}{N}\right) \left(\frac{\gamma_{N-1}}{\gamma_N}\right)^2 - 2 \frac{\gamma_{N-2}}{\gamma_N} \right]\,\,\,.
%- 2 N (N-2)\, \frac{a_{N-2}}{a_N}\,=\, - 2 \frac{\gamma_{N-2}}{\gamma_N} \,\,\,.
\label{variance}
\ee

\vspace{3mm}
\noindent 
{\bf Rules of thumb}. Let's now see how the three indicators introduced above, used for instance together with the Enestr\"{o}m bounds (\ref{Kakeyabounds1}), can help us in discriminating between different types of zero distributions. We are of course interested in the behaviour of the zeros for large $N$. To this aim let's consider in particular the cases\footnote{This table is not at all exhaustive of all possible zero distributions but it rather points out few significant cases.} gathered together in Table 1 in which the symbol $N$ means that the relative quantity scales as $N$ or even with higher power of $N$, while $\xi$ and $c$ denotes finite quantities, independent of $N$, and $\epsilon$ finally denotes an infinitesimal quantity. So, for example, the case $B$ corresponds to a situation in which both the arithmetic and geometrical means of the zeros are finite while the variance $\mu^2$ diverges when $N$ goes to infinity. In the following we will analyse and illustrate the various cases of Table 1 by means of some explicit polynomials\footnote{For convenience we often choose the zeros of the polynomials to be along the positive real axis but of course they can be placed along the negative axis if one has to comply the condition of dealing with polynomials with real positive coefficients.} whose means and variance satisfy the values in the table: the role of these polynomials is to guide us in understanding, if possible, more general situations. Notice that, from a statistical physics point of view, some of the polynomials shown below may be considered rather pathological.

\begin{table}[t]\label{casesa1}
\begin{center}
\begin{tabular}{||c|c|c|c||} \hline 
 & $  \langle z \rangle_{\rm geom} $ & $   \langle z \rangle_{\rm arith} $ & $ \mu^2 $ \\ 
 \hline \hline
A &  $\xi$ & c& $\epsilon$  \\
\hline
B & $\xi$ & $c$ & $N$   \\
 \hline
C & $\xi$ & $N$ & $N$   \\
\hline
D &$\xi$ & $N$  &  $\epsilon$  \\
\hline
E & $N$ &  $c$ & $\epsilon$  \\
\hline
F & $N$  & $c$ & $N$  \\
\hline
G & $N$  & $N$ & $\epsilon$  \\
\hline
H & $N$  &  $N$ & $N$  \\
\hline
\hline
\end{tabular}
\end{center}
Table 1: Cases relative to the various behavior of the arithmetic mean, the geometrical mean and the variance. 

\end{table}

\vspace{3mm}
\noindent
{\bf Case A}. Without losing generality, if the arithmetic mean is finite we can assume to be zero, since we can always shift the variable of 
the relative polynomial as $z \rightarrow z - \langle z \rangle_{\rm arith}$. So this case concerns with 
\be 
 \mid {\bar z}_{\rm geom}\mid \simeq \xi
\,\,\,\,\,\,\,
,
\,\,\,\,\,\,\,
\mid {\bar z}_{\rm arith}\mid \,=\,0 
\,\,\,\,\,\,\,
,
\,\,\,\,\,\,\,
\mu^2 \simeq \epsilon
\ee
The fact that the arithmetic mean vanishes obviously implies that the barycenter of the zeros is the origin of the complex plane; since the geometrical mean of this case goes instead as a constant for large $N$, this can be interpreted as the fact that the module of all the zeros (except probably few) cannot grow with $N$. Finally, since the variance vanishes for $N \rightarrow\infty$ this seems to imply a certain localization of the zeros, so most probable the zeros of this case satisfy a perimeter law. Notice all these features are realised by zeros placed along a circle of radius $\xi$, as for instance  
\be 
A(z) \,=\, z^N + \xi^N \,\,\,. 
\ee
In this case, in fact $\langle z \rangle_{geom} = \xi$, $\langle z \rangle_{arith} = 0$ and $\mu^2 = 0$. The distribution of the zeros of this polynomial satisfies 
indeed the perimeter law. Notice that for this polynomial the value $\xi$ sets the upper bound of the module of zeros, as seen applying the Fujiwara bound (\ref{Fujiwarabound}). More generally, all polynomials whose coefficients $\gamma_k$ scale as a power of $k$, i.e. $\gamma_k \simeq k^{\delta}$, 
where $\delta$ is either positive or negative real number, have asymptotically as Enestr\"{o}m bounds $R_d = R_u =1$ and  
for $N \rightarrow \infty$ 
\begin{eqnarray}
&& \langle z \rangle_{geom} \,=\, (N^{\delta})^{1/N} \rightarrow 1\,\,\,, \nonumber \\
&& \langle z \rangle_{arith} \,=\,-\frac{1}{N} \left(\frac{N-1}{N}\right)^k \rightarrow 0 
\,\,\,, \\ 
&&\mu^2 \,=\,\frac{1}{N} \left[\left(\frac{N-1}{N}\right)^{2k} - 2 \left(\frac{N-2}{N}\right)^k \right] 
\rightarrow 0 \,\,\, ,\nonumber
\end{eqnarray} 
namely all of them belong to the class $A$. Therefore we expect that the polynomials of this class will satisfy the perimeter law, apart 
eventually few zeros displaced somewhere else. Few experiments, in addition to the one shown in Figure \ref{Distributions}, are reported in Figure \ref{newexperiments}, and indeed they show that the zeros of these polynomials satisfy the perimeter law. 

\begin{figure}[t]
\centering
\includegraphics[width=0.48 \textwidth]{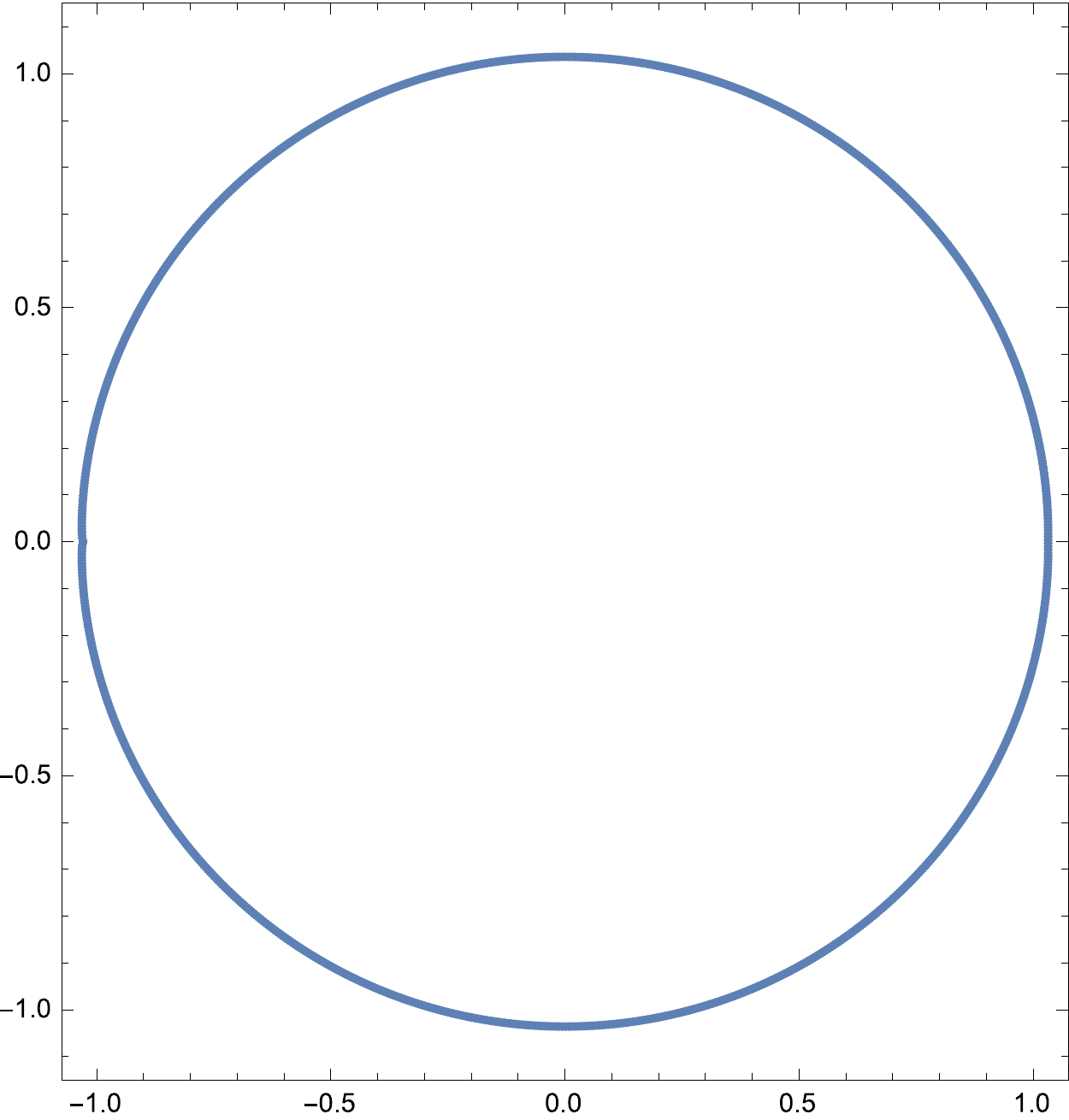}
\includegraphics[width=0.48 \textwidth]{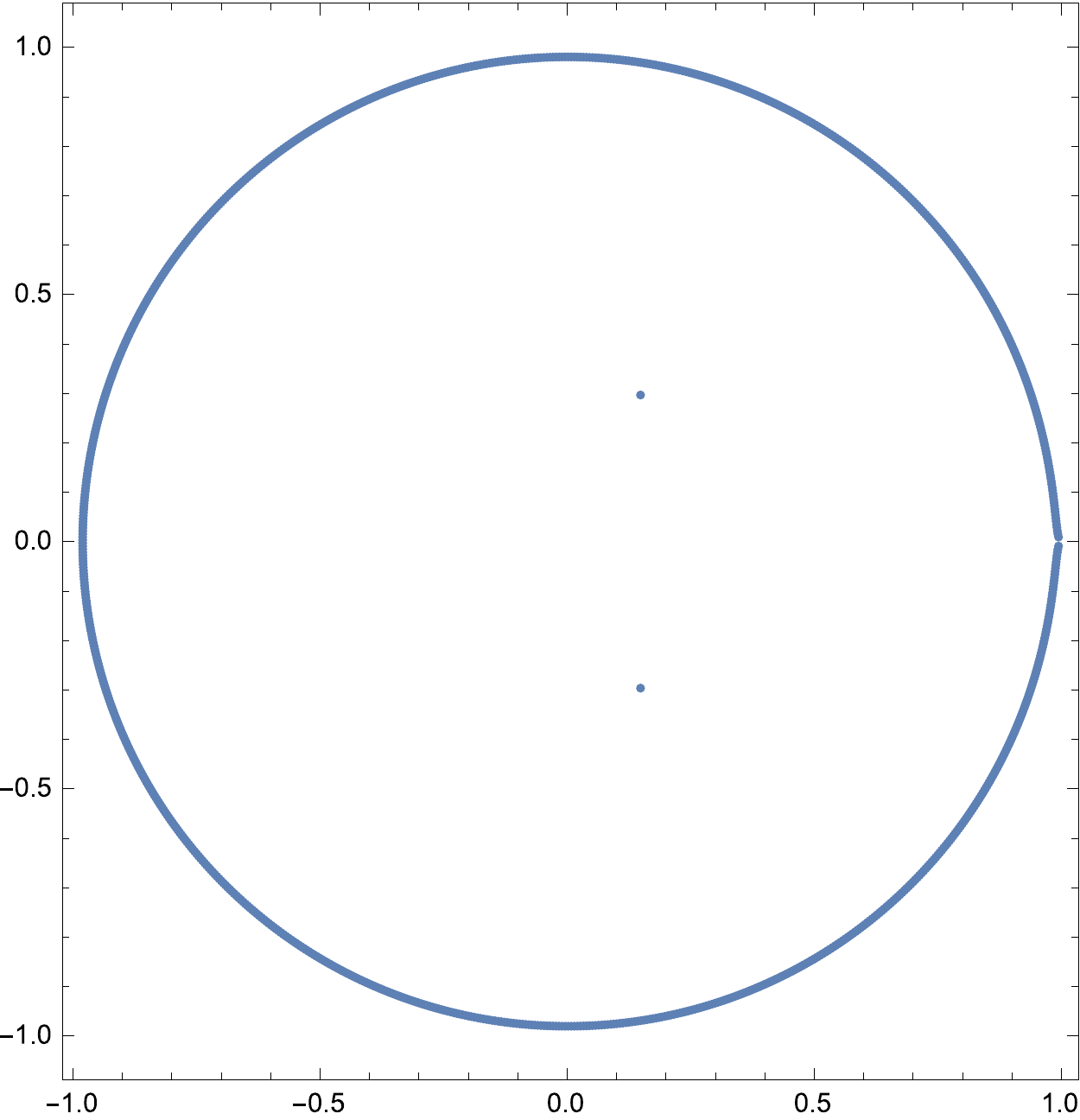}
\caption{a) Distribution of zeros of the partition function whose coefficients $\gamma_k$ go as $1/k^5$. b) Distribution of zeros of the partition function whose 
coefficients $\gamma_k$ go as $k^3$. Notice that in both cases, all zeros, apart few, are placed along a circle, i.e. they satisfy the perimeter law.}
\label{newexperiments}
\end{figure}

\vspace{3mm}
\noindent
{\bf Case B}. This case concerns with the values 
\be 
 \mid {\bar z}_{\rm geom}\mid \simeq \xi
\,\,\,\,\,\,\,
,
\,\,\,\,\,\,\,
\mid {\bar z}_{\rm arith}\mid \,=\,c 
\,\,\,\,\,\,\,
,
\,\,\,\,\,\,\,
\mu^2 \simeq N
\ee
A polynomial which has such values is 
\be
B(z) \,=\, z^N + N z^{N-1} + 1 \,\,\,, 
\label{Bpoly}
\ee
(plus all other lower coefficients which however scale with lower powers in $N$) since 
\begin{eqnarray}
&& \langle z \rangle_{geom} \,=\, (1)^{1/N} \,=\, 1\,\,\,, \nonumber \\
&& \langle z \rangle_{arith} \,=\,- 1\  
\,\,\,, \\ 
&&\mu^2 \,=\,\frac{1}{N} \left(1-\frac{1}{N}\right) N^2\  
\rightarrow N \,\,\,. \nonumber
\end{eqnarray} 
Notice that, in general, to have a finite geometrical mean not all zeros can grow as $N$, the same is also true in order to have a finite value for the arithmetic mean.
However, to have a variance which grows as $N$ it is enough that some of the zeros are far off from $\langle z \rangle_{arith}$, order $N$. This is indeed the case
for the roots of the polynomial (\ref{Bpoly}), where one of them is far away from the arithmetic mean, $z_* \sim - N$, while all the others are essentially along a circle of radius 1 and center at $\langle z \rangle_{arith} = - 1$. Therefore for polynomials of this class we expect that, apart few zeros, the others satisfy a perimeter law. 
Notice that in this case a bound such as the Fujiwara bound, taken alone, is pretty loose since it states that all roots satisfy $|z_i| < 2 N$, a condition which is 
indeed true but in this case only one root is order $N$.  

\vspace{3mm}
\noindent
{\bf Case C}. This case concerns with the values 
\be 
 \mid {\bar z}_{\rm geom}\mid \simeq \xi
\,\,\,\,\,\,\,
,
\,\,\,\,\,\,\,
\mid {\bar z}_{\rm arith}\mid \,\sim N
\,\,\,\,\,\,\,
,
\,\,\,\,\,\,\,
\mu^2 \simeq N
\ee
A polynomial with these properties is given by 
\be
C(z) \,=\, (z - N)^{\frac{N}{2}} \left( z - \frac{1}{N}\right)^{\frac{N}{2}}
\ee
Assume $N$ to be an even number. This polynomials has $N/2$ zeros at $z = N$ and $N/2$ zeros at $z = 1/N$. Therefore the geometrical mean 
is equal to 1 while the arithmetic mean is equal to $\bar{z}_{\rm arith} = (N + 1/N)/2$ and its variance $\mu^2 = (N - 1/N)^2$. In this case 
we are in presence of a bunch concentration of zeros (in this case placed at reciprocal positions) which essentially do not satisfy neither area or perimeter laws.

\vspace{3mm}
\noindent
{\bf Case D}. This case concerns with 
\be 
 \mid {\bar z}_{\rm geom}\mid \simeq \xi
\,\,\,\,\,\,\,
,
\,\,\,\,\,\,\,
\mid {\bar z}_{\rm arith}\mid \,=\,N 
\,\,\,\,\,\,\,
,
\,\,\,\,\,\,\,
\mu^2 \simeq \epsilon
\ee
A corresponding polynomial with these features is given by 
\be
D(z) \,=\, 1+ (N^4 - N^3 + N) \, z^{N-2} + 2 N \, z^{N-1} + 2 z^{N} \,\,\,
\label{Dpoly}
\ee
plus lowest order coefficients which scale with lower powers in $N$. Such a polynomial, apart few of its zeros which have large modules (order N) so that their arithmetic average goes with $N$, has the rest of the zeros with bounded modules. Since their spread is small, most probably we are in presence of a perimeter law. In the case of the polynomial (\ref{Dpoly}), its bounded zeros are indeed all around a circle. 

\vspace{3mm}
\noindent
{\bf Case E}. This case concerns with 
\be 
 \mid {\bar z}_{\rm geom}\mid \simeq N
\,\,\,\,\,\,\,
,
\,\,\,\,\,\,\,
\mid {\bar z}_{\rm arith}\mid \,\simeq \,c 
\,\,\,\,\,\,\,
,
\,\,\,\,\,\,\,
\mu^2 \simeq \epsilon
\ee
A representative polynomial with these features is 
\be
E(z) \,=\, 1 + \frac{N(N+1)}{2 N!}\, z^{N-2} + \frac{1}{(N-1)!} \, z^{N-1} + \frac{1}{N!} \, z^N\,\,\,, 
\label{epoly}
\ee
In order to compute the various statistical quantities we need the Stirling formula for the factorial 
\be 
N! \simeq \sqrt{2 \pi N} \,\left(\frac{N}{e}\right)^N \,\,\,. 
\ee
The divergence with $N$ of the geometrical mean implies that almost all zeros has a module which increases with $N$ although their barycenter remains at a finite distance. The infinitesimal value of the variance once again suggests that we may be in presence of a perimeter law. As a matter of fact, almost all roots of the polynomial (\ref{epoly}), varying $N$, are along a circle of increasing radius. 

\vspace{3mm}
\noindent
{\bf Case F}. This case concerns with 
\be 
 \mid {\bar z}_{\rm geom}\mid \simeq N
\,\,\,\,\,\,\,
,
\,\,\,\,\,\,\,
\mid {\bar z}_{\rm arith}\mid \,\simeq \,c 
\,\,\,\,\,\,\,
,
\,\,\,\,\,\,\,
\mu^2 \simeq N
\ee
This class may have as a significant representative a polynomial which employs the prime numbers $P_k$ 
\be 
F(z) \,=\, 1 + P_1 z + \frac{1}{2!} P_2 z^2 + \cdots \frac{1}{N!} P_N z^N
\label{primepoly}
\ee
In order to compute the various statistical quantities, in addition to the Stirling formula for the factorial, we also need  
the approximate formula for the $N$-the prime $P_N$ 
\be 
P_N \simeq N\, \log N \,\,\,.
\ee
Therefore 
\begin{eqnarray}
&& |\bar{z}_{\rm geom} | \,=\, \left(\frac{1}{\gamma_N}\right)^{1/N} \sim N \,\,\,,\nonumber \\
&&\bar{z}_{\rm arith}  \,=\, - \frac{1}{N} \frac{\gamma_{N-1}}{\gamma_N} \,\sim -1 \\
&& \mu^2 \sim N \nonumber 
\end{eqnarray}
Applying either the Fujiwara or the Enestr\"{o}m bound, one sees that the radius of the disc which includes all zeros increases linearly with $N$. All these 
behaviors suggest that the zeros of the polynomial (\ref{primepoly}) satisfy the area law, which is indeed the case, as shown in Figure \ref{Distributions}. 

\vspace{3mm}
\noindent
{\bf Case G}. This case concerns with 
\be 
 \mid {\bar z}_{\rm geom}\mid \simeq N
\,\,\,\,\,\,\,
,
\,\,\,\,\,\,\,
\mid {\bar z}_{\rm arith}\mid \,\simeq \,N 
\,\,\,\,\,\,\,
,
\,\,\,\,\,\,\,
\mu^2 \simeq \epsilon
\ee
A polynomial with these features is given by 
\be
G(z) \,=\, (z - N)^N \,\,\,.
\label{A}
\ee
The zeros of this polynomial are obviously all at $z_* = N$. Both their arithmetic and geometric means are equal to $N$ and therefore diverge when $N\rightarrow \infty$, however their variance is identically zero. When we are in presence of these values of the indicators it is quite probable that the distribution of zero is sharply 
peaked around a value $z_*$ that grows with $N$. Also in this case the zeros do not satisfy neither area or perimeter laws. 

\vspace{3mm}
\noindent
{\bf Case H}. This case concerns with 
\be 
 \mid {\bar z}_{\rm geom}\mid \simeq N
\,\,\,\,\,\,\,
,
\,\,\,\,\,\,\,
\mid {\bar z}_{\rm arith}\mid \,\simeq \,N 
\,\,\,\,\,\,\,
,
\,\,\,\,\,\,\,
\mu^2 \simeq N
\ee
A representative polynomial may be 
\be
H(z) \,=\, (z - N) (z - N +1) (z - N +2 ) \cdots (z - 2 N) \,\,\,.
\label{H}
\ee
The zeros of this polynomial are $z = N, N+1, \ldots, 2 N$. Their arithmetic mean is $\bar{z}_{\rm arith} = 3(N +1)/2$, its geometrical mean 
goes as $\bar{z}_{\rm geom} \sim N$ and its variance goes as $\mu^2 \sim N^2/12$. The geometrical shape of these zeros consists of a 
cluster of $N$ isolated zeros, all of order $N$, which move away from the origin enlarging their spreading. This kink of zeros do not satisfy neither 
area nor perimeter law but are rather discrete isolated points which move at $z = \infty$ by increasing $N$. 

\vspace{3mm}

Let's close our discussion on the patterns of zeros by commenting on the information one can extract from the statistical indicators. To this aim let's consider an another example of zeros which satisfy the area law, as the case F analysed above: this example consists of a polynomials whose coefficients $p_m$ are given not by the prime numbers but by their approximation $p_m = [m \log m]$, where $[x]$ denotes the integer part of the real number $x$. In this case all the three statistical indicators (as well as the bounds) coincide with those of the polynomial based exactly on the prime numbers and therefore also for this example we expect that its zeros satisfy the area law. They do indeed but their actual distribution has a finer structure, with arcs and other curves along which the zeros are placed, as evident from their plot in the complex plane shown in Figure \ref{NLOGNFIGURE}. This example shows explicitly that the arithmetic and geometrical means of the zeros and their variance may help in capturing the gross structure of the zeros but of course not the detailed features of their actual distribution. 
\begin{figure}[t]
\begin{center}
%\vspace{8mm}
\psfig{figure=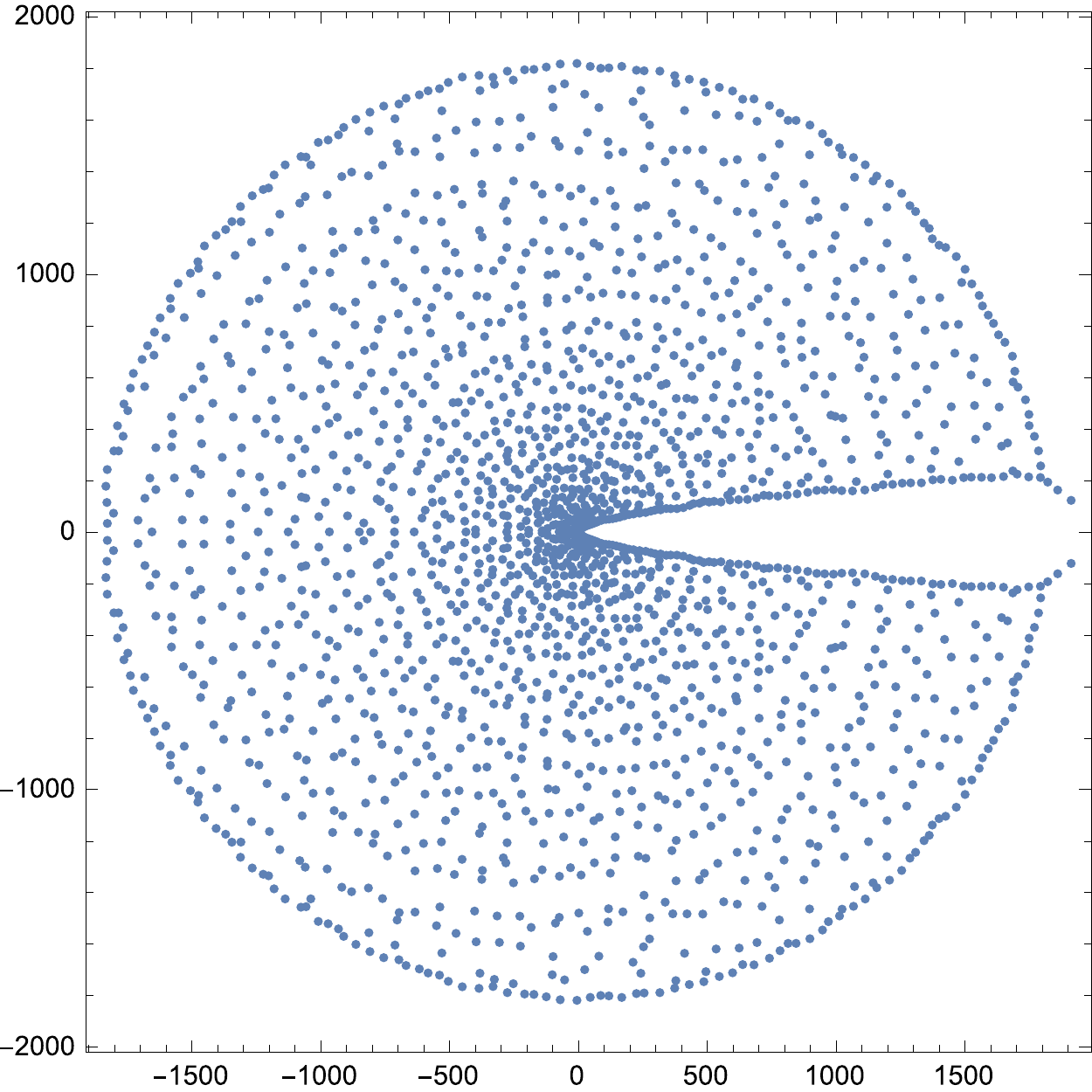,height=8cm,width=8cm}
\caption{An example of area law: distribution of zeros of the partition function where the coefficients are given by 
$a_n = [n \log n]$.}
\label{NLOGNFIGURE}
\end{center}
\end{figure}

\section{Free energy of the free bosonic and fermionic theories}\label{quarta}
\noindent
In this Section we consider quantum theories in which there is no interactions among the particles. For these theories 
we can study the distribution of the zeros and learn some interesting facts about these zeros. The theories that we consider are  
\begin{enumerate}
\item non-relativistic free bosonic and fermionic theories in $d$ dimensions;
\item non-relativistic bosonic and fermionic theories in a (spherical symmetrical) harmonic trap in $d$-dimensions;
\item relativistic free bosonic and fermion theories in $d$ dimensions. 
\end{enumerate}
As shown below, all these cases can be treated altogether. The starting point is the grand-canonical partition function $\Omega$ for bosons and fermions 
\citep{Huang} 
%(see Huang, formula 8.62)
\begin{equation}
\Omega(z) \,=\, \prod_{p} \left(1\pm z \, e^{-\beta \epsilon_p}\right)^{\pm 1} \,\,\,,
\label{directformula}
\end{equation}
where $+$ refers to fermion and $-$ to boson, with the infinite product extended to all values $p$ of the momentum which 
parameterises the energies of the excitation. Taking the logarithm of both sides we end up in
\begin{equation}
\log \Omega(z) \,=\,\pm \,\sum_p \log \left(1 \pm z \, e^{-\beta \epsilon_p}\right)\,\equiv F_{\pm}(z,\beta) \,\,\,,
\end{equation}
where 
\begin{equation}
F_{\pm}(z,\beta) =\pm \int d\epsilon \, g(\epsilon) \, \log \left(1 \pm z \, e^{-\beta \epsilon}\right) \,\,\, . 
\label{generalfreeenergy}
\end{equation}
This universal way to express the free energy simply employs the {\em density of states} $g(\epsilon)$ of each system, whose definition is 
\begin{equation}
g(\epsilon) \,=\,\int \frac{d{\bf r} \, d{\bf p}}{(2 \pi \hbar)^d} \, \delta(\epsilon - H(r,p)) \,\,\,.
\end{equation}
%Hence, each situation is just characterised by its own density of states. 
We report hereafter the expression of the various densities of states (got by a straightforward calculation) and the expression of the relative free-energies for the various cases enumerated above. 

\begin{enumerate}
\item {\bf Non-relativistic free bosonic and fermionic theories in $d$ dimensions}. Both these theories have the Hamiltonian 
\be 
H = \frac{p^2}{2m} \,\,\,, 
\ee 
and their density of states is given by  
\begin{equation} 
g_{nr}(\epsilon) \,=\, V \left(\frac{m}{2 \pi \hbar^2}\right)^{\frac{d}{2}} \,\frac{1}{\Gamma\left(\frac{d}{2}\right)} \, \epsilon^{\frac{d}{2}-1} \,\,\,.
\label{freenonrel}
\end{equation}
Substituting this expression in eq.\,(\ref{generalfreeenergy}) and expanding the logarithm for $| z | < 1$,  we have 
\begin{eqnarray}
F_{\pm}(z,\beta) &\,=\,&  \sum_{k=1}^\infty \int_0^{\infty} d\epsilon g(\epsilon) (\mp 1)^{k+1} \frac{z^k}{k} \,e^{-k \beta \epsilon} \nonumber \\
& = & \frac{V}{\lambda_T^d} \,f_{\pm}(z; d) \,\,\,,
\label{finalexpressionnr}
\end{eqnarray}
where $\lambda_T$ is the thermal wave-length 
\begin{equation}
\lambda_T \,=\, \left(\frac{2 \pi \hbar^2}{m k T}\right)^{1/2} \,\,,
\label{thermallength}
\end{equation}
while the functions $f_{\pm}(z; d)$ depend upon {\em only} $z$ and the dimensionality $d$, but not from the temperature $T$
\begin{equation}
f_{\pm}(z; d) \,=\, \sum_{k=1}^\infty (\mp 1)^{k+1} \frac{z^k}{k^{\frac{d}{2} +1}} \,\,\,.
\end{equation}
These functions can be expressed in terms of poly-logarithmic functions, defined by 
\begin{equation}
L_s(z) \,=\,\sum_{k=1}^\infty \frac{z^n}{n^s} \,\,\,.
\label{polylogarithm}
\end{equation}
More precisely, for the boson, we have 
 \begin{equation}
 f_-(z;d) \,=\, L_{\frac{d}{2} +1}(z) \,\,\,, 
 \end{equation}
while for the fermion 
\begin{equation}
 f_+(z;d) = - L_{\frac{d}{2} +1}(-z) \,\,\,.
\end{equation}
\item 
{\bf Non-relativistic bosonic and fermion theories in a (spherical symmetrical) harmonic trap in $d$-dimensions.}The Hamiltonian in this case is 
given by 
\be 
H = \frac{p^2}{2 m} + \frac{1}{2} m \omega^2 r^2 \,\,\,, 
\ee
and the density of states is 
\begin{equation} 
g_{h}(\epsilon) \,=\,  \left(\frac{1}{\hbar \omega}\right)^{d} \,\frac{1}{\Gamma\left(d\right)} \, \epsilon^{d - 1} \,\,\,.
\label{harmonic}
\end{equation}
Notice that, apart from the prefactors, the density of states for the harmonic trap in $d$-dimension is the {\em same} of the one of the free theories 
but in a dimension twice as large! Therefore, repeating the same computations as before, the expression of the free energies is 
\begin{equation}
F_{\pm}(z,\beta) \,=\, \left(\frac{1}{\hbar \omega}\right)^d \, f_{\pm}(z; 2 d) \,\,\,. 
\end{equation}
\item {\bf Relativistic free bosonic and fermionic theories in $d$ dimensions}. The Hamiltonian of these cases is 
 \be 
 H = \sqrt{p^2 + m^2} \,\,\,, 
 \ee
and for the density of states we have 
\begin{equation}
g_r(\epsilon) \,=\, 2 V \left(\frac{1}{4 \pi \hbar^2}\right)^{\frac{d}{2}} \, \epsilon \, \left(\epsilon^2 - m^2)\right)^{\frac{d}{2} -1} \,\,\,.
\label{densitystaterel}
\end{equation} 
Repeating the same steps as before, we end up in the following expression of the free energies
\begin{equation}
F_{\pm}(z,\beta) \,=\,  \frac{V}{\lambda_T^d} \sqrt{\frac{2 m \beta}{\pi}}  \, H_{\pm}(z;d;\beta) \,\,\,,
\label{relativistic1}
\end{equation}
where 
\begin{equation}
H_{\pm}(z;d;\beta) \,=\, \sum_{n=1}^\infty \frac{(\mp 1)^{n+1}}{n^{\frac{d+1}{2}}} \, K_{\frac{d+1}{2}}(n \beta m) z^n \,\,\,,
\label{relativistic2}
\end{equation}
and $K_\nu(z)$ is the modified Bessel function. 
In these relativistic cases the dependence on the temperature $T$ no longer 
factorises as in the non-relativistic cases, but also enters explicitly the cluster coefficients through the Bessel function. These distributions 
correspond to free bosonic ($-$ subscript) and fermionic fields ($+$ subscript). In particular, for $d=1$ the fermionic distribution is relevant for the two-dimensional classical Ising model, regarded as a one-dimensional quantum model of free Majorana fermions (see, for instance \citep{GM,KS,YZ,Fendley,KM}). 

\end{enumerate}

\section{Distribution of the zeros for non-relativistic free theories}\label{quinta}
\noindent
In this Section we study the distribution of the zeros relative to the non-relativistic free theories. Surprisingly enough, there are I
two alternative ways to address the problem that end up in two different distributions of the zeros. We denote the first method as {\em Truncated Series Approach} (TSA) while the second method as {\em Infinite Product Approach} (IPA). With the TSA, we will get an expression $F_1(z)$ of the free energy of the systems which is valid in the disk $|z| <1$ while, with the IPA, we will get instead an expression $F_2(z)$ which extends to the entire complex plane, providing in particular  the 
analytic continuation of $F_1(z)$ outside its disk of convergence. It is important to notice, though, that both distributions of zeros share the same value 
of all the moments of the zeros. Let's discuss these two methods in more detail. 

\subsection{Truncated Series Approach}\label{TSA}
\noindent
As shown in Section \ref{quarta}, there is a closed expression for the free energies $F_{\pm}(z,\beta)$ for the the non-relativistic free theories of bosons and fermions, see eq.\,(\ref{finalexpressionnr}). However, to determine the Yang-Lee zeros we need to consider the zeros of the partition function itself, alias of a sequence of polynomials $\hat\Omega^{(N)}_\pm(z,\beta)$ in the limit $N \rightarrow \infty$. This sequence of polynomials can be constructed by {\rm truncating} the series relative to $F_{\pm}(z,\beta)$ up to a given order $N$ and then expanding the exponential exactly up to that order. In more detail: 
\begin{enumerate} 
\item 
Let's define the truncated expression of the free energies up to the order $N$ as 
\be 
F_{\pm}^{(N)}(z,\beta) \,\equiv \, \frac{V}{\lambda_T^d}\,\sum_{k=1}^N (\mp 1)^{k+1} \frac{z^k}{k^{\frac{d}{2} +1}}\,\,\,.
\ee
\item
Let's also define 
\be 
\hat \Omega^{(N)}_{\pm}(z,\beta) \,\equiv \, e^{F_{\pm}^{(N)}(z,\beta)}\,\,\,, 
\label{hatomega}
\ee
\end{enumerate}
where $\hat\Omega^{(N)}_{\pm}(x,\beta)$ is given by the Taylor series 
\be 
\hat\Omega^{(N)}_{\pm}(x,\beta) \,\equiv\,\sum_{k=0}^N\frac{p_k}{k!} \,z^k\,=\,  
\sum_{k=0}^N \frac{z^k}{k!} \frac{d^k \hat\Omega^{(N)}_{\pm}}{d z^k}(0,\beta) \,\,\,, 
\label{truncatedpolyn}
\ee
truncated at the order $N$. With the coefficients of the truncated polynomial  $\hat\Omega^{(N)}_{\pm}(x,\beta)$ determined as in eq.\,(\ref{truncatedpolyn}), 
it is easy to see (compare with Section \ref{SectionZeros}) that -- by construction -- the negative moments of the zeros of this polynomial coincide exactly with those given in eq. \,(\ref{bmsm})
\be
\hat s_k \,=\, - \frac{V}{\lambda_T^d}\, (\mp 1)^{k+1} \frac{1}{k^{\frac{d}{2}}}\,\,\,.
\ee
Let's underline that in eq.\,(\ref{truncatedpolyn}) the coefficients $p_k$ relative to the bosonic case are all positive, while those of the fermionic case have alternating sign. In order to have always a positive and increasing value of $F(z)$ by increasing $z$, in the case of fermions it is convenient to always truncate at an even power of $N$.  Let's observe, moreover, that in all the expressions given above of the free energy the dependence upon the volume $V$ is always factorised. So, it is convenient to divide by $V$ and consider the intensive part of all quantities. Equivalently we can just work putting $V=1$, which is what we have done in the rest of the paper for the distributions of the zeros associated to the actual partition functions. The numerical determination of the zeros of the polynomials (\ref{truncatedpolyn}) are shown in Figure \ref{zeronrpartition}, where the module of the roots are expressed in unit of $V/\lambda_T^d$. 
%\begin{figure}[t]
%\centering\
%\includegraphics[width=0.38 \textwidth]{Figure/bose80.eps}
%\includegraphics[width=0.38 \textwidth]{Figure/fermi80.eps}
%\caption{a) Distribution of zeros of the partition function of the boson (left) and fermion (right) for $d=2$ and with% the partition function of eq.\,(\ref{truncatedpolyn}) 
%truncated to the first $80$ terms. }
%\label{zeronrpartition}
%\end{figure}
\begin{figure}[t]
\centering
\includegraphics[width=0.38 \textwidth]{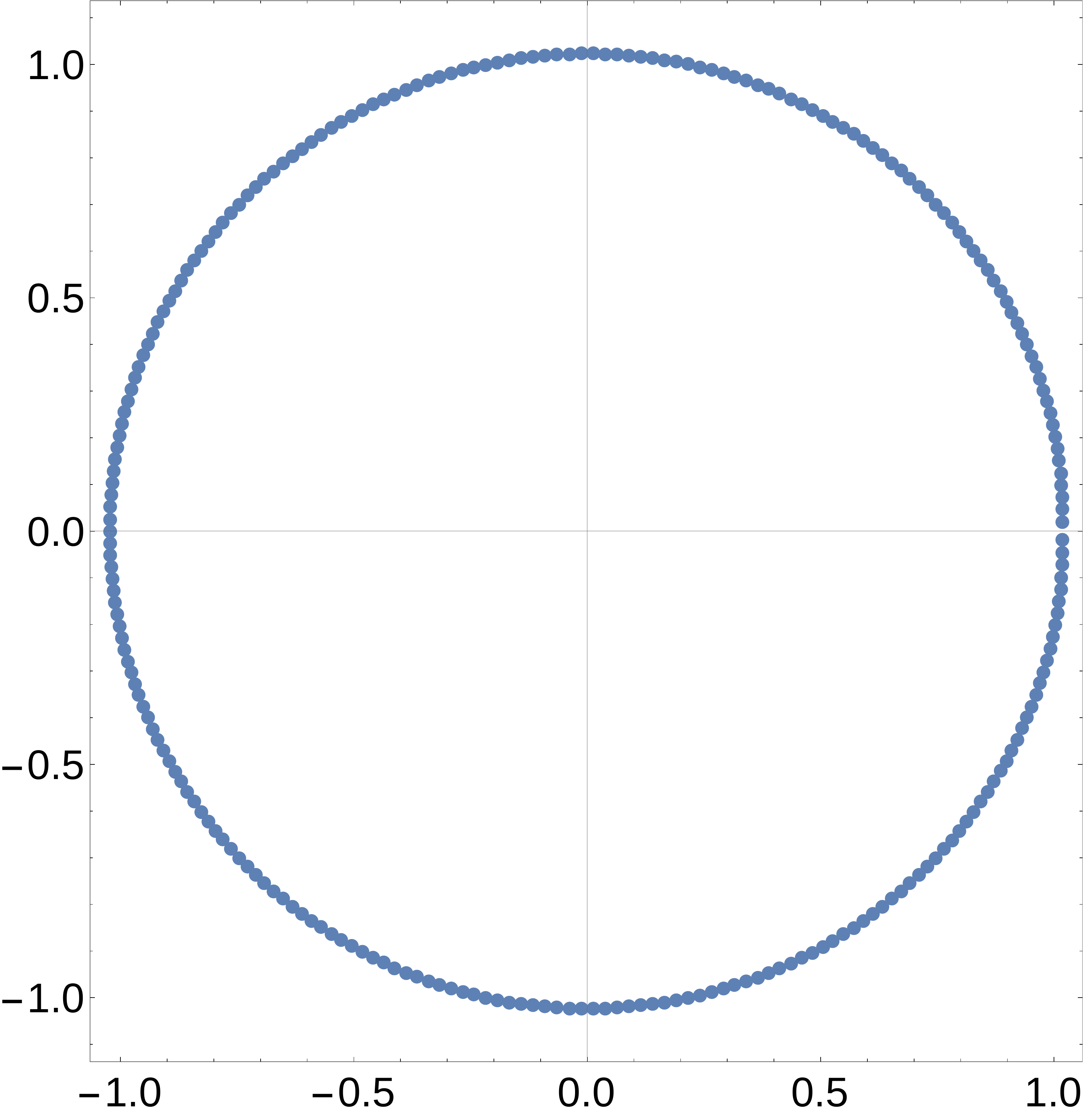}
\includegraphics[width=0.38 \textwidth]{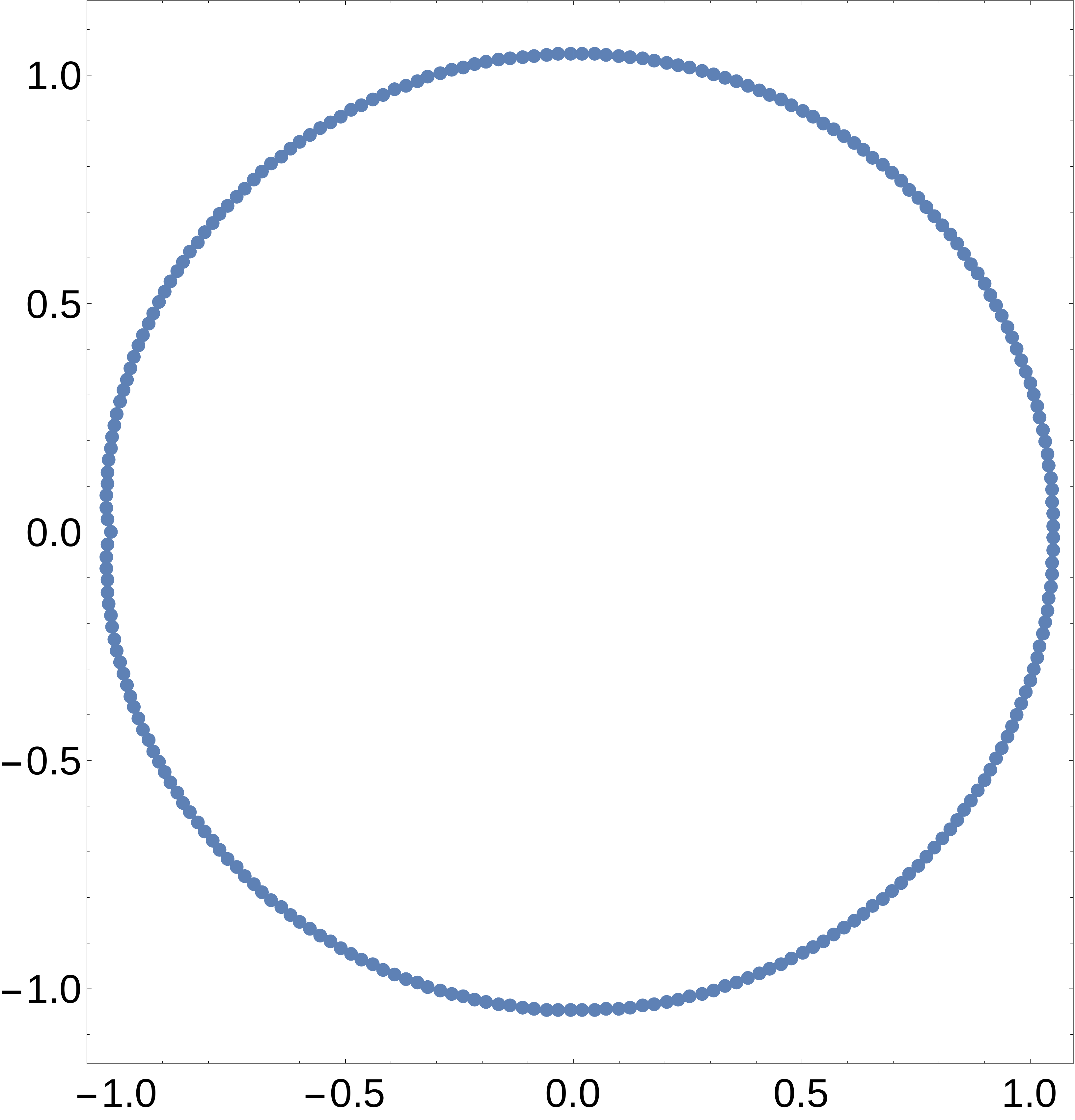}
\caption{a) Distribution of zeros of the partition function of the boson (left) and fermion (right) for $d=2$ and with 
the partition function of eq.\,(\ref{truncatedpolyn}) 
truncated to the first $300$ terms. The module of the roots are in units of $V/\lambda_T^d$.
}
\label{zeronrpartition}
\end{figure}
%\begin{figure}[htbp]
%\begin{center}
%\vspace{8mm}
%\includegraphics[width=0.38\textwidth]{NRF.png}
%\hspace{22mm}
%\includegraphics[width=0.35\textwidth]{NRB.png}
%\caption{a) Distribution of zeros in the fermionic nonrelativistic free theory. Even if they seems to distribute along a unit circumference, the radius of the distribution varies withs the angle that individuates the roots in the complex plane, as showed in the plot in the upper-right part of the figure. b) The same for bosonic nonrelativistic free case.}
%\end{center}
%\label{pippoferm}
%\end{figure}
%\begin{figure}[t]
%\centering
%\includegraphics[width=0.40 \textwidth]{Figure/rdb.eps}
%\,\,\,\,\,\includegraphics[width=0.40 \textwidth]{Figure/rdf.eps}
%\caption{Radius $\rho$ versus the angle $\theta$ for the 80 zeros shown in Figure \ref{zeronrpartition}. The boson case on the left, the fermion case on the right.} 
%\label{radiustheta}
%\end{figure}
\begin{figure}[b]
\centering
\includegraphics[width=0.42 \textwidth]{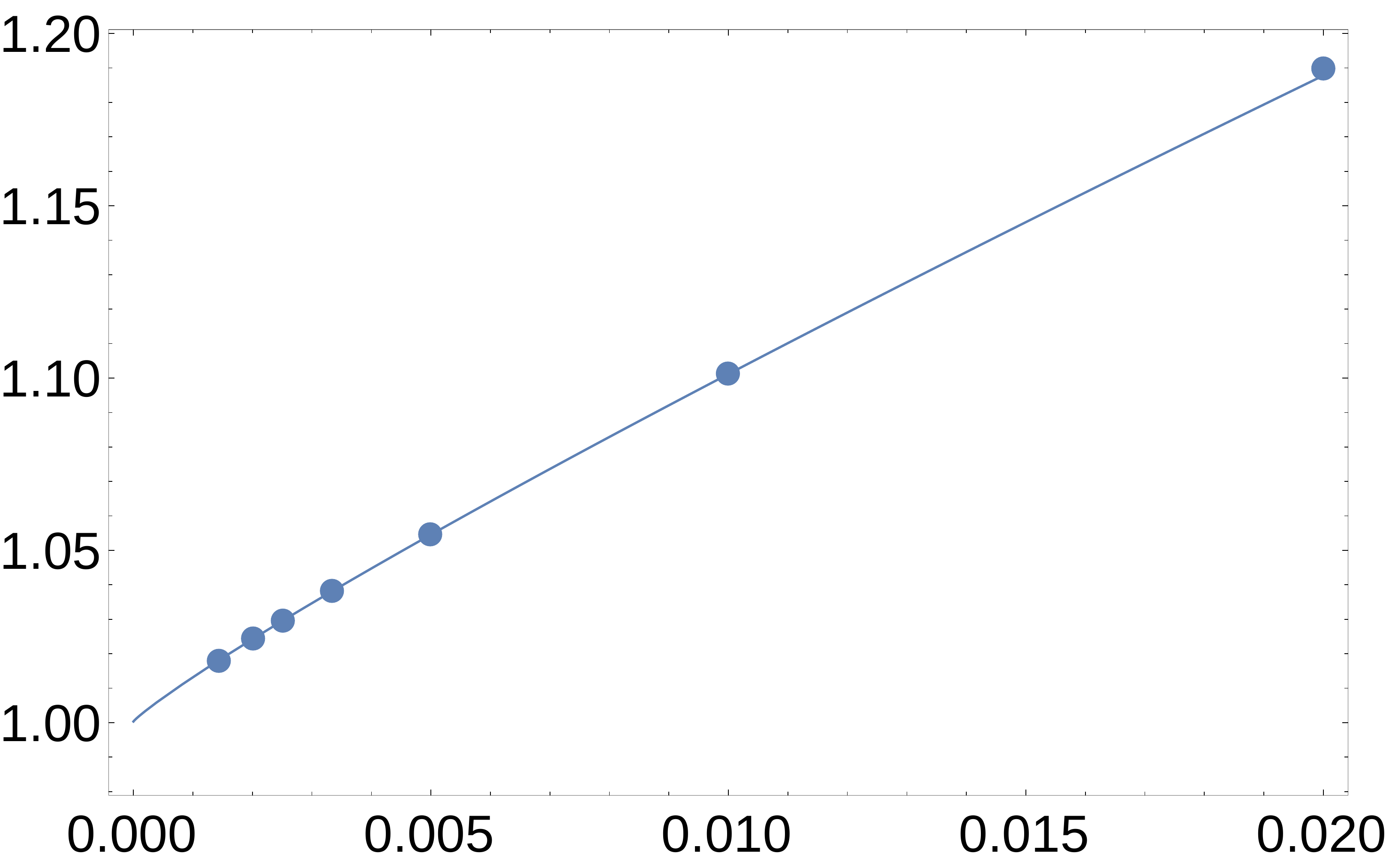}
\caption{
Scaling of the radius of the distribution of zeros of the non-relativistic fermionic free theory for increasing number of terms $N$ of the series at $\beta =1$. The values of the radius $R_N$ are plotted together a function of $1/N^\alpha$, where the best fit of the parameter $\alpha$ in this case is given by $\alpha=0.91 \pm 0.04$. For $1/N \rightarrow 0$ the average radius $R_N$ goes to $1$.}
\label{pipposppp}
\end{figure}
\begin{figure}[t]
\centering
\includegraphics[width=0.42 \textwidth]{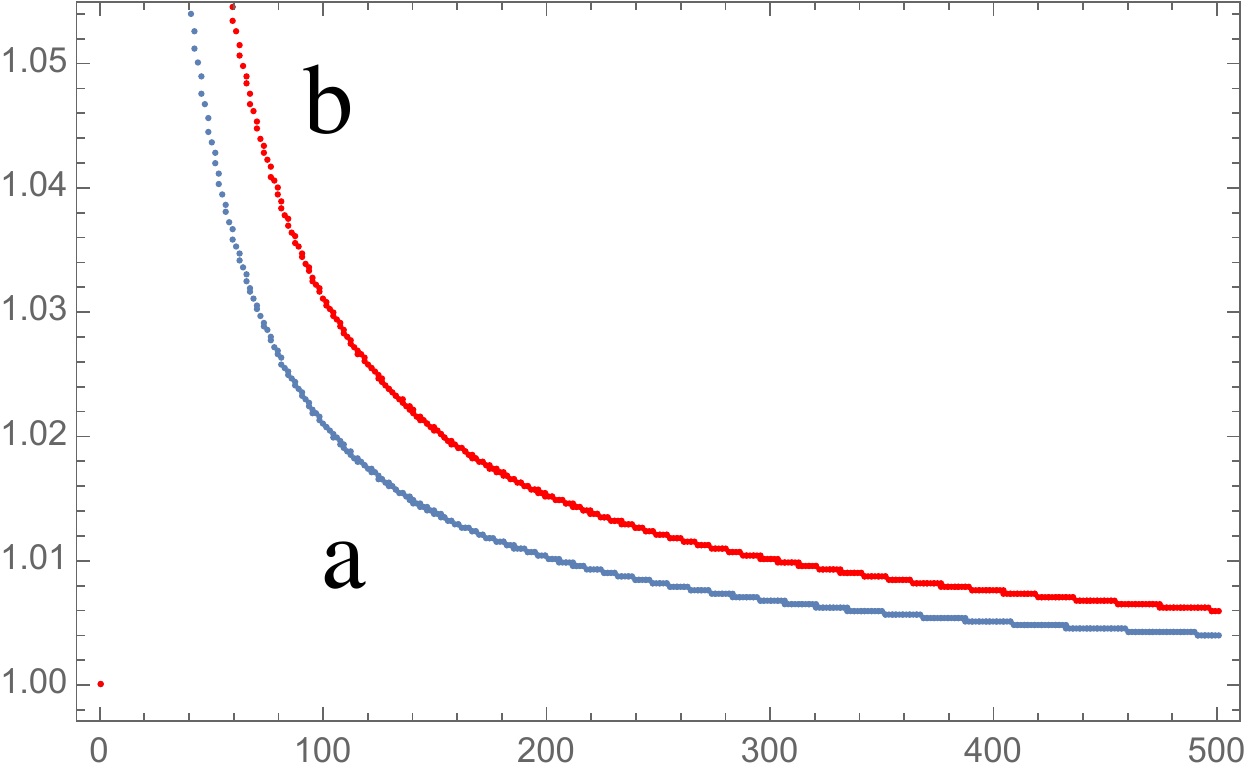}
\caption{
Ratio $r_n = \gamma_n/\gamma_{n-1}$ versus n of the coefficients of $\Omega_N(z)$ for bosonic system with $d=2$ (curve a) and $d=4$ (curve b). 
The minimum and the maximum values of these curve do not coincide: while the minimum tends to 1, the maximum of these sequence is always larger than 1.}
\label{ratiocoefff}
\end{figure}
\begin{figure}[b]
\centering
\includegraphics[width=0.40 \textwidth]{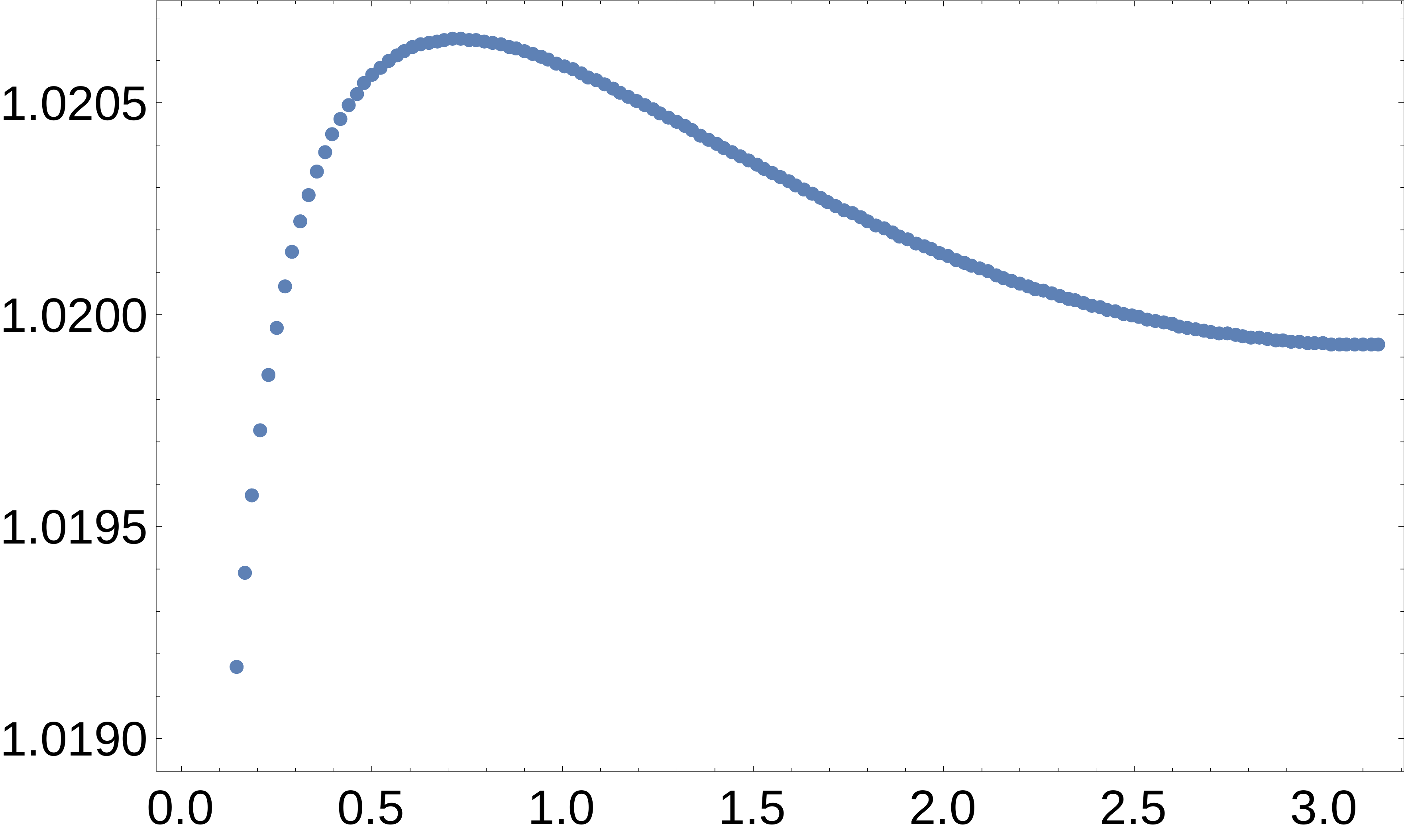}
\,\,\,\,\,\includegraphics[width=0.40 \textwidth]{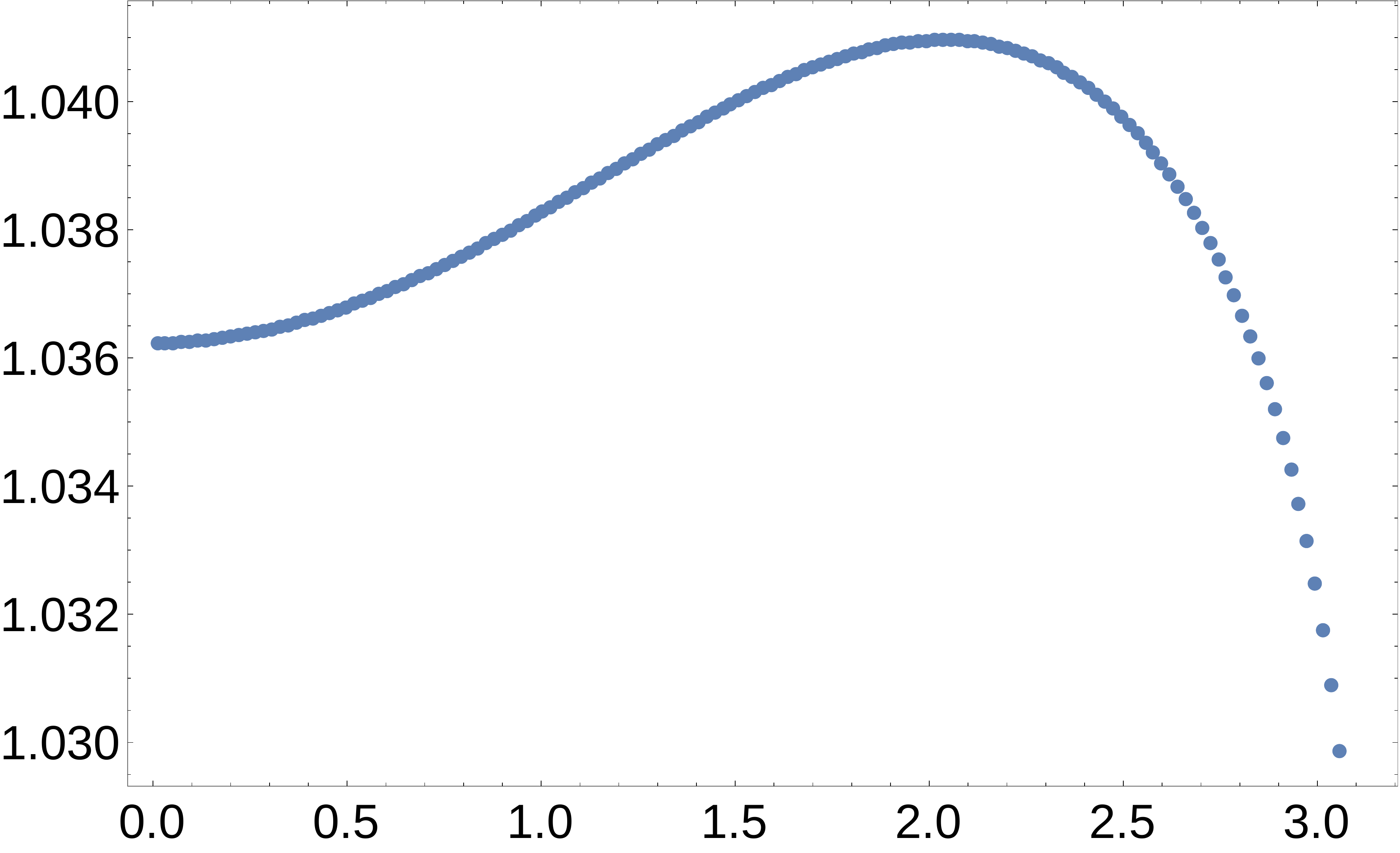}
\caption{Radius $\rho$ versus the angle $\theta$ for the $300$ zeros shown in Figure \ref{zeronrpartition}. The boson case is on the left, the fermion case on the right.} 
\label{radiustheta}
\end{figure}

There are many interesting features of the distribution of these zeros: 
\begin{itemize} 
\item for a finite value of $N$, the zeros are placed on a curve which looks to be a circle but as a matter of fact is not! A fit of their 
location with a circle (i.e. with fit parameters given by the mean radius $R_N$ and the center $z_N^{(0)}$) gives values of the mean radius 
$R_N$ typically larger than $1$ and a non-zero value of $z_0$, therefore the circle determined by the fit is anyhow displaced from the origin. 
However these are just finite-size effects. In fact increasing the values of $N$ and extrapolating the asymptotic values of  $R_N$ and $z_N^{(0)}$ 
we have (see Figure \ref{pipposppp})
\be
\lim_{N\rightarrow \infty} R_N \rightarrow 1 
\,\,\,\,\,\,
,
\,\,\,\,\,\,
\lim_{N\rightarrow \infty} z_N^{(0)} \rightarrow 0 \,\,\,.
\ee 
Notice that the ratio $r_n = \gamma_n/\gamma_{n-1}$ of two consecutive coefficients of $\Omega_N(z)$ is not a flat curve and the minimum $R_d$ and 
the maximum $R_u$ of this sequence do not coincide, see Figure \ref{ratiocoefff}. According to the Enestr\"{o}m bounds, the roots then generally lay in the annulus 
\be 
R_d \leq |z_i|  \leq R_u 
\label{annulus}
\ee 
  
\item However, despite the fact that the asymptotic value of the mean radius being $R_{\infty} =1$ and the apparent circular shape of the zeros as shown in Figure \ref{zeronrpartition}, the asymptotic distribution of the zeros is not on a circle! This can be proved in two different ways. The first consists in parameterising the various zeros $z_k$ as $z_k = \rho_k e^{i \theta_k}$ and plotting the dependence of $\rho_k$ from the angle $\theta_k$: if the distribution was a circle, we shall find a flat function for the radius $\rho(\theta)$ while the actual functions for boson and fermion are the ones shown in Figure \ref{radiustheta}. This figure shows that the radius of the zeros is not constant and this feature persists also in the limit $N \rightarrow \infty$. However, in addition to this numerical evidence, there is a simple but incisive analytic argument to show that in general, for the free non relativistic theories (and for the harmonic trap as well) the zeros are {\em not} placed on a circle. Indeed, if they were on a circle, the coefficients of the expansion of the functions $f_{\pm}(z)$ would be simply related to the Fourier coefficients of the distribution function $\eta(\theta)$ of the zeros along the unitary circle, parameterised by the angle $\theta$ \citep{YL2}. Namely, if 
\begin{equation}
\left(\frac{\lambda_T}{V}\right) \, \log \Omega \,=\, \sum_{n=1}^\infty b_n z^n \,\,\,,
\end{equation}
then 
\begin{equation}
b_n \,=\, -\frac{1}{n} \sum_{l=1}^N \left(\frac{1}{z_l}\right)^n \,=\, -\frac{1}{n} \int_{-\pi}^{\pi} \eta(\theta) \, \cos(n \theta) \, d\theta\,\,\,.
\end{equation}
Hence, the distribution function $p(\theta)$ of the zeros would be given by the Fourier series 
\begin{equation}
\eta(\theta) \,=\, \frac{1}{2\pi} - \frac{1}{\pi}\,\sum_{n=1}^\infty n \, b_n \, \cos(n \theta) \,\,\,.
\label{definitiongtheta}
\end{equation}
The interesting point is now that, for various values of $d$ and both for boson and fermion, one knows how to sum the Fourier series! 
For instance 
\begin{equation}
\sum_{n=1}^\infty \frac{(-1)^{n+1}}{n^2} \cos n \theta \,=\, \frac{1}{4} \left(\frac{\pi^2}{3} - \theta^2\right) \,\,\,,
\label{unoo}
\end{equation}
\begin{equation}
\sum_{n=1}^\infty \frac{1}{n^2} \cos n \theta \,=\, \frac{1}{4} \left(\pi - \mid \theta\mid\right)^2 - \frac{\pi^2}{12} \,\,\,,  
\label{duee}
\end{equation}
\begin{equation}
\sum_{n=1}^\infty \frac{(-1)^{n+1}}{n} \cos n \theta \,=\, \log\left(2 \cos\frac{\theta}{2}\right) \,\,\,,
\label{tree}
\end{equation}
\begin{equation}
\sum_{n=1}^\infty \frac{1}{n} \cos n \theta \,=\, -\log\left(2 \sin\frac{\theta}{2}\right) \,\,\,.
\label{quattroo}
\end{equation}
For all these cases (which corresponds to fermion or boson in different dimensions), one can easily see that the density function $\eta(\theta)$ of the 
zeros becomes {\em negative} in some intervals (see Figures \ref{putative1} and \ref{putative2}). But this is a contradiction, since $\eta(\theta)$ has to be intrinsically a positive function. So, for free non-relativistic theories, this proves that their Yang-Lee zeros cannot be exactly on a circle.
 
\item The distribution of the $N$ zeros, both in the bosonic and in the fermionic case, are dense nearby $z=1$. While this may signal an actual phase transition 
in the bosonic case (and this, we know, only happens for a finite density of the bosonic gas for $d > 2$), for the fermionic case we know that there cannot be a phase transition at $z=1$, therefore this singularity for the fermionic systems simply emerges from the finite radius of convergence of the series (\ref{finalexpressionnr}) employed in the method. The difference between the two cases of course would become evident by employing the analytic continuation of the expression of the pressure given by the original formula (\ref{generalfreeenergy}) in which it will be present the singularity at $z=1$ for the boson and its absence for the fermion. 
\end{itemize}

\begin{figure}[t]
\centering
\includegraphics[width=0.38\textwidth]{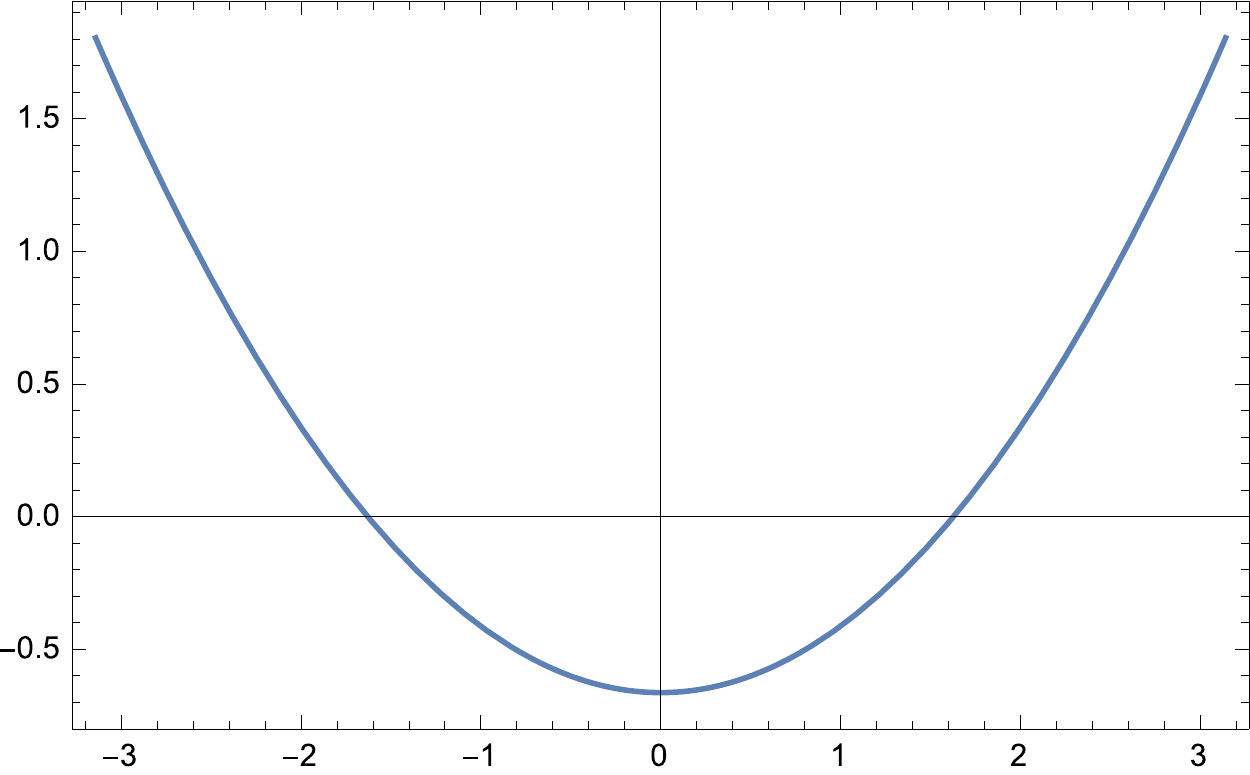}
\,\,\,\,\,\,\,
\includegraphics[width=0.38\textwidth]{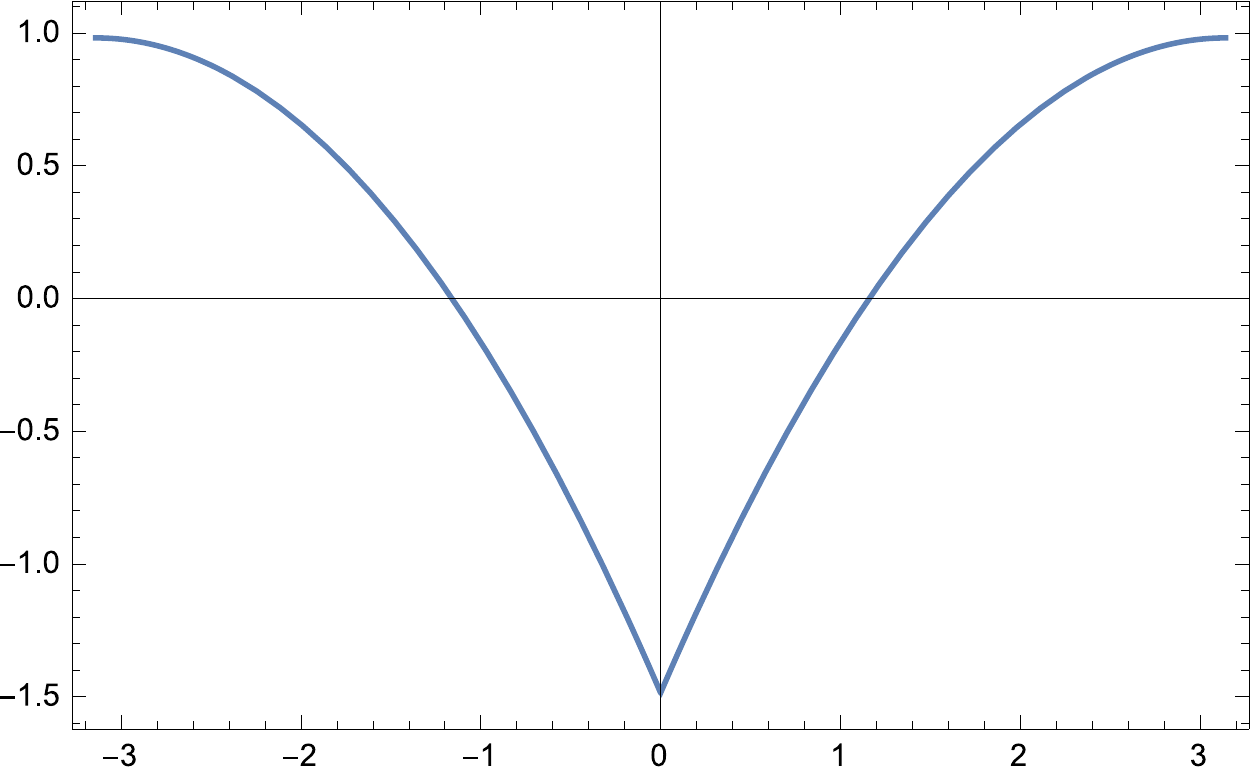}
\caption{Plot of the function $\eta(\theta)$ defined in eq.\,(\ref{definitiongtheta}) with the Fourier series given by eq.\,(\ref{unoo}) (left-hand side plot) and 
eq.\,(\ref{duee}) (right-hand side plot). In both cases the function $\eta(\theta)$ has an interval in which it is negative.}
\label{putative1}
\end{figure}
 
\begin{figure}[b]
\centering
\includegraphics[width=0.38\textwidth]{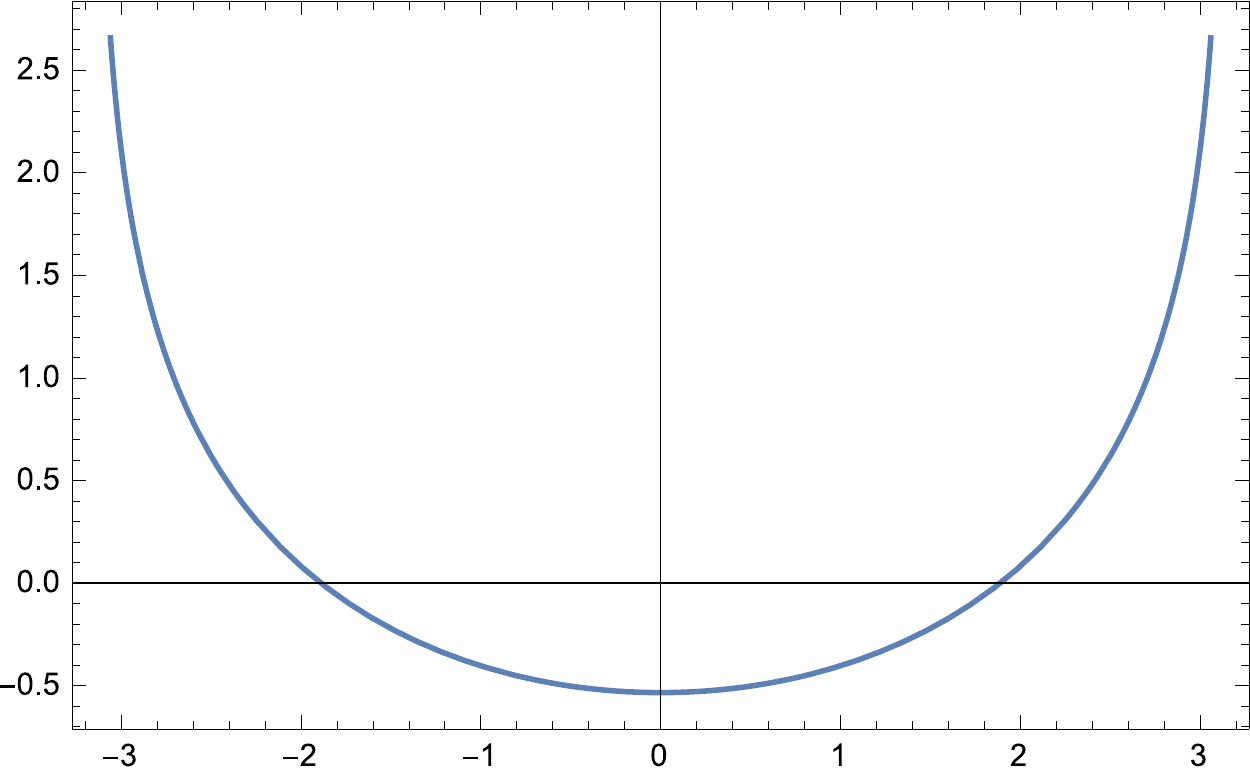}
\,\,\,\,\,\,\,
\includegraphics[width=0.38\textwidth]{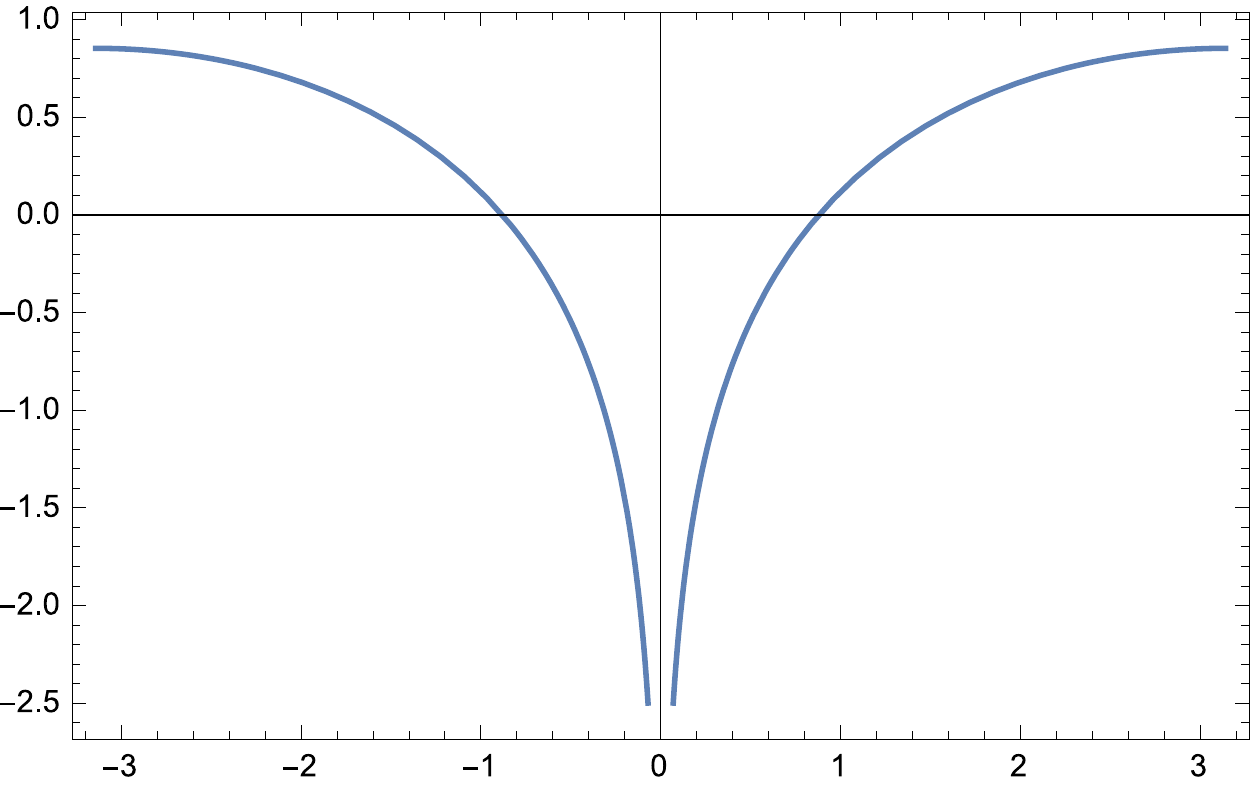}
\caption{Plot of the function $\eta(\theta)$ defined in eq.\,(\ref{definitiongtheta}) with the Fourier series given by eq.\,(\ref{tree}) (left-hand side plot) and 
eq.\,(\ref{quattroo}) (right-hand side plot).In both cases the function $\eta(\theta)$ has an interval in which it is negative.}
\label{putative2}
\end{figure}

\subsection{Infinite Product Approach}\label{IPAIkeda} 
\noindent
Ikeda, in a series of papers \citep{IkedaF,IkedaB}, has worked out the distribution of the zeros of the grand-canonical partition function of free bosons and fermions starting directly from the factorised formula (\ref{directformula}). The particles of these gas do not have hard core and consequently the grand-canonical partition function has always an infinite number of terms, even at a finite volume. Notice that the fermionic grand-canonical partition function has genuine zeros whose physical origin may be traced back to the effective repulsion of particles due to Fermi-Dirac statistics. The bosonic grand-canonical partition function has instead  
poles rather than zeros: however, for what the free-energy of the system is concerned (where it matters taking the logarithm of the partition function), these poles play essentially the same role as the zeros. In light of this remark, let's initially focus the attention on the computation of the density of the zeros for the fermionic case.  

Contrary to the TSA, it is important to stress that in the IPA there is no consistent way of getting a sequence of polynomials of increasing order $N$. Indeed, there are infinitely many polynomials of order $N$ that could be defined: to do so, simply pick up $N$ arbitrary terms out of the infinite product (\ref{directformula}). Therefore in this approach we must necessarily consider at once the totality of all the infinite zeros. 

\begin{figure}[t]
\centering
\includegraphics[width=0.40\textwidth]{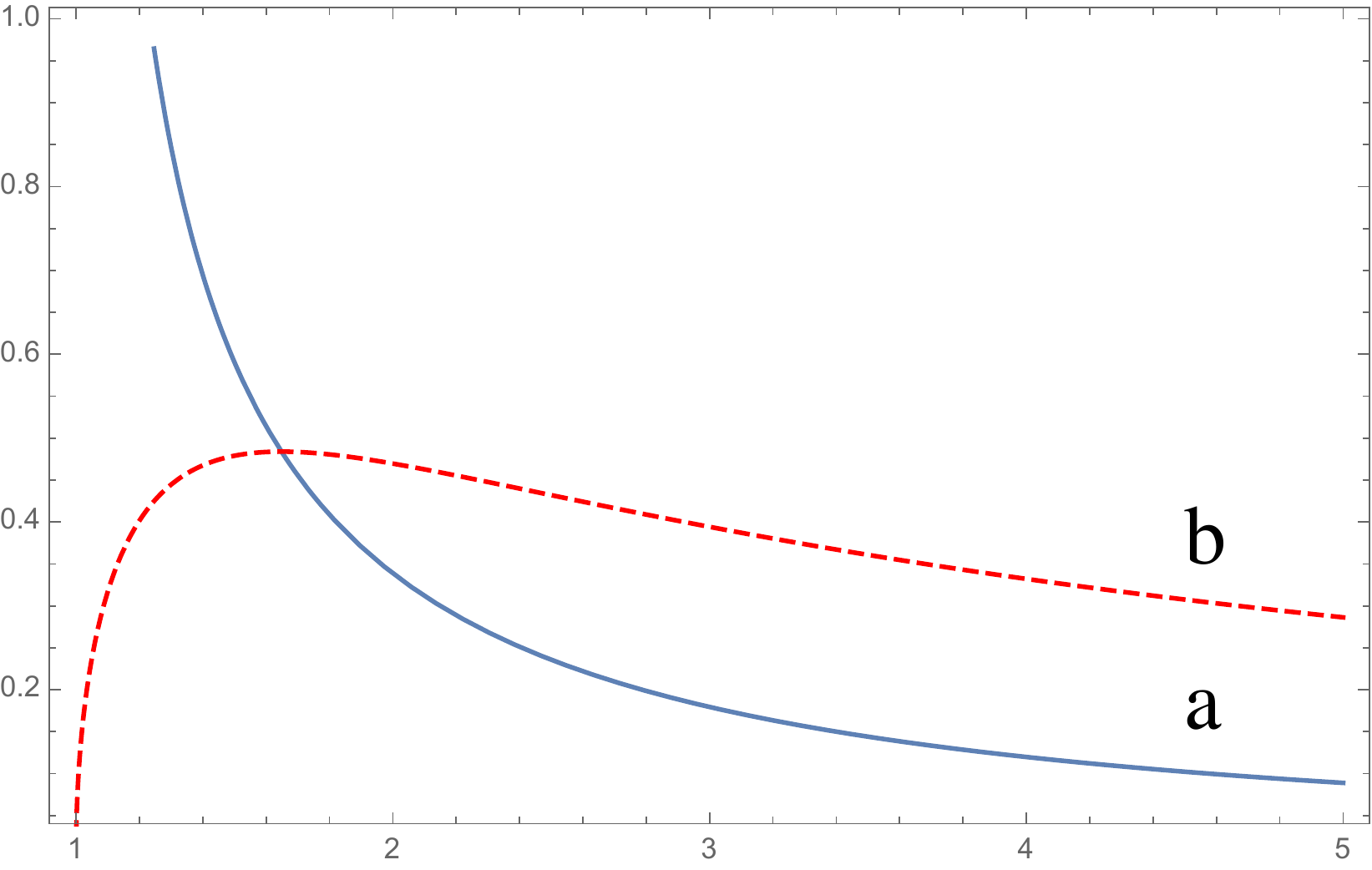}
\caption{Plot of the density of zeros $\eta(-x)$ defined in eq.\,(\ref{dddfff}) along the real axis of the fermionic free theories, where the curve (a) refers to $d=1$ while the curve (b) to $d=3$. }
\label{Ikedaplot}
\end{figure}
 
\vspace{3mm}
\noindent 
{\bf Fermionic case}. From the explicit factorised formula (\ref{directformula}), it is obvious that for the fermionic case the zeros are on the negative real axis, from 
$z = -1$ to $ \infty$, and localised at the positions 
\be 
z_p \,=\, - e^{\beta \epsilon_p} \,\,\,.
\label{equationabove}
\ee
Their density $\eta(z)$ can be explicitly computed using the density of states $g(\epsilon)$. Indeed, from 
eq.\,(\ref{equationabove}), solving for $\epsilon_p$, we have 
\be
\epsilon_p \,=\,k T \, \log(- z_p) \,\,\,,
\ee
and therefore 
\be 
\frac{d\epsilon}{dz} \,=\, \frac{k T}{z} \,\,\,.
\ee
Since the number of energy states with energy values in the shell $(\epsilon, \epsilon + d\epsilon)$ is given by $g(\epsilon) d\epsilon$, the density of zeros 
$\eta(x)$ in the interval between $x$ and $x + dx$ along the negative real line is 
\be
\eta(x)  \,=\, g(\epsilon) \left| \frac{d\epsilon}{d x} \right| \,\,\,. 
\ee
Let's first consider the density of zeros for the non-relativistic fermionic system in $d$ dimensions. Employing eq.\,(\ref{freenonrel}) (with $V =1$), we have 
\be
\eta(x) \,=\,- \left(\frac{m}{2 \pi \hbar^2}\right)^{\frac{d}{2}} \,\frac{1}{\Gamma\left(\frac{d}{2}\right)} \,\frac{k T}{x} (k T \log (-x))^{\frac{d}{2}-1} 
\,\,\,\,\,\,\,
,
\label{dddfff}
\,\,\,\,\,\,\,
-\infty < x < -1 
\ee
This function changes its behavior by varying the dimensionality $d$ of the system, in particular for $d > 2$, it vanishes at $x = -1$, while for 
$d < 2$ it diverges (see Figure \ref{Ikedaplot}). Notice that the density of the zeros given in eq.\,(\ref{dddfff}) is not normalisable 
\be 
\int_{-\infty}^{-1} \eta(x) \,=\, \infty \,\,\,,
\ee
since the number of states of the fermionic gas is infinite, independently from the volume $V$. One can easily compute the negative moments of the 
zeros and see that they correctly coincide with the cluster coefficients of the free energy of the fermionic system 
\be 
b_m \,=\,-\frac{1}{m} \,\int_{-\infty}^{-1} \eta(x) \, x^{-m} \, dx \,=\, (-1)^{m-1} m^{-d/2 -1} \,\,\,.
\label{integralmoments}
\ee
Finally, let's comment that it is also easy to compute the density of zeros also in the relativistic case, the only thing that changes in this case is the density of states employed by the formula which in this case is given in eq.\,(\ref{densitystaterel}). Hence 
\be
\eta(x)  \,=\,- 
2  \left(\frac{1}{4 \pi \hbar^2}\right)^{\frac{d}{2}} \, k T \, \log(- x) \, \left((k T \, \log(- x))^2 - m^2)\right)^{\frac{d}{2} -1} \, \frac{k T}{x} \,\,\,.
\ee
This distribution is defined on an interval determined by the mass, $x \in (-\infty,  - e^{m/kT})$. 

\vspace{3mm}
\noindent
{\bf Bosonic case}. Let's now briefly discuss the bosonic case and the phase transition which this system may have at $z=1$. Repeating the 
same considerations as in the fermionic case, we arrive to the density of zeros given by this distribution 
\be
\eta(x) \,=\, \left(\frac{m}{2 \pi \hbar^2}\right)^{\frac{d}{2}} \,\frac{1}{\Gamma\left(\frac{d}{2}\right)} \,\frac{k T}{x} (k T \log (x))^{\frac{d}{2}-1} 
\,\,\,\,\,\,\,
,
\label{dddbbb}
\,\,\,\,\,\,\,
1 < x < \infty 
\ee
which extends on the positive interval $(1,\infty)$. Hence, this positive interval is made of singular values for the pressure, the smaller singular value being $z=1$. This seems to signal a phase transition of the bosonic system for any value of its dimensionality $d$. Let's compute however the density of the system which, for $| z | < 1$, is given by 
\be 
\rho(z) \,=\, \sum_{k=1}^\infty \frac{z^k}{k^{\frac{d}{2}}}\,\,\,,
\ee
and, at $z=1$, by the Riemann zeta-function $\zeta(s)$ computed at $s=d/2$
\be
\rho(1) \,=\,\zeta\left(\frac{d}{2}\right) \,\,\,.
\label{rhovalues}
\ee
Since for $d \leq 2$, $\rho(1)$ diverges while is finite for $d > 2$, the conclusion is that the bosonic system have a phase transition at a finite 
density of the gas only for $d > 2$, while for lower values of $d$ the phase transition at $z=1$ is only realised at the price to have an infinite density of the 
gas, in other words, correctly it will never happen at a finite density of the gas.

\subsection{Discussion}
\noindent
For the partition function of the free non-relativistic fermionic theories\footnote{The same is also true for the bosonic theories.} we have seen that it can be characterised by two different sets of zeros: the first consisting of zeros approximatively placed along a circle of radius $1$, the second consisting of zeros placed on the negative real axis from $z = -1$ to $- \infty$, see Figure \ref{ipatsazeros}. 
\begin{figure}[b]
\begin{center}
%\vspace{8mm}
\psfig{figure=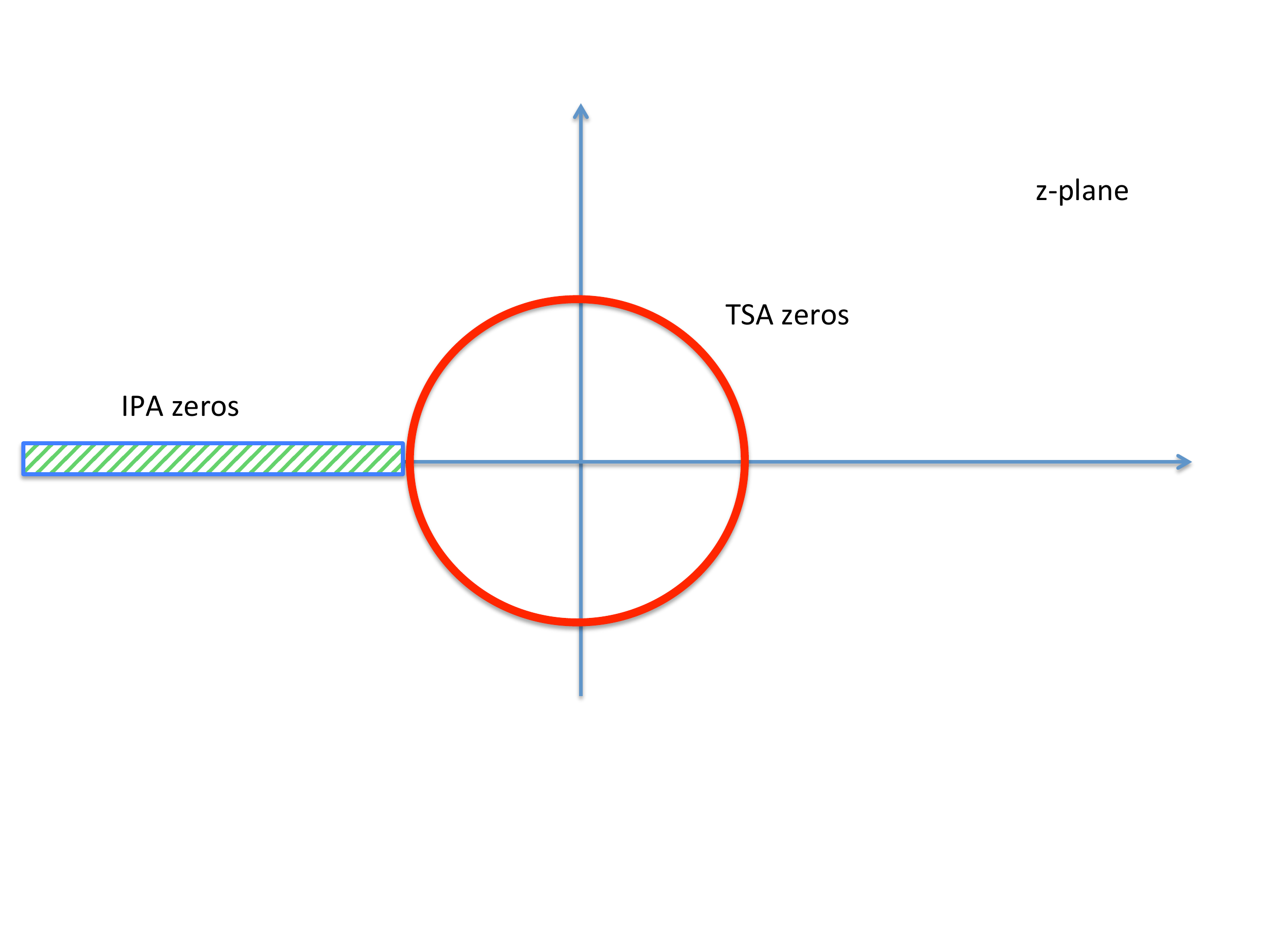,height=8cm,width=10cm}
\caption{Distribution of the Yang-Lee zeros in the Infinite Product Approach (IPA) and in the Truncated Series Approach (TSA).}
\label{ipatsazeros}
\end{center}
\end{figure}
Although these sets of zeros are quite different, both give rise to the {\em same} values of the negative moments of the zeros: in the TSA, this is true by construction, since the coefficients of the truncated polynomials come directly from the cluster coefficients; in the IPA, this was shown by the integrals (\ref{integralmoments}). Imagine now that we know the zeros coming from TSA and imagine we employ eq.\, (\ref{pressurezeros}) to compute the pressure of the gas: notice that in this case we can define the pressure $p(z)$ {\em only} for $| z| <1$, because this was, from the start, the domain of validity of this approach: indeed, it is based on the expansion of the logarithm present in eq.\,(\ref{generalfreeenergy}), so that all the following series have inherently a finite radius of convergence coming from the radius of convergence of the series expansion of the logarithmic function. This state of matter is similar to the case discussed in Appendix A. Concerning the IPA, it is worth to notice that it may be be simply interpreted as a change of variable done in such a way to bring the original expression (\ref{generalfreeenergy}) of the free energy in a "Yang-Lee" form (\ref{curvezerosss1}): hence, the corresponding integral on the set of zeros ranging in the interval $(-1,-\infty)$ provides a function which is also defined for $| z| >1$, which is of course the {\em analytic continuation} of free-energy defined by the TSA. In the real interval $0 < z < 1$, the two functions obtained by employing the two different sets of zeros of course coincide. Even though in the free theories the IPA seems to perform better than TSA, in the interactive theories discussed in the next Sections we will see however that the the proper Yang-Lee zeros can be identified only using the TCS.

\begin{figure}[b]
\centering
\includegraphics[width=0.38\textwidth]{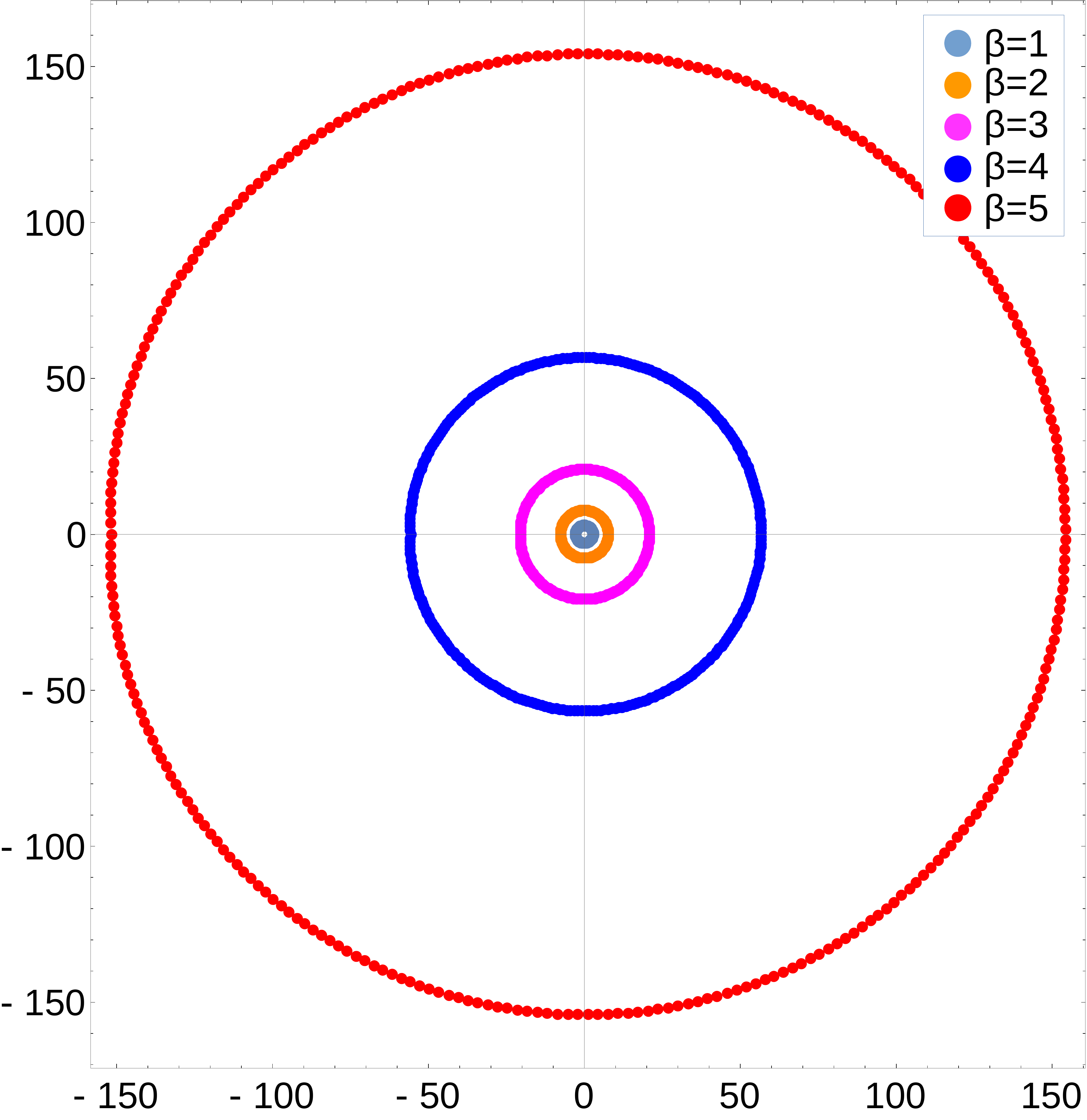}
\caption{Yang-Lee zeros of the relativistic free fermionic case for $d=1$ for various values of $\beta$.} 
\label{cerchiconcentrici}
\end{figure} 
    
\section{Distribution of the zeros for relativistic free theories}\label{freerel}
\noindent
Let's now employ the TSA to determine the zeros of the relativistic free theories whose free energy was given in eqs.\,(\ref{relativistic1}) and 
(\ref{relativistic2}). Particularly important for later comparisons is to work out in some detail the bosonic/fermionic case for $d=1$: for this value of $d$, for instance, 
the fermionic case can be directly related to the partition function of the two-dimensional classical Ising model \citep{GM,KS,YZ,Fendley,KM}
and moreover it provides the proper limit of the interactive integrable model analysed in the next Sections when $\beta \rightarrow \infty$. 

For $d=1$ the pressure of boson/fermion is given by 
\begin{equation}
F_{\pm}(z,\beta) \,=\,  \frac{V}{\lambda_T} \sqrt{\frac{2 m \beta}{\pi}}  \, \hat H_{\pm}(z; \beta) \,\,\,,
\label{relativisticIsi1}
\end{equation}
where 
\begin{equation}
\hat H_{\pm}(z; \beta) \,=\, \sum_{n=1}^\infty \frac{(\mp 1)^{n+1}}{n} \, K_{1}(n \beta m) z^n \,\,\,.
\label{relativisticIsi2}
\end{equation}
Notice that, with the dependence of the thermal length $\lambda_T$ from the temperature $T=1/\beta$ given in 
eq.\,(\ref{thermallength}), the pre-factor in front of $\hat H_{\pm}(z; \beta)$ in eq.\,(\ref{relativisticIsi1}) is completely 
independent on the temperature and therefore for $d=1$ the only dependence on temperature comes from the argument 
of the Bessel functions. From now on we put the value of the mass to 1, $m=1$. 

For any finite value of $\beta$, using the TSA we can determine the zeros of the truncated partition function $\hat\Omega^{(N)}_\pm(z)$  
and they essentially lay along circles (see Figure \ref{cerchiconcentrici}) whose radius $R(\beta)$ has an exponential dependence on $\beta$, 
$D \simeq e^{\beta}$ as shown in Figure \ref{scaling}. For large $\beta$ this exponential behaviour has an easy explanation. Consider the 
bosonic case, where all coefficients of $\hat\Omega^{(N)}_+(z)$ are positive (or equivalently one can consider the fermionic case computed 
at $-z$) so that we can apply the Enestr\"{o}m bounds (\ref{Kakeyabounds1}) and (\ref{Kakeyabounds2}) given by the ratio of the consecutive 
coefficients of  $\hat\Omega^{(N)}_+(z)$: it easy to see that the maximum of the sequence $\left\{ \frac{\gamma_k}{\gamma_{k+1}}\right\}$ is obtained for $k=0$ and 
therefore is 
\be 
{\rm max} \left\{\frac{\gamma_k}{\gamma_{k+1}}\right\} \,=\, \frac{\gamma_0}{\gamma_{1}}
\,=\,\frac{1}{K_1(\beta )} \,\sim \sqrt{\frac{2 \beta}{\pi}}\,\exp(\beta ) \,\,\,,
\ee
Concerning the minimum of the same sequence, it is also easy to see that scale with the same exponential law: in fact 
from the exponential expansion of (\ref{relativisticIsi1}) the general expression of the $k$-th term $\gamma_k$ is given by \citep{LeclairMussardo} 
\be 
\gamma_k \,=\,\alpha_1 K_1^k(\beta) + \alpha_2 K_1^{k-2}(\beta) K_1(2 \beta) + \alpha_3 K_1^{k-3}(\beta) K_1(3 \beta) + \alpha_3 
K_1^{k-4}(\beta) K_1^2(2 \beta) + 
\cdots \alpha_p \, K_1( k \beta)
\,\,\,,
\ee
where all terms, for large $\beta$, scale as $e^{- k \beta}$ and therefore $\gamma_k \simeq A_k e^{- k \beta}$. Hence, taking the ratio 
of two consecutive coefficients we have that 
\be 
{\rm min}\left\{ \frac{\gamma_k}{\gamma_{k+1} }\right\}\,=\, \left({\rm min}\left\{\frac{A_k}{A_{k+1}}\right\}\right) \,\, e^{\beta} \,\,\,. 
\ee
Putting $B =\lim_{k\rightarrow \infty} \frac{A_k}{A_{k+1}}$ (and it possible to prove that this number is of order 1), we have that the module of the 
all zeros satisfy the bounds 
\be 
B \, e^{\beta} \leq | z_i | \leq \sqrt{\frac{2 \beta}{\pi}} \, e^{\beta} \,\,\,.
\label{qed}
\ee
\begin{figure}[t]
\centering
\includegraphics[width=0.42\textwidth]{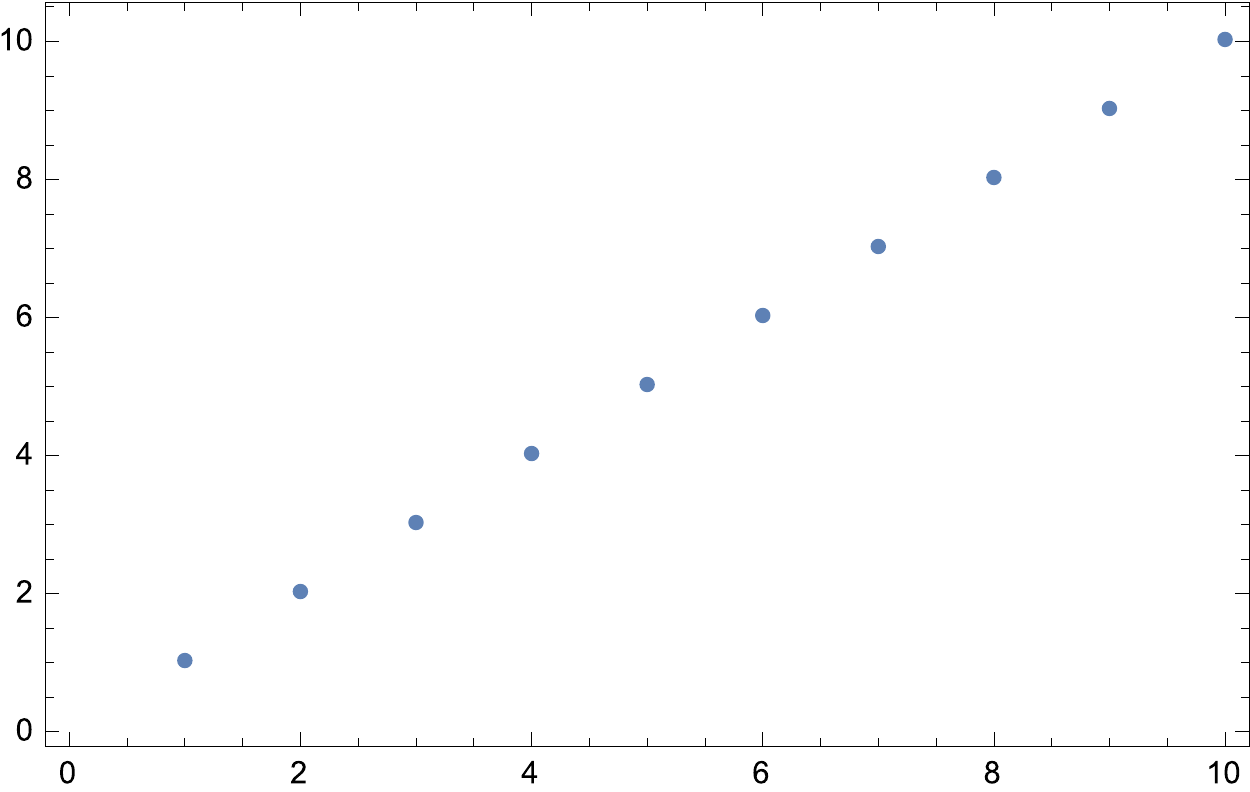}
\caption{Logarithm of the radius $R$ versus $\beta$.} 
\label{scaling}
\end{figure} 
More subtle is the behavior of the radius $R(\beta)$ of the zeros when $\beta \rightarrow 0$. Indeed, we can prove that this function must have a discontinouos 
behavior at $\beta =0$, namely arbitrarily close to this value, we have $R =1$ while at exactly $\beta = 0$, we have $R(0) = 0$. To show this discontinuity, 
we follow two different routes: the first consists of computing the series (\ref{relativisticIsi2}) for a small value of $\beta$: it is clear that in this case the first coefficients may be quite large (since they are computed near the origin of the Bessel function where $K_1(x) \simeq \frac{1}{x}$ if $ x\rightarrow 0$) but for all indices $n$ larger that a certain index $\tilde n$ (with $\tilde n \beta > 1$), all the remaining coefficients become once again small since they refer to large value 
of the Bessel function. Under such circumstances, the sequence $\left\{ \frac{\gamma_k}{\gamma_{k+1}}\right\}$ is no longer monotonically decreasing as it was for large $\beta$ but nevertheless has a min and a max of order $1$, so that the circle of the zeros has also a radius $D \simeq 1$. 

The second route consists instead in taking directly the limit $\beta \rightarrow 0$ in the series (\ref{relativisticIsi2}): if we do so, we can substitute 
each Bessel function with its short distance asymptotic behavior and therefore one has
\be 
\hat H_{\pm}(z; \beta) \,=\, \frac{1}{\beta} \,\sum_{n=1}^\infty \frac{(\mp 1)^{n+1}}{n^2} \, z^n \,=\,
\mp \frac{1}{\beta}  \,L_2(\pm z) \,\,\,,
\ee
where $L_2(z)$ is the dilogarithmic function, obtained by eq.\, (\ref{polylogarithm}) with $s=2$ (the interested reader may consult the references \citep{Zagierdilo} for the remarkable mathematical properties of this function!). It is now clear why at $\beta =0$ the radius of the zeros vanishes. Indeed in this case the partition function is given by (consider for simplicity the bosonic case where we can easily apply the Enestr\"{o}m bounds)
\be
\Omega_+(z) \,=\,e^{\frac{1}{\beta} L_2(z)}\,\,\,. 
\label{Omegabeta0}
\ee
Strictly speaking this result just applies at $\beta = 0$\footnote{This means that the proper mathematical statement is that $\beta \log \Omega(z)$ has a well-defined value in $\beta =0$ given by $L_2(z)$.} but if we assume the validity of eq.\,(\ref{Omegabeta0}) even for infinitesimal values of $\beta$, this implies that the ratio of two consecutive coefficients of $\Omega_+(z)$ is always proportional to $\beta$ and therefore for all roots we have the bounds 
\be 
\mu \, \beta \leq |z_i | \leq \, \nu\, \beta \,\,\,,
\ee 
(for some constanst $\mu$ and $\nu$) and therefore when $\beta =0$ the module of the roots vanishes. The discontinuous behavior of $R(\beta)$ at 
$\beta =0$ comes because eq.\,(\ref{Omegabeta0}) in free theories actually holds only at $\beta =0$.

\section{Integrable Quantum Field Theories}\label{sesta}
\noindent

\subsection{Elastic $S$-matrix}
\noindent
Relativistic quantum integrable models have the distinguished property that their $S$-matrix is elastic and factorizable in terms of the two-body $S$-matrix amplitudes \citep{ZamZam,Zam1,GM}. Hereafter, for simplicity, we will only consider quantum integrable models in $(1+1)$ dimensions with the spectrum consisting of just one neutral particle $A$ of mass $m$. Our main example is the Yang-Lee model \citep{CM,ZamFFYL,YZ,ZamTBA}. 

In order to write down explicitly the scattering amplitude of such simple integrable theory it is convenient to parameterise the energy $E$ and the momentum $p$ of the particle in terms of the so-called rapidity variable $\theta$ 
as 
\be 
E \,=\, m\, \cosh\theta 
\,\,\,\,\,\,\,
,
\,\,\,\,\,\,\,
p\,=\, m\, \sinh\theta 
\,\,\,.
\label{dispersionrelation}
\ee
The two-body $S$-matrix amplitude is then defined by 
\be
|A(\theta_1) \, A(\theta_2) \rangle \,=\, S(\theta_1 - \theta_2) \, |A(\theta_2) \, A(\theta_1) \rangle \,\,\,.  
\label{Smatrix}
\ee 
It depends on the difference of the rapidities $\theta \equiv \theta_1 - \theta_2$ and satisfies the unitarity and crossing equations \citep{ZamZam,Zam1,GM}
\begin{eqnarray}
&& S(\theta) \, S(-\theta) \,=\, 1 \,\,\,,
\label{unitarity}
\\
&& S(\theta) \,=\, S(i \pi - \theta) 
\label{crossing}\,\,\,.
\end{eqnarray}
As evident from eq.\,(\ref{Smatrix}), the $S$-matrix also plays the role of braiding matrix, interchanging the rapidities of the two particles. Notice that the unitarity condition (\ref{unitarity}) forces the value of the $S$-matrix at $\theta = 0$ to be $S(0) = \pm 1$. Consistent models have been found only for $S = -1$ (see, for instance, \citep{SM}): models whose $S$-matrix satisfies $S(0) = -1$ are called {\em fermionic models} because, looking at eq.\,(\ref{Smatrix}), interchanging two excitations of equal rapidities, their wave-function changes sign. As discussed below, this property is important for the thermodynamics of the integrable models. Let us notice that for such simple quantum integrable systems the $S$-matrix is 
a pure phase 
\be
S(\theta) \,=\, e^{i \delta(\theta)} \,\,\,,
\label{purephase}
\ee
where $\delta(\theta)$ is the phase shift. For later use it is convenient to introduce the kernel $\varphi(\theta)$, given by the 
derivative of the phase shift
\be
\varphi(\theta) \,=\,\frac{d \delta(\theta)}{d \theta} \,\,\,.
\label{kernel}
\ee

\subsection{Thermodynamic Bethe Ansatz}
\noindent
Al.B. Zamolodchikov \citep{ZamTBA} derived the basic equations which allows us to compute the free-energy of a quantum integrable model 
defined on a cylinder of width $\beta$ and length $L$. The radius $\beta$ of the cylinder is directly associated to the temperature $T$ of the system 
through the relation $\beta = 1/T$ while $L$ is the volume of the system. The formalism can be generalised to include also the chemical potential 
\citep{Fendley,KM} and the final equations read as follow: the pressure $p(z)$ is expressed by 
\be 
\beta \,p(z) \,=\,\int_{-\infty}^{\infty} \frac{d\theta}{2\pi} \, \cosh\theta \, \log\left(1 + z \, e^{-\epsilon(\theta,z)}\right) \,\,\,,
\label{prsinteg}
\ee
where the function $\epsilon(\theta,z)$ (called pseudo-energy) satisfies the non-linear integrable equation 
\be
\epsilon(\theta,z) \,=\, m \beta \,\cosh\theta - \int_{-\infty}^{\infty} \frac{d\theta'}{2\pi} \, \varphi(\theta - \theta') \, \log\left(1 + z \, e^{-\epsilon(\theta',z)}\right)
\,\,\,. 
\label{epsilonint}
\ee
Therefore the partition function of the system is given by  
\be
\Omega(z) \,=\,\exp \left[m L \,\int_{-\infty}^{+\infty} \, \cosh \theta \log \left(  1 + z\, e^{-\epsilon(\theta,z)}  \right)\frac{d \theta}{2 \pi} \right]\,\,\,.
\label{partitionint}
\ee
Notice that $\Omega(z)$ refers already to the thermodynamic limit where both $L$ and the number of particles $N$ go to infinity with 
their ratio given by the density $\rho(z) = z \frac{d\Omega}{dz}$ at a fixed value of $z$. It is worth noticing that the pressure has formally the same expression as in  the free theories, the only difference is that while in free theories the pseudo-energy $\epsilon(\theta)$ coincides with the energy of the particle\footnote{If the theory is free the $S$-matrix is a constant and therefore the kernel $\varphi(\theta)$ vanishes. Therefore, in this case, eq.(\ref{epsilonint}) states that $\epsilon(\theta) = m \beta\,\cosh\theta$} multiplied by $\beta$, in an interactive integrable theory it satisfies instead the non-linear integral equation (\ref{epsilonint}). Observe that the pseudo-energy $\epsilon(\theta,z)$ of course also depends on $\beta$ but we avoid to write this explicitly to keep the notation simpler. The partition function 
(\ref{partitionint}) does not have a singularity for positive value of $z$, in agreement with the purely massive behavior of the theory and its fermionic nature.  

In summary, for a a quantum integrable model, it is sufficient to know its $S$-matrix for computing its partition function (\ref{partitionint}) at a finite temperature and a chemical potential through the equations (\ref{prsinteg}) and (\ref{epsilonint}).  In the next Section we will study the zeros of this partition function for one of the simplest integrable quantum field theories, the Yang-Lee model.
 
\section{Partition function of the Yang-Lee model}\label{settima}
\noindent
In this Section, using the TSA described in Section \ref{TSA}, we are going to determine the distribution of the Yang-Lee zeros of the 
Yang-Lee model in (1+1) dimensions. To proceed, let's briefly recall the exact $S$-matrix of this model.
  
\subsection{$S$-matrix of the Yang-Lee model}
\noindent
An example of the $S$-matrix which satisfy all the requirements discussed above is 
\be
S_{YL}(\theta) \,=\, \frac{\tanh\frac{1}{2} \left(\theta + \frac{2\pi i}{3} \right)}{\tanh\frac{1}{2} \left(\theta - \frac{2\pi i}{3} \right)} \,\,\,.
\label{SYL}
\ee
In \citep{CM} it has been identified with the exact scattering amplitude of the massive theory that comes from the deformation of the non-unitarity 
conformal minimal model with central charge $c = - 22/5$ \citep{CardyYL} made by its only relevant field $\phi$ of conformal dimension $\Delta = - 1/5$. 
Apart being a fermionic $S$-matrix, the amplitude (\ref{SYL}) has its own peculiarities: first of all, it has a pole at $\theta = \frac{2\pi i}{3}$ and another at $\theta = i \frac{\pi}{3}$ which correspond to the same particle $A$ that appears as a bound state of itself both in the $s$ and the $t$-channel. Secondly, the residue of the $S$-matrix at the pole in the $s$-channel has the "{\em wrong}" sign with what usually expected in an hermitian theory. Generally, the residue in the $s$-channel pole  at $i u$ of an $S$-matrix is associated to the Feynamn diagram shown in Figure \ref{couplingresidueYL} 
\be
S(\theta) \,=\, i \frac{g^2}{\theta - i u} \,\,\,,
\ee
and therefore it corresponds to the square of the {\em on-shell } three-point coupling constant $g$. Since this quantity in an hermitian theory is real, the residue is positive. But in Yang-Lee model the coupling constant is purely imaginary, as evident from its action (\ref{YLaction})
%\be 
%{\cal A} \,=\, \int d^2x \left[\frac{1}{2} (\partial \phi)^2 +  i (h-h_0) \phi + i  g\,\phi^3 \right] \,\,\,,
%\label{YLaction2}
%\ee
and therefore the residue of this theory at the $s$-channel pole $\theta = \frac{2\pi i}{3}$ is negative, as confirmed by its explicit computation 
\EQ
i\,g^2\,=\, i \, \tan\frac{2 \pi}{3} \,=\,- \sqrt{3} \, i \,\,\,.
\EN
\begin{figure}[t]
\centerline{
\vspace{1mm}
\psfig{figure=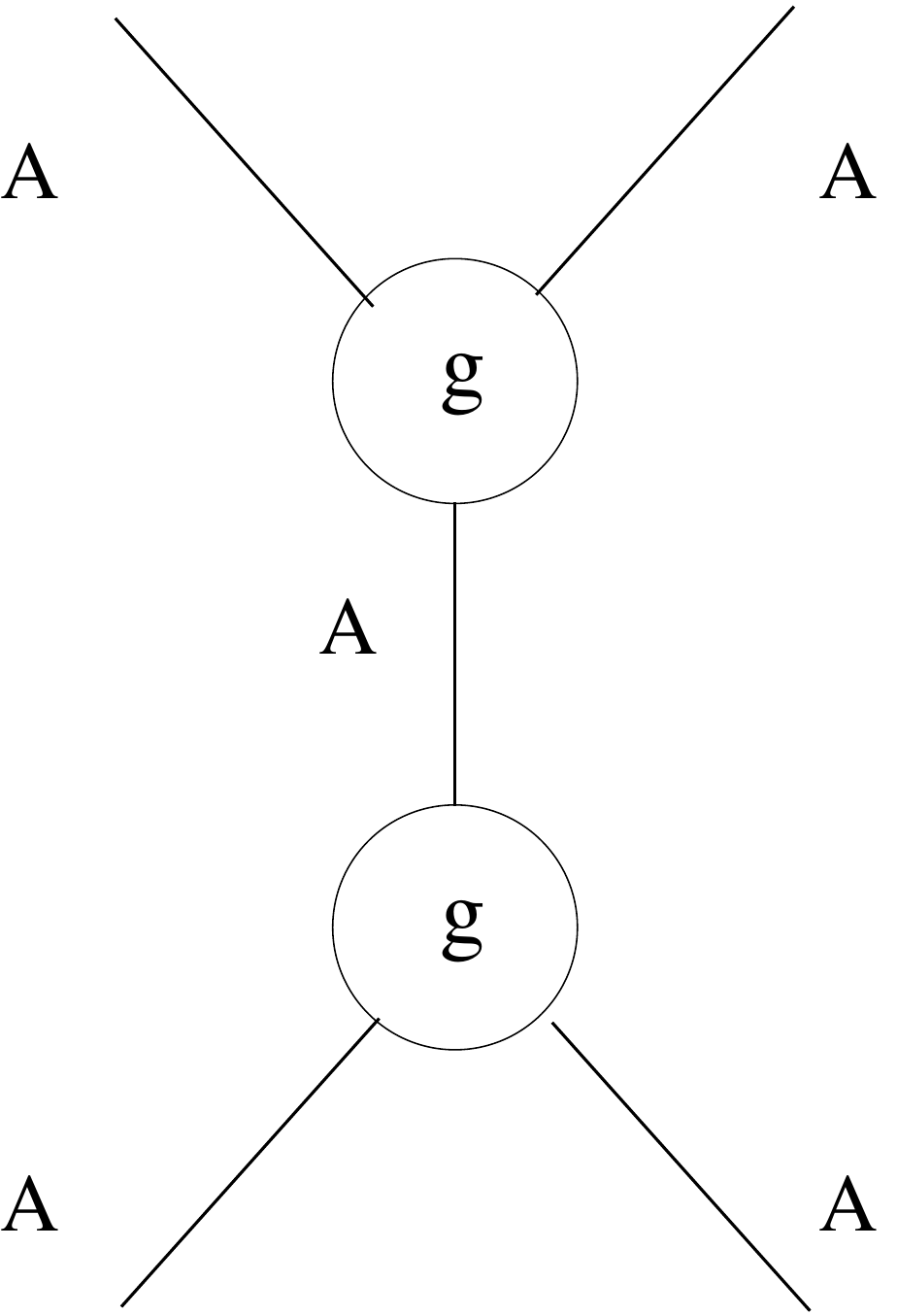,height=4.5cm,width=3.45cm}}
\vspace{1mm}
\caption{Residue at the pole expressed in terms of the {\em on-shell} coupling constant.}
\label{couplingresidueYL}
\end{figure}

\noindent
The action (\ref{YLaction}) is not hermitian but nevertheless CP invariant. The way to see this is to introduce the operator $CP$ which changes $\phi$ in $- \phi$, under which ${\cal A}^\dagger = (CP) \, {\cal A} \, (CP)$. This relation ensures the reality of the spectrum but implies a modification in the completeness relation 
given by the multi-state sum, for the Yang-Lee model given by 
\be
\sum_{N=0} (-1)^N | N \rangle \, \langle N | \,=\, 1 \,\,\,.
\label{completenessYL}
\ee

\subsection{Solution of the Bethe Ansatz equations}
\noindent
The non-linear integral equation for the pseudo-energy of the Yang-Lee model is given by (we put the mass $m=1$)
\beq
\epsilon(\theta, z)\,=\, \beta \,\cosh\theta + \int^{\infty}_{-\infty} \frac{\sqrt{3}   \cosh(\theta)}{\left( \sinh^2(\theta) + \frac{3}{4}  \right)}\ln(1+ze^{-\epsilon(\theta', z)}) 
\frac{d \theta'}{2\pi}\,.
\label{epsYL}
\eeq
To find the Yang-Lee zeros of this model we have used the TSA implemented as follows.   
Firstly we express $\epsilon$ as a series in the variable $z$
\beq
\label{YLeps2}
\epsilon(\theta, z) \,=\, \sum_{n=0}^{\infty}\epsilon_n(\theta)\,z^n\,\,\,,
\eeq
where the functions $\epsilon_n(\theta)$ can be obtained by substituting this expression in (\ref{epsYL}), expanding the logarithm in the integral in powers of $z$ and comparing the various order of this expansion. Therefore 
\begin{eqnarray}
\epsilon(\theta, z) &=& \beta \cosh\theta - \int^{\infty}_{-\infty} \varphi(\theta-\theta') \log\left(1 + z e^{-\epsilon(\theta', z)}\right) \frac{d \theta'}{2\pi}\\
&=& \beta \cosh(\theta) - \int^{\infty}_{-\infty} \varphi(\theta - \theta') \sum_{k=0}^{\infty} \frac{z^k}{k!} \frac{\partial^k}{\partial z^k} \left.\left[\ln\left(1+z
e^{-\epsilon(\theta', z)}\right)\right]\right|_{z=0} \frac{d \theta'}{2\pi}\,.
\end{eqnarray}
Let's now calculate the coefficients of the expansion of the logarithm by evaluating its various derivatives in $z=0$:
\begin{itemize}
\item $k=0$: 
\beq
\ln\left.\left(1+ze^{-\epsilon_0(\theta')-\epsilon_1(\theta')z-\epsilon_2(\theta') z^2- \ldots}\right)\right|_{z=0}=0
\eeq
\item $k=1$: 
\begin{eqnarray}
&&\frac{\partial}{\partial z} \left.\ln\left(1+ze^{-\epsilon_0(\theta')-\epsilon_1(\theta')z-\epsilon_2(\theta') z^2- \ldots}\right)\right|_{z=0}=\nonumber\\
&& \quad =\left[\frac{e^{-\epsilon_0(\theta')-\epsilon_1(\theta')z-\epsilon_2(\theta') z^2- \ldots}}{\left(1+ze^{-\epsilon_0(\theta')-\epsilon_1(\theta')z-\epsilon_2(\theta') z^2- \ldots}\right)} \,\, + \right.\nonumber\\ 
&& \quad\quad\,\,\,\,+ \left.\left.\frac{ze^{-\epsilon_0(\theta')-\epsilon_1(\theta')z-\epsilon_2(\theta') z^2- \ldots} (-\epsilon_1 (\theta')z-\epsilon_2(\theta') z^2- \ldots)}{\left(1+ze^{-\epsilon_0(\theta')-\epsilon_1(\theta')z-\epsilon_2(\theta') z^2- \ldots}\right)}\right]\right|_{z=0}= \nonumber\\
&&\quad = e^{-\epsilon_0(\theta')}
\end{eqnarray}
\item $k=2$: 
\begin{eqnarray}
&&\frac{\partial^2}{\partial z^2}\ln\left.\left(1+ze^{-\epsilon_0(\theta')-\epsilon_1(\theta')z-\epsilon_2(\theta') z^2- \ldots}\right)\right|_{z=0}=\nonumber\\
&&\quad = \left[\frac{e^{-\epsilon_0(\theta')-\epsilon_1(\theta')z-\epsilon_2(\theta') z^2- \ldots} \left[(-\epsilon_1(\theta')\! -\!  \epsilon_2(\theta')z \!-\! \ldots\!-\epsilon_1(\theta') \!-\!  \epsilon_2(\theta')z- \ldots)\right.}{{\left(1+ze^{-\epsilon_0(\theta')-\epsilon_1(\theta')z-\epsilon_2(\theta') z^2- \ldots}\right)}^2} \right.\nonumber\\
&&\qquad \quad\,\, \frac{\left. z(-\epsilon_1(\theta') -  \epsilon_2(\theta')z- \ldots)^2 \right](1+ze^{-\epsilon_0(\theta')-\epsilon_1(\theta')z-\epsilon_2(\theta') z^2- \ldots})}{{\left(1+ze^{-\epsilon_0(\theta')-\epsilon_1(\theta')z-\epsilon_2(\theta') z^2- \ldots}\right)}^2} -\nonumber\\
&&\left.\left.\qquad\,\,\,-\frac{[e^{-\epsilon_0(\theta')-\epsilon_1(\theta')z-\epsilon_2(\theta') z^2- \ldots}(1+z(-\epsilon_1(\theta') -  \epsilon_2(\theta')z- \ldots))]^2}{(1+ze^{-\epsilon_0(\theta')-\epsilon_1(\theta')z-\epsilon_2(\theta') z^2- \ldots})^2}\right]\right|_{z=0}=\nonumber \\
&&\quad = -2\epsilon_1(\theta') e^{-\epsilon_0(\theta')}-e^{-2\epsilon_0(\theta')}=\nonumber\\
&&\quad=- e^{-\epsilon_0(\theta')}\left[2 \epsilon_1+e^{-\epsilon_0(\theta')}\right]
\end{eqnarray}
and so on.
\end{itemize}
By comparing now the various terms which appear with the same power in $z$ in eq.\,(\ref{epsYL}), we can then identify the 
functions $\epsilon_n(\theta)$ as:
\begin{eqnarray}
\epsilon_0(\theta) &=& \beta \cosh(\theta)\\
\epsilon_1(\theta) &=& - \int_{-\infty}^{\infty} d\theta' \varphi(\theta-\theta')e^{-\epsilon_0(\theta')}\\
\epsilon_2(\theta) &=& \frac{1}{2!}\int_{-\infty}^{\infty} d\theta' \varphi(\theta-\theta')[2 \epsilon_1e^{-\epsilon_0(\theta')}+e^{-2\epsilon_0(\theta')}]\\
\epsilon_3(\theta) &=& -\frac{1}{3!} \int_{-\infty}^{\infty} d\theta' \varphi(\theta-\theta')\bigg[2e^{-3\epsilon_0(\theta')} + 6\epsilon_1(\theta')e^{-2\epsilon_0(\theta')}+\\
&&\hphantom{-\frac{1}{3!} \int_{-\infty}^{\infty}d\theta' \varphi(\theta-\theta')\bigg[ }+ 3 \epsilon_1(\theta')^2e^{-\epsilon_0(\theta')} - 6\epsilon_2(\theta')e^{-\epsilon_0(\theta')}\bigg]
\end{eqnarray}
and so on. 

Let's observe that, once we fix the inverse temperature $\beta$, the coefficients $\epsilon_n(\theta)$ are obtained in terms of convolutions between the kernel $\varphi(\theta)$ and a function $G$ involving the previous coefficients, $G\left(\epsilon_1(\theta'), \ldots, \epsilon_{n-1}(\theta')\right)$. The actual calculation of these coefficients consists in solving numerically this hierarchical set of integral equations.   
\begin{figure}
\centering
\includegraphics[width=0.48\textwidth]{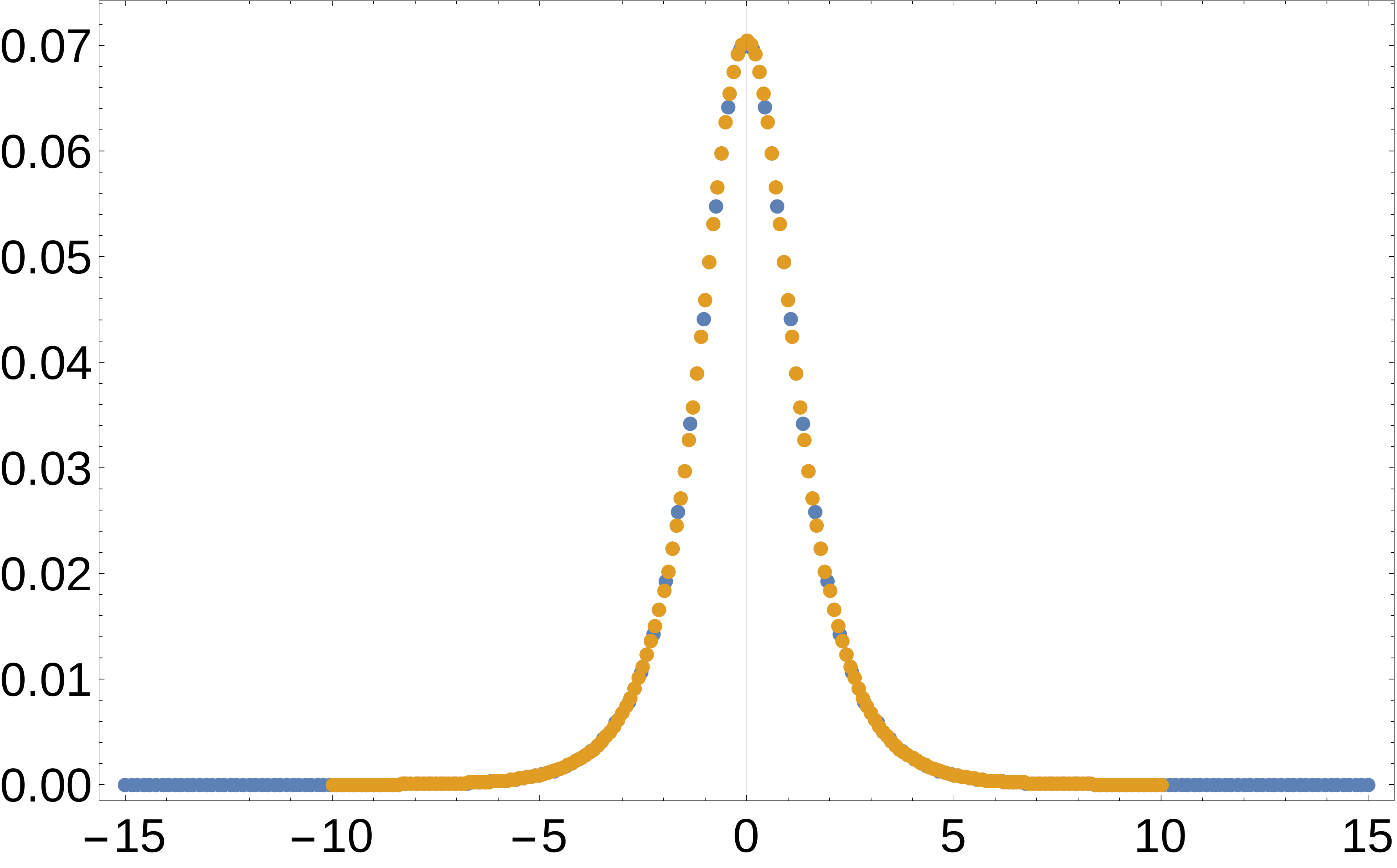}
\includegraphics[width=0.48\textwidth]{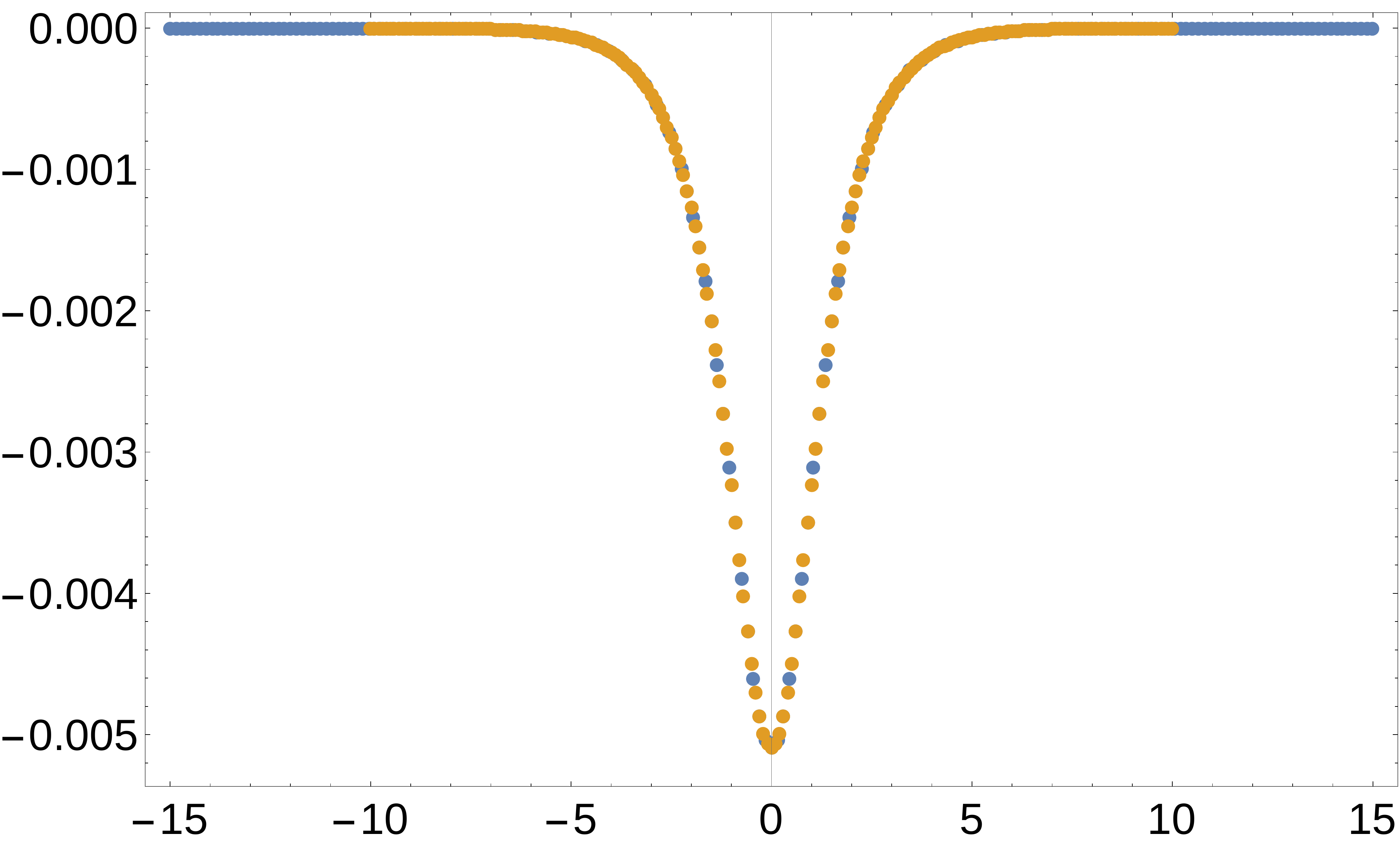}
\caption{Plots of the functions $\epsilon_1$ and $\epsilon_2$, evaluated in the grid of the $\theta$ (blue dots) and of the $\theta '$ (yellow dots), for $\beta = 4$.}
\label{plotfun51}
\end{figure}
Notice that, except for the term $\epsilon_0 (\theta)=  \beta \cosh(\theta)$ that is positive, the sign of the other coefficients oscillates, and in particular it is positive for the odd index coefficients, and negative for the even index ones. This will induce an expansion of the free-energy in terms of a series of alternating terms and ultimately a partition function with positive and negative coefficients, as in the case of free fermionic theory. 

\subsection{Calculation of the free energy}
\noindent
We can now compute the pressure using the same method adopted for the pseudo-energy. In fact we can write it as an expansion:
\beq
p(\beta, z)\equiv \sum_n f_n(\beta)\,z^n
\label{przz}
\eeq
with the coefficients $f_n(\beta)$ obtained by expanding the integrand of the eq.\,\ref{prsinteg})
\beq
p(\beta, z)\, =\, \frac{1}{\beta}\int_{-\infty}^{\infty} \frac{d \theta}{2 \pi} \sum_{k} \frac{z^k}{k!} \frac{\partial^k}{\partial z^k} \left.\left[\ln(1+ze^{-\epsilon(\theta)}) \right]\right|_{z=0}
\eeq
The coefficients have the following expression:
\begin{eqnarray}
f_0(\beta) &=&0\\
f_1(\beta) &=&  \frac{1}{\beta}\int_{-\infty}^{\infty} \frac{ d\theta}{2\pi} \cosh(\theta)e^{-\epsilon_0(\theta)}\\
f_2(\beta) &=& -\frac{1}{2! \beta }\int_{-\infty}^{\infty} \frac{ d\theta}{2\pi} \cosh(\theta)[2 \epsilon_1e^{-\epsilon_0(\theta')}+e^{-2\epsilon_0(\theta')}]\\ 
f_3(\beta)&=& \frac{1}{3! \beta} \int_{-\infty}^{\infty} \frac{ d\theta}{2\pi} \cosh(\theta)\bigg[2e^{-3\epsilon_0(\theta')}+6\epsilon_1(\theta')e^{-2\epsilon_0(\theta')}+\\ 
&&\hphantom{-\frac{1}{3! \beta} \int_{-\infty}^{\infty} \frac{ d\theta}{2\pi} \cosh(\theta)\bigg[ }+3\epsilon_1(\theta')^2e^{-\epsilon_0(\theta')}-6\epsilon_2(\theta')e^{-\epsilon_0(\theta')}\bigg]
\end{eqnarray}
and so on. 
Notice that also in this case we have alternating sign for the coefficients, but the negative ones are those with an even index. 

\subsection{Calculation of the zeros of the partition function}
\noindent
We can now determine the zeros of the partition function, written as polynomial in the fugacity $z$ using the expansion:
\beq
\Omega(z) \,=\, \exp \left[  \beta \sum_i f_i(z)\, z^i  \right] \equiv \sum_k p_k z^k
\eeq
where:
\beq
p_k=\frac{1}{k!} \left.\frac{\partial^k \exp \left[  \beta \sum_i f_i(z,\beta)z^i  \right]  }{\partial z^k}\right|_{z=0}
\eeq

We can verify that also in this case, as for the fermionic free theory, the coefficients of the expansion in $z$ of the partition function, except for the zero order term, have alternating signs, in particular the even index coefficients are negative and the odd index coefficients are positive. 

\subsection{Results for finite $\beta$}
\noindent
Once we obtain the expression of the partition function as a polynomial in the fugacity $z$, we can study the distribution of its zeros. For each value of $\beta$, 
we find that the zeros lay along an approximate circumference, whose radius increases with $\beta$ as shown in Fig. \ref{Results}. As a matter of fact, given the enormous number of terms generated by the various series expansions described above, the analytic computation can be handle only up to $24$ terms. However, we have pushed further the order of the polynomial by using a scaling law satisfied by the coefficients $f_n(\beta)$ in eq.\,(\ref{przz})
\be
\frac{f_n(\beta)}{f_{n-1}(\beta)} \simeq \frac{1}{z_c(\beta)} \,\left(1 + \frac{\alpha(\beta) -1}{n}\right) \,\,\,,
\label{scalinglawcoef}
\ee
where both the parameters $z_c(\beta)$ and $\alpha(\beta)$ depends on $\beta$. This scaling law is shown in Figure \ref{scalingaa} for a particular value of 
$\beta$. Using the scaling law (\ref{scalinglawcoef}) we study the zeros of polynomials up to order $500$. Increasing the order of the polynomial we observe that the 
average values of the radius of the circle distribution of the zeros shows a convergence to final values which depend on $\beta$, as  shown in Fig.~\ref{expbeh}. Contrary to the free cases, for the interactive Yang-Lee model the radius $R(\beta)$ of its zeros goes continuously to zero when $\beta \rightarrow 0$ with an overall law which a best fit of the data fixes to be 
\be
R(\beta) \,=\, e^\beta -1\,\,\,.
\label{bestfitt}
\ee 
We now show that we can easily predict the behavior of $R(\beta)$ 
at $\beta \rightarrow \infty$ and at $\beta\rightarrow 0$. 

\begin{figure}[t]
\includegraphics[width=0.4\textwidth]{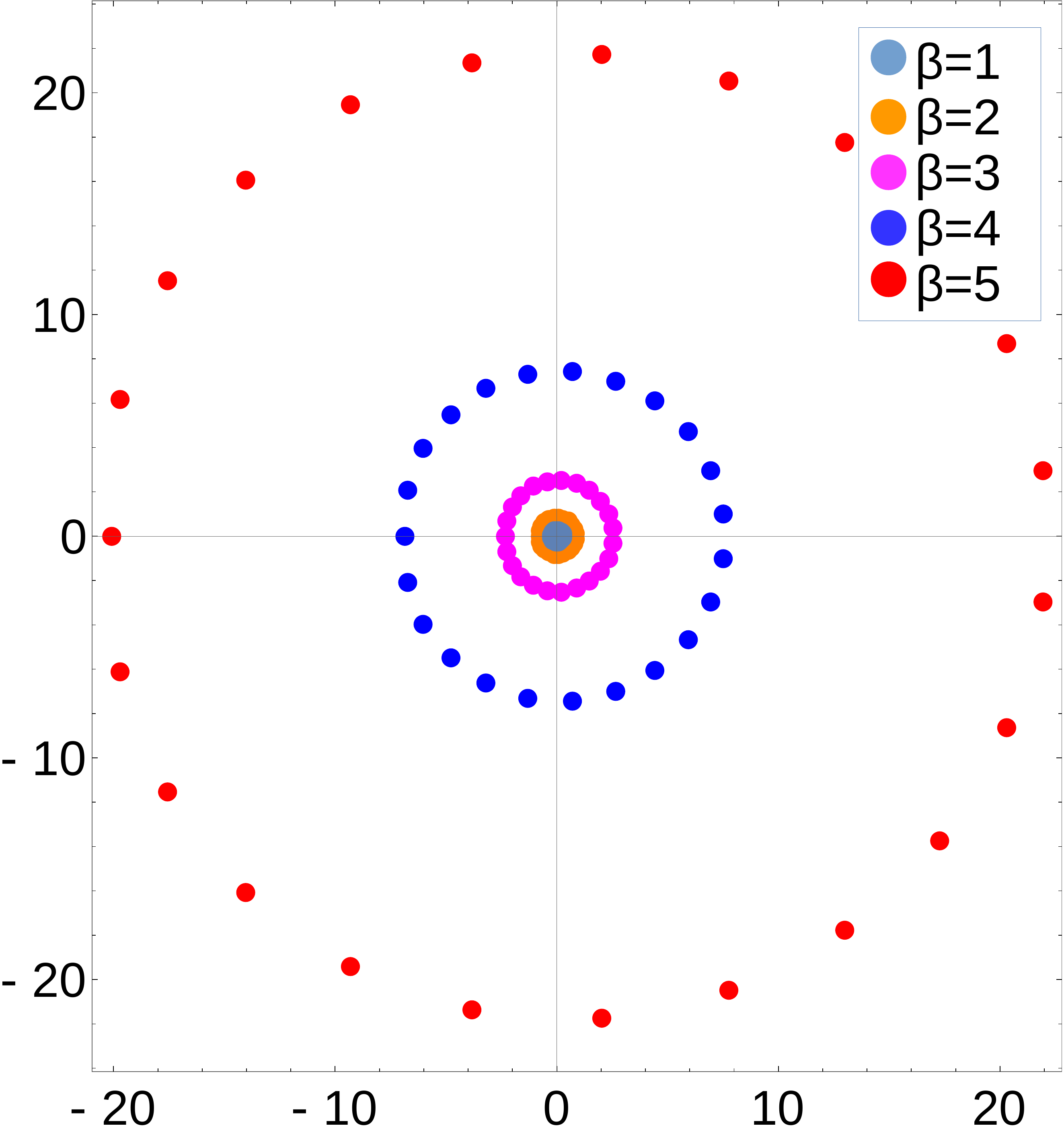}
\hspace{10mm}
\includegraphics[width=0.4\textwidth]{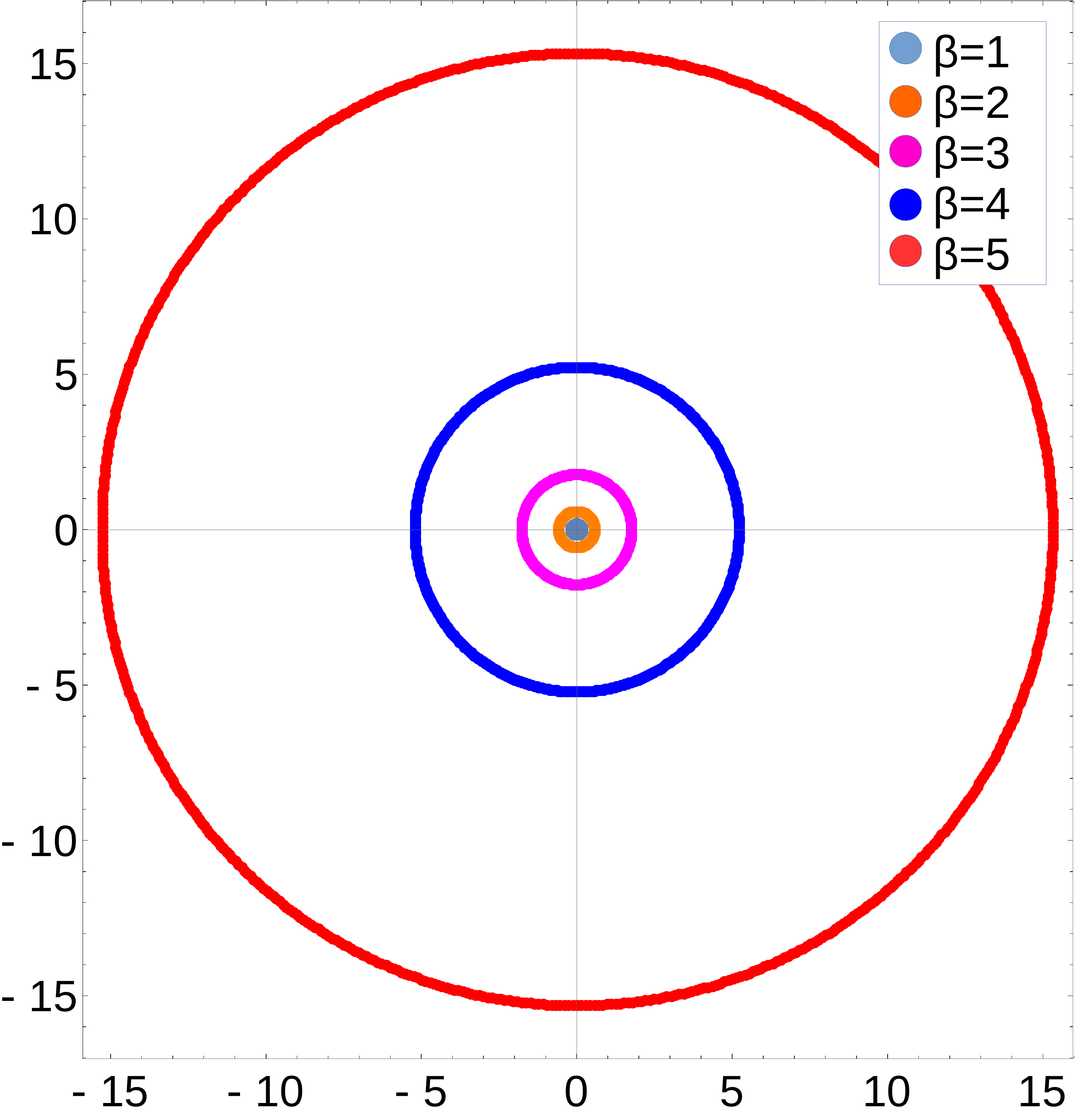}
\caption{a) Distribution of zeros of the partition function for fermionic Yang-Lee model for different values of $\beta$, obtained with the analytical calculations of the 
first 24 coefficients of the expansion; b) Distribution of zeros of the partition function for fermionic Yang-Lee model for different values of $\beta$, obtained with 500 coefficients obtained by using their scaling law (\ref{scalinglawcoef}).}
\label{Results}
\end{figure}

\begin{figure}[t]
\centering
\includegraphics[width=0.4\textwidth]{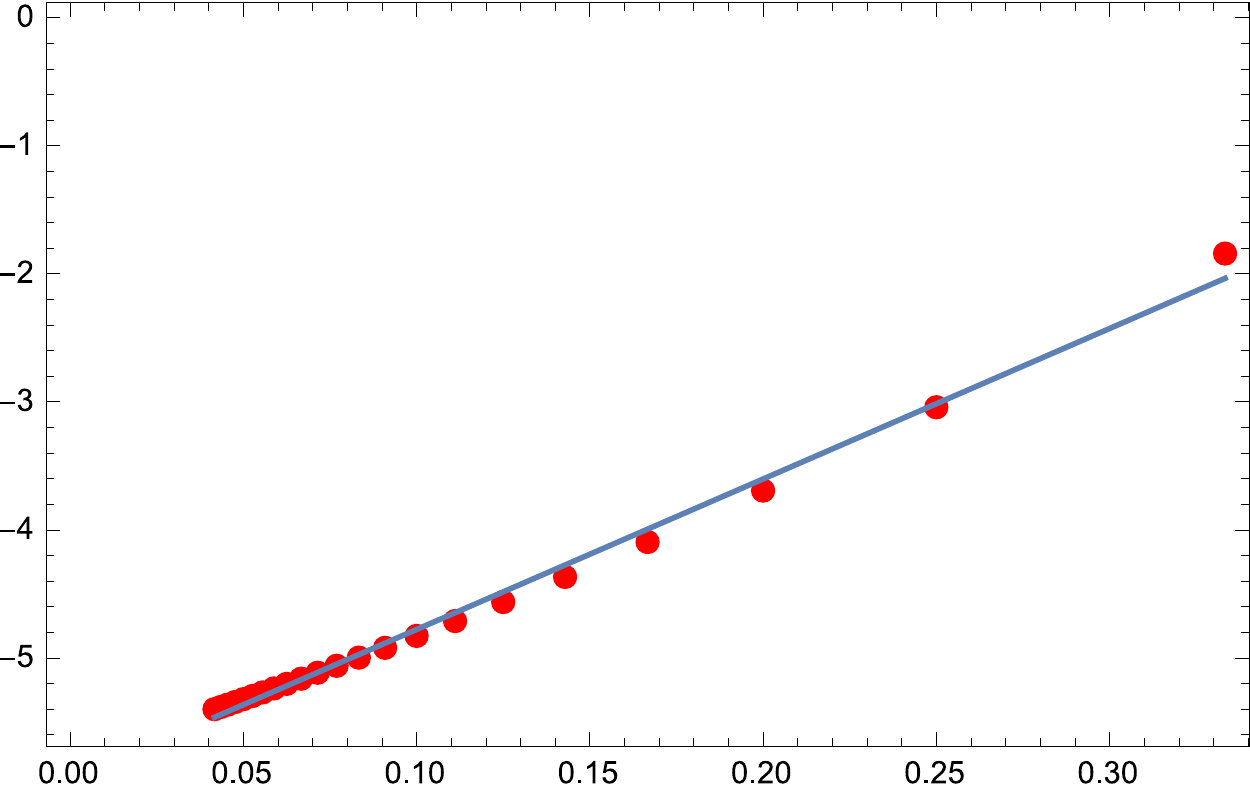}
\caption{Linear scaling law of the coefficients $f_n(\beta)$ in the variable $x = 1/n$. The plot refers to $\beta =1$.}
\label{scalingaa}
\end{figure}

%\begin{figure}[htbp]
%\centering
%\includegraphics[width=0.65\textwidth]{Figure/Results.png}
%\caption{ Dependence on $R=1/T$ of the radius of the curves.  }
%\label{Results}
%\end{figure}

\begin{figure}[b]
\includegraphics[width=0.44\textwidth]{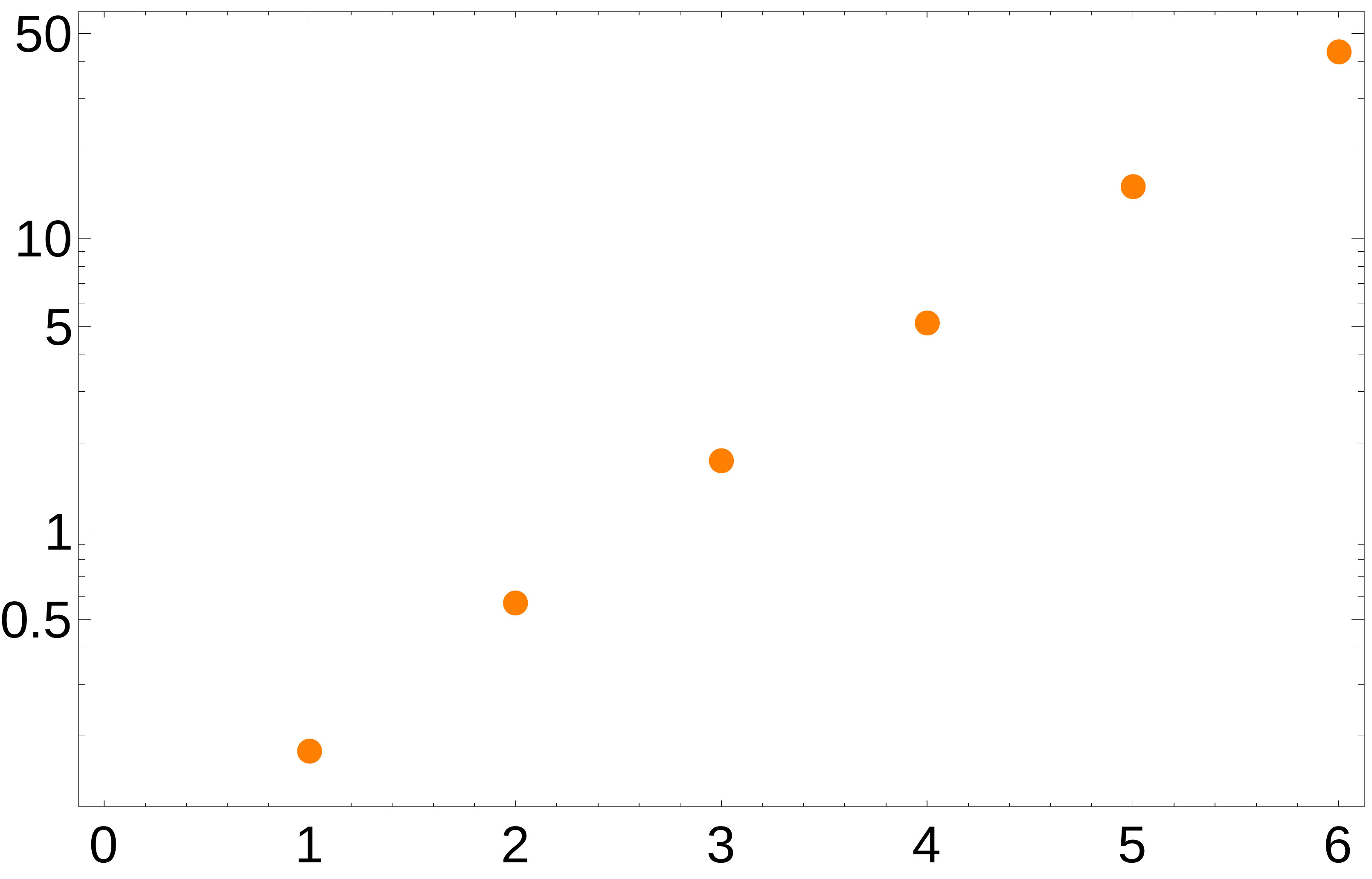}
\caption{Logarithm of the radius $R$ of the zeros versus the inverse temperature $\beta$.}
\label{expbeh}
\end{figure}

\subsection{The $\beta \rightarrow \infty $ limit}
\noindent
In the limit $\beta \rightarrow \infty$, the distribution of zeros of the partition function of the Yang-Lee model simply reduces to that of the free fermionic theory. 
In fact, for large value of $\beta$, the pseudo-energy becomes the one of free theory $\epsilon(\theta) \simeq \beta \cosh\theta$ and also its 
partition function 
\be 
\Omega(z) \simeq \,\exp \left[ L \,\int_{-\infty}^{+\infty} \, \cosh \theta \log \left(  1 + z\, e^{-\beta \cosh\theta)}  \right)\frac{d \theta}{2 \pi} \right]\,\,\,.
\label{partitioninttt}
\ee
The zeros of the Yang-Lee model in this limit thus coincide with those of the one-dimensional free fermionic model and therefore the 
radius $R(\beta)$ of the zeros scales for $\beta \rightarrow \infty$ as $R(\beta) \simeq e^{\beta}$.

\begin{figure}[t]
\centering
\includegraphics[width=0.4\textwidth]{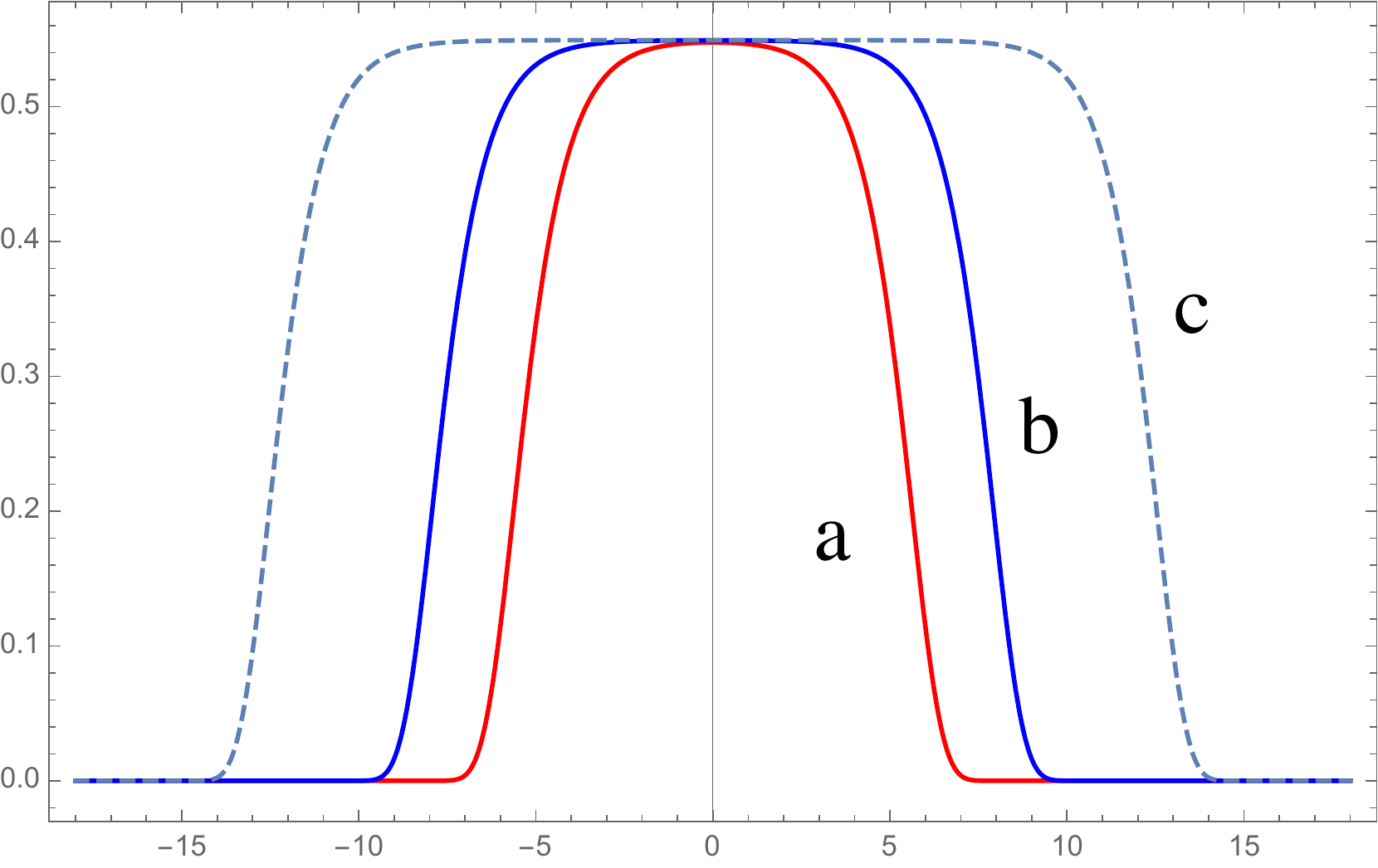}
\caption{Plot of the function $L(\theta,z)$ versus $\theta$ at $z=0.5$ and for various values of $\beta$: the curve {\bf a} refers to $\beta =0.01$, the curve {\bf b} 
to $\beta =0.001$ and the curve {\bf c} to $\beta =0.0001$. Going to smaller value of $\beta$, the function $L(\theta,z)$ develops a larger interval for it 
plateau value.}
\label{PlotTBA}
\end{figure}

\subsection{The limit case $\beta \rightarrow 0$}
\label{Rzero}
\noindent
When $\beta \rightarrow 0$, the Thermodynamics Bethe Ansatz solution for interactive theories has a qualitatively different behavior with respect to 
the behavior present in free theories, namely in an interactive theory such as the Yang-Lee model, the pseudo-energy $\epsilon(\theta,z)$ flattens in the central region $-\Lambda(\beta) \ll \theta \ll \Lambda(2/\beta)$, where $\Lambda(\beta) =  \log(2/\beta)$ \citep{ZamTBA}. 
This means that the function 
\be 
L(\theta,z) \,=\, \log\left( 1+ e^{-\epsilon(\theta,z)}\right) \,\,\,,
\label{Lfunction}
\ee
shows a plateau in the same region, with a constant value which depends on $z$ and with a double falloff outside (see Figure \ref{PlotTBA}). For $\beta \rightarrow 0$ the shape of the right and left edge becomes universal and gives rise to the so-called {\em kink-solution} of the Thermodynamics Bethe Ansatz equation. 
As shown in the Appendix B, this kink-solution allows us to compute the constant value $\epsilon_0(z)$ in the plateau region, solution of the 
non-linear equation
\be
\epsilon_0(z) \,=\,\log\left(1 + z \, e^{-\epsilon_0(z)}\right) \,\,\,,
\label{trascendental}
\ee
and the exact partition function in this limit
\be
\Omega(z) \,=\,e^{\frac{1}{\beta} f(z)} \,\,\,,
\label{exactf0}
\ee
where 
\beq
f(z)\,=\, -\frac{1}{\pi}  {\cal L} \left(\frac{z}{z+\frac{1 + \sqrt{1+4z}}{2}} \right)- \frac{1}{2 \pi} \ln (z) \ln\left(  \frac{\frac{1 + \sqrt{1+4z}}{2}}{\frac{1 + \sqrt{1+4z}}{2}+z} \right)\,\,\,,
\eeq
and ${\cal L}(x)$ is the the Rogers Dilogarithmic function
\be
{\cal L}(x) \,=\, -\frac{1}{2} \,\int_0^x dy \left[\frac{\log y}{1-y} + \frac{\log (1-y)}{y}\right] \,\,\,.
\ee
As in the free case (see eq.\,(\ref{Omegabeta0}) and the relative discussion there), eq.\,(\ref{exactf0}) strictly speaking applied just at $\beta =0$ but 
the qualitative difference with respect to the free case is that in an interactive theory as the Yang-Lee model the logarithm of the partition function $\Omega(z)$ presents a $1/\beta$ behavior also at a finite small values of $\beta$! This is a direct consequence of the plateau behavior of the pseudo-energy: indeed, we can first use the parity of the Bethe Ansatz equation and the quantity $\Lambda(\beta)$ that sets the scale of the plateau behavior to split the integral of the free-energy as 
\be
 \int_{0}^{+\Lambda} \, \cosh \theta \log \left(  1 + z\, e^{-\epsilon(\theta,z)}  \right)\frac{d \theta}{2 \pi} 
+ \int_{-\Lambda}^{+\infty} \, \cosh \theta \log \left(  1 + z\, e^{-\epsilon(\theta,z)}  \right)\frac{d \theta}{2 \pi}  \,\,\,. 
\label{splittingtwo}
\ee
It is easy now to argue that the first term, for small but not zero value of $\beta$, is essentially proportional to $1/\beta$. Indeed, in the interval 
$(0, \Lambda(\beta))$, the pseudo-energy $\epsilon(\theta,z)$ is essentially constant while a term as $\beta \cosh\theta$ is significantly different from zero in an interval around $\Lambda(\beta)$. So, multiplying and dividing for $\beta$, for the first term in eq.\,(\ref{splittingtwo}) we have 
\be
\frac{1}{\beta}  \int_{0}^{+\Lambda} \, \beta \cosh \theta \log \left(  1 + z\, e^{-\epsilon(\theta,z)}  \right)\frac{d \theta}{2 \pi} 
\simeq \frac{1}{\beta} \log \left(  1 + z\, e^{-\epsilon_0(z)}\right) \,=\,\frac{\epsilon_0(z)}{\beta} \,\,\,,
\ee
where in the last step we have used the identity (\ref{trascendental}). Concerning the second term in (\ref{splittingtwo}), for $\beta \rightarrow 0$ it gives 
essentially a negligible contribution. 

Having established a $1/\beta$ behavior at small but finite value of $\beta$ in the logarithm of the partition function, the presence of this factor 
implies that the module of the zeros of $\Omega(z)$ satisfies the bounds 
\be 
\tilde\mu \, \beta \leq |z_i | \leq \, \tilde\nu\, \beta \,\,\,,
\ee 
(for some constans $\tilde\mu$ and $\tilde\nu$) and therefore for $\beta \rightarrow 0$, it vanishes, in agreement with the scaling law (\ref{bestfitt}). 

\subsection{Remarks}
\noindent
In the previous Sections we have found the Yang-Lee zeros of the Yang-Lee model by using the TSA and these zeros could be in principle used to reconstruct the free energy of the model inside the disk of the complex plane limited by their radius. But could we in principle find another set of zeros which allows us to define an analytic continuation of the free-energy in the entire complex plane of $z$? In other word, is there a way to implement the IPA for an interactive theory such as the 
Yang-Lee model? In practise, to implement the IPA is equivalent to firstly make the change of variable $\beta \rightarrow t $ in (\ref{partitionint}), where the variable $t$ is solution of the equation
\be 
t \,=\, - e^{\epsilon(\beta,z)} \,\,\,,
\label{tteqs}
\ee
and secondly to find the corresponding density of the solutions of this equation. If this procedure could be implemented, the logarithm of the partition function (\ref{partitionint}) could be formally put in the form (\ref{curvezerosss1}), i.e. an integral on the zeros of the theory. All these steps seem to follow closely the IPA 
implemented for the free theories (see Section \ref{IPAIkeda}) but for interactive theories there is an important difference: the variable $t$ that solves eq.(\ref{tteqs}) depends itself on $z$! This implies that, in the interactive case, in order to implement the IPA one is forced to introduce an infinite number of distributions of zeros, according to the values of $z$ for which we want to evaluate the free-energy. Although nothing wrong with this procedure, of course its implementation is not particularly practical. This essentially seems to leave the TSA as the most convenient method to identify the Yang-Lee zeros in interacting integrable models.   
 
\section{Conclusions}\label{finale}
\noindent
In this paper we have discussed the Yang-Lee zeros for a quantum integrable field theory particularly simple, the Yang-Lee model. To compute such zeros, 
we have relied on the exact expression of the free-energy of the model expressed by the Thermodynamics Bethe Ansatz equations based on the 
exact $S$-matrix of the theory. We have used in particular the so-called Truncated Series approach to define a sequence of polynomials, each of them  
regarded as  an approximation of the exact partition function of the model at a given order $N$ in the fugacity $z$: increasing $N$, the zeros of these polynomials identify the Yang-Lee zeros of the model. As a matter of fact, they are distributed around approximate circles whose radius depends exponentially on the inverse temperature $\beta$. This feature may be also present in other integrable quantum field theories and it would be interesting to confirm or disprove this fact. 

In addition to satisfy the curiosity to see, at least in one explicit example, where the Yang-Lee zeros are located for quantum integrable field theories, in our opinion 
there are other interesting questions that come out from our analysis and which deserve future attention. The most important one concerns the possibility 
to invert the logic behind the theory of Yang-Lee zeros, namely to see whether it would be possible to use the distributions of zeros for defining 
well defined statistical models and what properties they would have. Is any distribution of zeros which leads to positive polynomials a good one? 
For instance, what is (if any) the statistical model behind the partition function (\ref{primepoly}) based on the prime numbers? Is it a local or non-local model? 
We denote such a kind of questions the {\em inverse Yang-Lee problem}.  

To make any progress in such a program, one shall be aware however of another important aspect: beside some very special cases, as for instance the one analysed in \citep{Abe2}, the knowledge of the Yang-Lee zeros alone  is of course not enough to address other important properties of the model, as for instance, the computation of its correlation functions. In light of this remark, as a first important step forward, it would be extremely interesting to identify what is the minimum set of quantities (in addition to the density of the Yang-Lee zeros) which would allow -- at least in principle -- to successfully deal with the 
inverse Yang-Lee problem.

\vspace{1cm}
\begin{flushleft}\large
\textbf{Acknowledgements}
\end{flushleft}

One of us (GM) is grateful to Don Zagier for very useful discussion and also acknowledges the Brazilian Ministry of Education (MEC) and the 
UFRN-FUNPEC for financial support during his visit to the International Institute of Physics in Natal, where this work was completed. GM also 
thanks Alvaro Ferraz and other members of IIP for the warm hospitality.

%would like to thank the International Institute of Physics of Natal, where this work started, for the warm hospitality.  

\newpage

\appendix

\section{Condensation of Yang-Lee zeros for a series expression}

\noindent 
In this Appendix we analyse the pattern of Yang-Lee zeros and, in particular, their condensation for a fictitious system 
whose grand-canonical partition function is given by the infinite series of alternating signs 
\be
\Omega(z) \,=\, 1 + \sum_{k=1}^\infty (-1)^{k+1} z^k \,\,\,.
\label{putative}
\ee
\begin{figure}[b]
\includegraphics[width=0.4\textwidth]{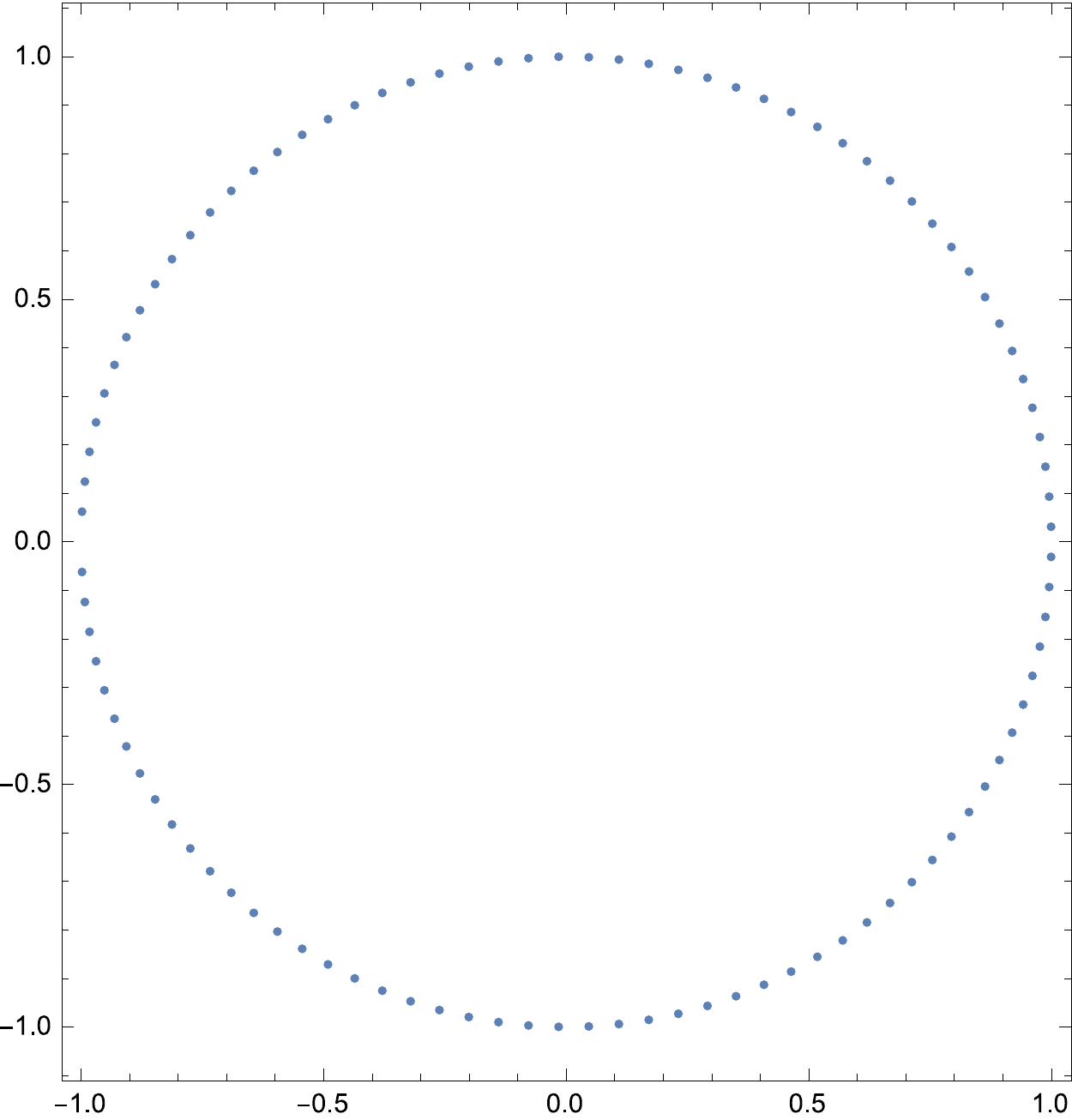}
\hspace{10 mm}
\includegraphics[width=0.4\textwidth]{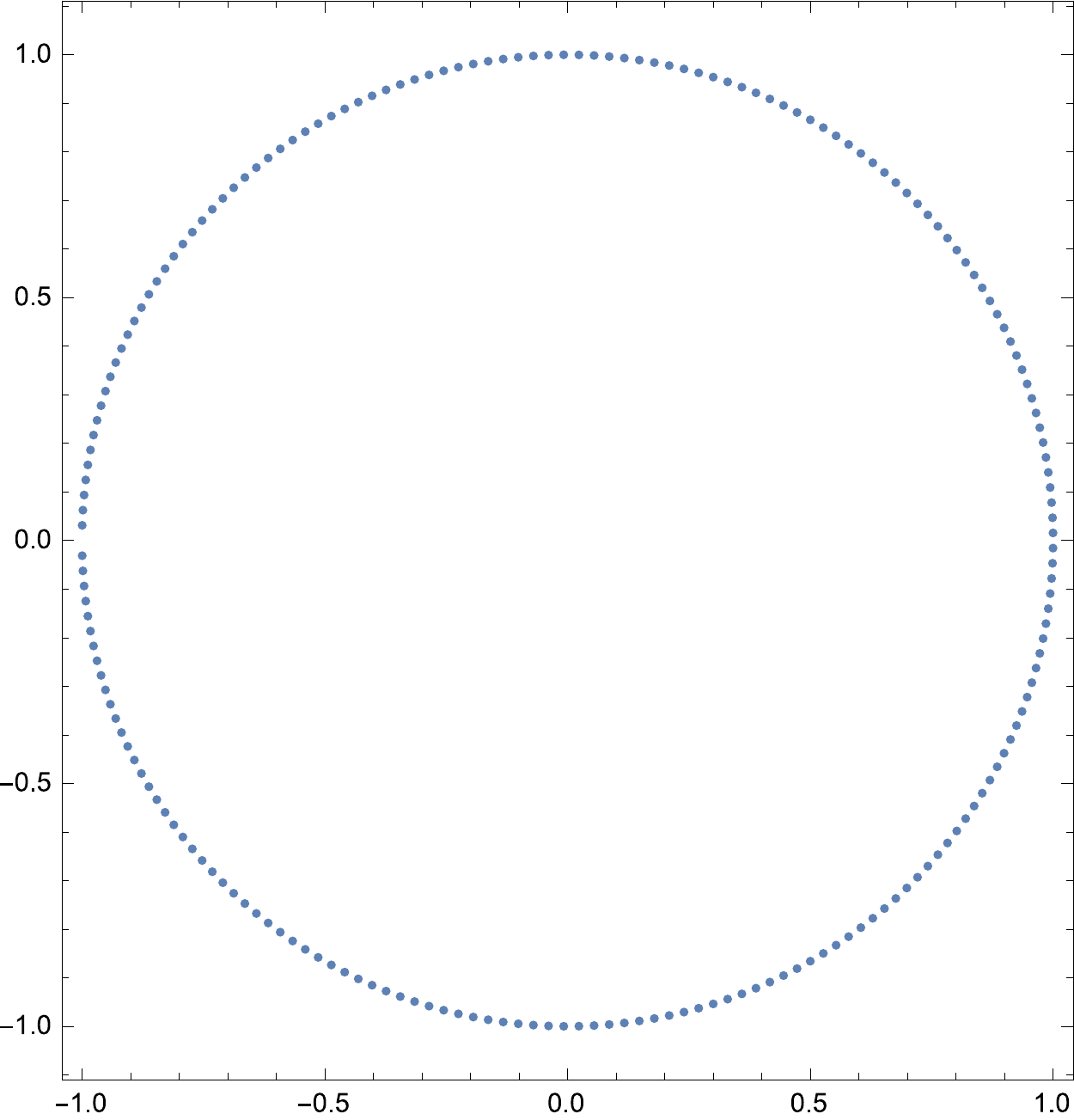}
\caption{Distribution of zeros of the polynomial $Y_N(z)$ for $N=100$ (left hand side figure) and for $N=200$ (righ hand side figure). Notice that the zeros 
are becomig equally placed on the circle except in an small arc nearby $-1$. }
\label{zeroputative}
\end{figure}
This series converges absolutely for $ | z | < 1$ and its unique singularity is at $ z = -1$, as can be seen by the exact resummation of 
the series
\be
1 + \sum_{k=1}^\infty (-1)^{k+1} z^k  \,=\, \frac{1 + 2 z}{1 + z} \,\,\,,
\label{resummationputative}
\ee
which of course provides the analytic continuation of $\Omega(z)$ in all the complex plane of $z$ except at $z = -1$. Hence, the exact expression of the pressure 
$p(z)$ is given in terms of the function which provides the analytic continuation of the original expression (\ref{putative})  
\be
\frac{p(z)}{k T} \,=\,\frac{1}{V} \,\log\frac{1 + 2 z}{1 + z} \,\,\,. 
\label{pressurecorrect}
\ee
This expression is obviously not singular at $z = 1$, which is indeed a smooth point of the pressure. However, this is not the conclusion to which one arrives using 
a different kind of route. To see this, let's instead compute the pressure $p(z)$ in terms of the expression (\ref{pressurezeros}) which involves a sum on the Yang-Lee zeros of $\Omega(z)$. To this aim, let's approximate $\Omega(z) \rightarrow \hat\Omega_N(z)$ in terms of the sequence of polynomials 
\be
\hat \Omega_N(z) \,=\,1 +  \sum_{k=1}^N (-1)^{k+1} z^k \,=\, (1+ 2 z) \,\sum_{k=0}^N (-1)^{k} z^k \,\equiv (1 + 2 z) Y_N(z) 
\label{truncated}
\ee
obtained by truncating the series (\ref{putative}) to its first $N$ terms. It is now easy to see that, apart of an isolate zero at $z = -1/2$, all the remaining zeros of $\hat \Omega_N(z)$, alias of the polynomials $Y_N(z)$, are along the unit circle. Consider in fact the transformation $ z\rightarrow - z$ in the polynomial $Y_N(z)$: 
this produces the polynomial $Y_N(-z)$ whose coefficients are all positive. Applying to this polynomial the Enestr\"{o}m bounds (\ref{Kakeyabounds1}), we see that the zeros of $Y_N(-z)$ are on the unit circle, and in particular they become dense near $z=1$, as evident by their numerical determination shown in Figure \ref{zeroputative} for two different values of $N$. 

Imagine now to compute the pressure by summing on the isolate zero at $z=-1/2$ and on the $N$ zeros of $Y_N(z)$, according to eq.\,(\ref{pressurezeros}).
This provides an approximation of the actual pressure and let's call this function $\tilde p_N(z)$, defined by 
\be
\frac{\tilde p_N(z)}{k T} \,=\,\frac{1}{V} \left[\log (1 + 2 x) + \sum_{l=1}^N \log\left(1 - \frac{z}{z_l}\right) \right]\,\,\,.
\label{approximatesum}
\ee
\begin{figure}[t]
\includegraphics[width=0.4\textwidth]{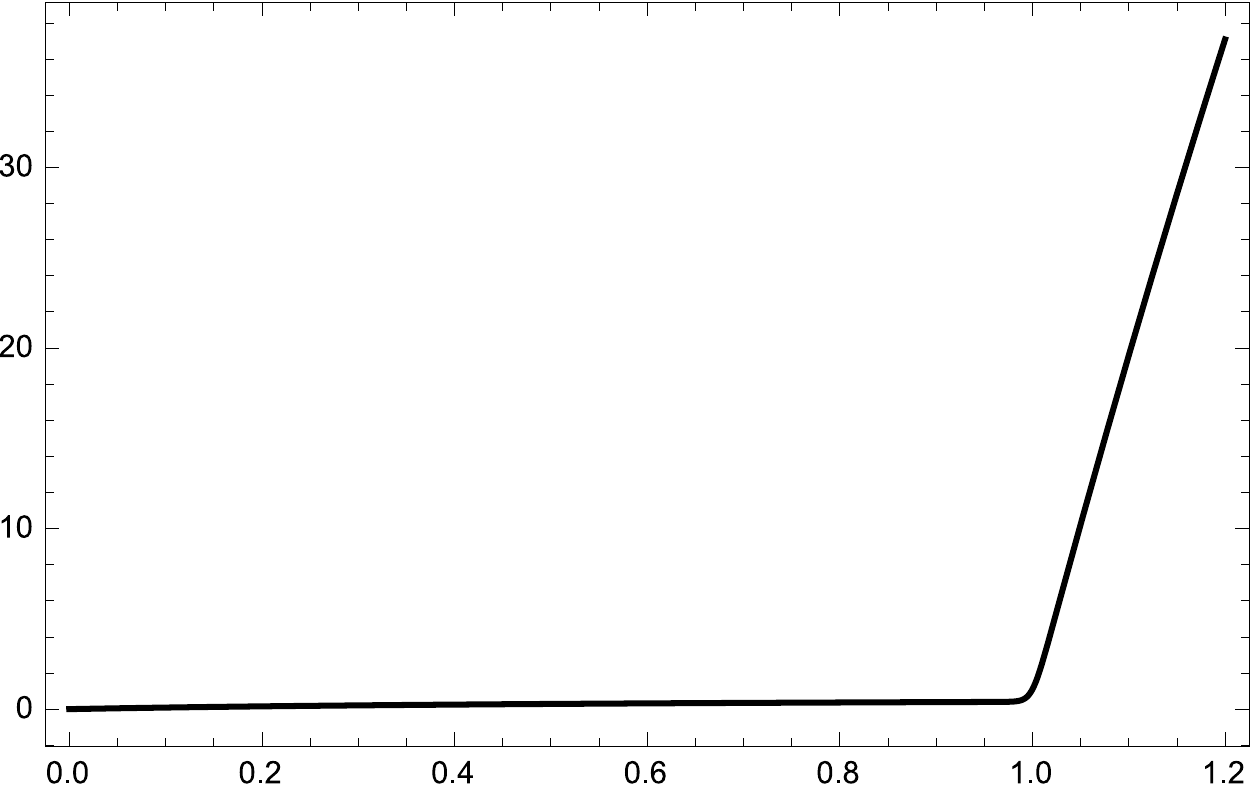}
\caption{Plot of the pressure $\tilde p_N(z)$ versus $z$ computed employing the first $N=100$zeros.}
\label{jumpder}
\end{figure}
Plotted versus $z$, for positive values of $z$, this function has a jump in its derivative at $z =1$ due to a finite density of the zeros nearby $z=1$, see Figure \ref{jumpder}.  While for $0 < z < 1$ summing on an increasing but finite number of zeros provides better and better approximation of the 
actual pressure $p(z)$ in this interval, the fact that we employ a finite number of zeros produces the unphysical jump of the approximate function $\tilde p_N(z)$. This behavior does not correspond to any phase transition but it is simple due to the finite radius of convergence of the original series (\ref{putative}). Proper physical behavior can be obtained only resumming the entire series. Notice that this resummation can be directly done using the density of zeros on the circle: indeed, using the cluster exponents coming from the expansion 
\be 
\log (1 + x) \,=\,\sum_{n=1}^\infty b_n \, z^n \,=\,\sum_{n=1}^\infty (-1)^{n+1}\frac{1}{n} \, z^n \,\,\,,
\ee 
we can get the density of zeros $\eta(\theta)$ by using the Fourier series 
\begin{eqnarray}
\eta(\theta) &\,=\,& \frac{1}{2 \pi} - \frac{1}{\pi} \sum_{n=1}^\infty n \,b_n \cos(n \theta)  \\
& \,= \,& \frac{1}{2 \pi} + \frac{1}{\pi} \sum_{n=1} (-1)^{n} \cos(n \theta) \,=\, \delta(\theta - \pi) \nonumber 
\end{eqnarray}
where we have used the Fourier transform of the periodic $\delta$-function 
\be
\delta(x - a) \,=\,\frac{1}{2\pi} + \frac{1}{\pi} \,\sum_{n=1}^\infty \cos(n (x - a))\,\,\,.
\ee
It is worth underlying that the infinite sum on the zeros has produced a density distribution for them which is peaked at $z = -1$ while is 
zero everywhere on the circle! Of course, quite the contrary of what happens at any finite $N$. Using now this exact distribution of the zeros and 
employing eq.\,(\ref{pressurezeroscircle}), one can easily see that the integral on the zeros gives rise to the correct expression (\ref{pressurezeroscircle}) of the pressure. 

Summarising: when the partition function $\Omega(z)$ is defined ab-initio by an infinite series, the truncation of this series at order $N$ 
produces a set of $N$ zeros which can be employed in the equation (\ref{pressurezeros}) to get the correct value of the pressure 
for values of $z$ {\em inside} the radius of convergence of the series. Outside the domain of convergence, the pressure may present 
a jump in its derivative at some positive value of $z$ which is not necessarily associated to a phase transition but it is rather a different manifestation of the radius of 
convergence of the original series. In absence of some physical principle which allows us to decide {\em a priori} on the nature of this jump,  
to resolve whether or not it is associated to an actual phase transition, one should in principle sum the series to get its analytic continuation and see the location of the actual singularities of the function.

\section{Plateau value of the Thermodynamics Bethe Ansatz equations}

In this Appendix we derive the plateau value $\epsilon_0(z)$ of the pseudo-energy by using the kink-solution of the Thermodynamics Bethe Ansatz equation and the value of the free energy at $\beta = 0$.  The kink solution emerges when $\beta \rightarrow 0$ and concerns the behavior of $\epsilon(\theta,z)$ nearby the edge $\theta \simeq \Lambda(\beta) = \log(2/\beta)$: it solves the non-linear integral equation 
\be
\epsilon^{kink}(\theta)\,=\, e^{\theta}-\frac{1}{2\pi} \int_{-\infty}^{\infty} d \theta '  \varphi (\theta-\theta ') \ln \left( 1+ z e^{-\epsilon^{kink}(\theta ')}  \right)\,,
\ee
and it allows to compute the free energy at $\beta =0$
\be
f(z)=-\frac{1}{\pi} \int_{-\infty}^{\infty}e^{\theta}  \ln \left( 1+ z e^{-\epsilon^{kink}(\theta ')}  \right)\,.
\label{fziniziale}
\ee
Using some standard steps (see for instance \citep{Fendley}), this expression can be also written as 
\be
f(z)\,=\,\frac{1}{2 \pi} \int_{\epsilon_0}^{\infty} d \epsilon \left[  \frac{\epsilon z e^{\epsilon}}{1+ z e^{-\epsilon}} + \ln \left( 1+ze^{-\epsilon} \right)\right].
\ee
The lower bound of the integral $\epsilon_0(z)$ is the value of the pseudo-energy at the plateau and for the Yang-Lee model it is solution of the non-linear equation 
\be
\epsilon_0(z) \,=\, \ln \left( 1+z e^{-\epsilon_0(z)} \right)\,\,\,.
\ee
If we define $x_0 \equiv e^{\epsilon_0}$, we obtain the equation:
\begin{eqnarray}
&&x_0 = 1+ \frac{z}{x_0}\nonumber\\
&&x_0^2-x_0-z=0\\
&&x_0= \frac{1 + \sqrt{1+4z}}{2}\nonumber
\end{eqnarray}
where we have taken the positive solution of the quadratic equation. We can now make a change of variable in the integral (\ref{fziniziale}) in terms of 
\be
y\,=\,\frac{z e^{-\epsilon}}{1+ze^{-\epsilon}}
\ee
so that the extrema of the integral become:
\begin{eqnarray}
y_-&=&\frac{z e^{-\epsilon_0}}{1+ze^{-\epsilon_0}}= \frac{z}{e^{\epsilon_0}+z}= \frac{z}{x_0+z}\,,\\
y_+&=&0 \,.
\end{eqnarray}
With this substitution we also have:
\be
\epsilon\,=\, \ln(1-y) - \ln y - \ln z \,\,\,,
\ee
so that
\be
d \epsilon = \left( \frac{1}{1-y}- \frac{1}{y} \right) dy= - \frac{dy}{y(1-y)}
\ee
Therefore 
\begin{eqnarray}
f(z)&=&\frac{1}{2 \pi} \int_{y_-}^0 \frac{d y}{y(1-y)} \left[ y \left( \ln(1-y)- \ln y - \ln z  \right) - \ln(1-y) \right]=\nonumber\\
&=& \frac{1}{2 \pi} \int_{y_-}^0 \frac{d y}{y(1-y)} \left[  (y-1) \ln(1-y) - y \ln y - y \ln z \right]=\nonumber\\
&=& \frac{1}{2 \pi} \left[ - \int_{y_-}^0 \frac{dy}{y} \ln(1-y) -\int_{y_-}^0 dy \frac{\ln y}{(1-y)} \right]- \frac{1}{2\pi} \int_{y_-}^0 \frac{\ln z}{(1-y)}dy=\nonumber\\
&=& -\frac{1}{ \pi} \left[ - \frac{1}{2}\int_0^{y_-} \frac{dy}{y} \ln(1-y) -\int^{y_-}_0 dy \frac{\ln y}{(1-y)} \right]- \frac{1}{2\pi} \int_{y_-}^0 \frac{\ln z}{(1-y)}dy
\end{eqnarray}
The function within the square parenthesis is the Rogers Dilogarithmic function ${\cal L}(y_-) = {\cal L} \left( z/(z+x_0) \right)$ and therefore 
\begin{eqnarray}
f(z)&=&-\frac{1}{\pi}  L \left(\frac{z}{z+x_0} \right)-\frac{1}{2\pi} \int_{y_-}^0 \frac{\ln z}{(1-y)}dy\nonumber\\
&=& -\frac{1}{\pi}  L \left(\frac{z}{z+x_0} \right)+\frac{1}{2 \pi} \ln z \left(\ln 
\left(1-y \right)\vert_{y_-}^0\right)= \nonumber\\
&=& -\frac{1}{\pi}  L \left(\frac{z}{z+e^{\epsilon_0}} \right)- \frac{1}{2 \pi} \ln (z) \ln\left(  \frac{x_0}{x_0+z} \right) = \nonumber \\
& = & -\frac{1}{\pi}  {\cal L} \left(\frac{z}{z+\frac{1 + \sqrt{1+4z}}{2}} \right)- \frac{1}{2 \pi} \ln (z) \ln\left(  \frac{\frac{1 + \sqrt{1+4z}}{2}}{\frac{1 + \sqrt{1+4z}}{2}+z} \right)\,.
\end{eqnarray}

\end{document}